\pdfoutput=1
\newcommand*{\ATLASLATEXPATH	}{latex/}
\documentclass[cernpreprint,UKenglish,texlive=2011,txfonts]{\ATLASLATEXPATH atlasdoc}

\usepackage{\ATLASLATEXPATH atlaspackage}
\usepackage{\ATLASLATEXPATH atlasbiblatex}
\usepackage{\ATLASLATEXPATH atlascontribute}
\usepackage{\ATLASLATEXPATH atlasphysics}

\usepackage{amsmath}
\usepackage{mathrsfs}
\usepackage{longtable}
\usepackage{caption}
\usepackage{mathrsfs}
\usepackage{afterpage}
\usepackage{amsmath}
\usepackage{url}
\usepackage{siunitx}

\graphicspath{{logos/}{figures/}}

\usepackage{latex/atlaspackage}
\usepackage{TTJetsPaper-def}
\AtlasTitle{Measurement of jet activity produced in top-quark events with an electron, a muon and two $b$-tagged jets in the final state in $pp$ collisions at $\sqrt{s}=13$ TeV with the ATLAS detector}

\author{The ATLAS Collaboration}

 \usepackage{authblk}
\author{The ATLAS Collaboration }

\PreprintIdNumber{CERN-EP-2016-218}

\AtlasJournalRef{Eur. Phys. J. C77 (2017) 220}
\AtlasDOI{doi:10.1140/epjc/s10052-017-4766-0}

\AtlasAbstract{%
Measurements of jet activity in top-quark pair events produced in proton--proton collisions are presented, using 3.2 fb$^{-1}$ of $pp$ collision data at a centre-of-mass energy of 13 TeV collected by the ATLAS experiment at the Large Hadron Collider. Events are chosen by requiring an opposite-charge $e\mu$ pair and two $b$-tagged jets in the final state. The normalised differential cross-sections of top-quark pair production are presented as functions of additional-jet multiplicity and transverse momentum, $p_{\text{T}}$. The fraction of signal events that do not contain additional jet activity in a given rapidity region, the gap fraction, is measured as a function of the $p_{\text{T}}$ threshold for additional jets, and is also presented for different invariant mass regions of the $e\mu b\bar{b}$ system.  All measurements are corrected for detector effects and presented as particle-level distributions compared to predictions with different theoretical approaches for QCD radiation. While the kinematics of the jets from top-quark decays are described well, the generators show differing levels of agreement with the measurements of observables that depend on the production of additional jets.}

\hypersetup{pdftitle={ATLAS draft},pdfauthor={The ATLAS Collaboration}}

\begin{document}
\maketitle
\tableofcontents
\newpage

\section{Introduction}
\label{sec:intro}
Top-quark pair production final states in proton--proton ($pp$) collisions at the Large Hadron Collider (LHC) often include additional jets not directly produced in the top-quark decays. The uncertainties associated with these processes are significant in precision measurements, such as the measurement of the top-quark mass\,\cite{Aaboud:2016igd} and the inclusive $t\bar{t}$ production cross-section\,\cite{Aaboud:2016pbd}.

These additional jets arise mainly from hard gluon emissions from the hard-scattering interaction beyond $t\bar{t}$ production and are described by quantum chromodynamics (QCD). The higher centre-of-mass energy of the $pp$ scattering process in LHC Run 2 opens a large kinematic phase space for QCD radiation. Several theoretical approaches are available to model the production of these jets in $t\bar{t}$ processes, including next-to-leading-order (NLO) QCD calculations, parton-shower models, and methods matching fixed-order QCD with the parton shower. The aim of this analysis
is to test the predictions of extra jet production in these approaches and to provide data to adjust free parameters of the models to optimise their predictions. 

The jet activity is measured in events with at least two $b$-tagged jets, i.e. jets tagged as containing $b$-hadrons, and exactly one electron and exactly one muon of opposite electrical charge in the final state. Additional jets are defined as jets produced in addition to the two $b$-tagged jets required for the event selection, without requiring any matching of jets to partons. In order to probe the \pt\ dependence of the hard-gluon emission, 
this analysis measures the normalised differential $t\bar{t}$ cross-sections as a function of the jet multiplicity for different transverse momentum (\pt) thresholds of the additional jets. The \pt\ of the leading additional jet is measured, as well as the \pt\ of the leading and sub-leading  jets initiated by $b$-quarks ("$b$-jets"),  which are top-quark decay products in most of the events. 

Furthermore, the gap fraction defined as the fraction of events with no jet activity in addition to the two $b$-tagged jets above a given \pt\ threshold in a rapidity region in the detector, is measured as a function of the additional jets' minimum \pt\ threshold as defined in Refs.~\cite{TOPQ-2011-21} and~\cite{TOPQ-2015-04}. The results are presented in a fiducial phase space in which all selected final-state objects are produced within the detector acceptance following the definitions in Ref.~\cite{TOPQ-2012-03}.

This paper provides a measurement of additional jets in \ttbar\ events in the dilepton channel for the new centre-of-mass energy of 13~\tev. Measurements similar to those presented in this paper were performed  by ATLAS at 7~\tev~\cite{TOPQ-2012-03, TOPQ-2011-21} and have been used to tune parameters in Monte Carlo (MC) generators for LHC Run 2~\cite{hdamp, ATL-PHYS-PUB-2015-011, ATL-PHYS-PUB-2015-002}. These earlier measurements were performed in the lepton+jets channel where the inclusive jet multiplicity was measured, since it is difficult to distinguish jets originating in $W$ decays from additional jets produced by QCD radiation. 
Recent measurements of jet multiplicity were performed in the single lepton channel by CMS at 13~\tev ~\cite{Khachatryan:2016mnb}  and
 in the dilepton channel, including also the gap fractions,  by ATLAS and CMS at 8~\tev\ ~\cite{TOPQ-2015-04,Khachatryan:2015mva}.

\section{ATLAS detector}
The ATLAS detector~\cite{PERF-2007-01} at the LHC covers nearly the entire solid angle\footnote{ATLAS uses a right-handed coordinate system with its origin at the nominal interaction point (IP) in the centre of the detector and the $z$-axis along the beam pipe. The $x$-axis points from the IP to the centre of the LHC ring, and the $y$-axis points upwards. Cylindrical coordinates $(r, \phi)$ are used in the transverse plane, $\phi$ being the azimuthal angle around the $z$-axis. The pseudorapidity is defined in terms of the polar angle $\theta$ as $\eta=-\ln\tan(\theta / 2)$. Angular distance is measured in units of $\Delta R = \sqrt{(\Delta\eta)^{2} + (\Delta\phi)^{2}}$.} around the interaction point. It consists of an inner tracking detector surrounded by a thin superconducting solenoid, electromagnetic and hadronic calorimeters, and a muon spectrometer incorporating three large superconducting toroid magnets. The inner-detector system is immersed in a \SI{2}{\tesla} axial magnetic field and provides charged-particle tracking in the range $|\eta| < 2.5$.

The high-granularity silicon pixel detector covers the interaction region and provides four measurements per track. The closest layer, known as the Insertable B-Layer (IBL)~\cite{Capeans:1291633}, was added in 2014 and provides high-resolution hits at small radius to improve the tracking performance. The pixel detector is followed by the silicon microstrip tracker, which provides four three-dimensional measurement points per track. These silicon detectors are complemented by the transition radiation tracker, which enables radially extended track reconstruction up to $|\eta| = 2.0$. The transition radiation tracker also provides electron identification information based on the fraction of hits (typically 30 in total) passing a higher charge threshold indicative of transition radiation.

The calorimeter system covers the pseudorapidity range $|\eta| < 4.9$. Within the region $|\eta|< 3.2$, electromagnetic calorimetry is provided by barrel and endcap high-granularity lead/liquid-argon (LAr) electromagnetic calorimeters, with an additional thin LAr presampler covering $|\eta| < 1.8$
to correct for energy loss in material upstream of the calorimeters. Hadronic calorimetry is provided by the steel/scintillator-tile calorimeter, segmented into three barrel structures within $|\eta| < 1.7$, and two copper/LAr hadronic endcap calorimeters. The solid angle coverage is completed with forward copper/LAr and tungsten/LAr calorimeter modules, which are optimised for electromagnetic and hadronic measurements, respectively.

The muon spectrometer comprises separate trigger and high-precision tracking chambers, measuring the deflection of muons in a magnetic field generated by superconducting air-core toroids. The precision chamber system surrounds the region $|\eta| < 2.7$ with three layers of monitored drift tubes, complemented by cathode strip chambers in the forward region, where the background is highest. The muon trigger system covers the range $|\eta| < 2.4$ with resistive plate chambers in the barrel, and thin-gap chambers in the endcap regions.

A two-level trigger system is used to select interesting events~\cite{PERF-2011-02, ATL-DAQ-PUB-2016-001}. The Level-1 trigger is implemented in hardware and uses a subset of detector information to reduce the event rate to a design value of at most \SI{100}{\kHz}. This is followed by the high-level software-based trigger (HLT), which reduces the event rate to \SI{1}{\kHz}.
\label{sec:detector}

\section{Data and simulation samples}
The proton--proton ($pp$) collision data used in this analysis were collected during 2015 by the ATLAS detector and correspond to an integrated luminosity of 3.2 fb$^{-1}$ at $\sqrt{s} = 13$~\tev. The data considered in this analysis were collected under stable beam conditions, requiring that all detectors were operational. Each selected event includes interactions from an average of 14 inelastic $pp$ collisions in the same proton bunch crossing, as well as residual signals from previous bunch crossings with a 25~ns bunch spacing. These two effects are collectively referred to as ``pile-up''. Events are required to pass a single-lepton trigger, either electron or muon. Multiple triggers are used to select events: either triggers with low lepton \pt\ thresholds of 24~\gev\ which utilise isolation requirements to reduce the trigger rate, or triggers with higher \pt\ thresholds  but looser isolation requirements to increase event acceptance. The higher \pt\ thresholds were  50~\gev\ for muons and 60~\gev\ or 120~\gev\ for electrons.

MC simulations are used to model background processes and to correct the data for detector acceptance and resolution effects. The nominal \ttbar\ sample is simulated using the NLO \pow-Box v2 matrix-element (ME)  generator \cite{powheg1,powheg2,powheg3}, referred to as \pow\ in the following, 
and \py6~\cite{pythia6} (v6.427) for the parton shower (PS), hadronisation and underlying event. \pow\ is interfaced to the CT10~\cite{CT10} NLO parton distribution function (PDF) set, while \py6 uses the CTEQ6L1 PDF set~\cite{CT6.1}. \py\ simulates the underlying event and parton shower using the P2012 set of tuned parameters (tune)~\cite{perugia2012}. The ``$h_{\text{damp}}$'' parameter, which controls the \pt\ of the first
additional emission beyond the Born configuration, is set to the mass of the top quark ($m_{t}$). The main effect of this is to regulate the high-\pt\ emission against which the \ttbar system recoils. The choice of this $h_{\text{damp}}$ value has been found to improve the modelling of the \ttbar\ system kinematics with respect to data in previous analyses~\cite{hdamp}. In order to investigate the effects of initial- and final-state radiation, alternative \pps\ samples are generated with the renormalisation and factorisation scales varied by a factor of 2 (0.5) and using low (high) radiation variations of the
Perugia 2012 tune and an $h_{\text{damp}}$ value of $m_{t}$ ($2m_{t}$), corresponding to less (more) parton-shower radiation~\cite{hdamp}. These samples are called RadHi and RadLo in the following. These variations are selected to cover the uncertainties in the measurements of differential distributions in 7~\tev\ data~\cite{Aad:2014zka}. Alternative samples are generated using \pow\ and \mcnlolong\,\cite{amcatnlo} (v2.2.1) with CKKW-L, 
referred to as \mcnlo\ hereafter, both interfaced to \hw~\cite{herwig} (v2.7.1), in order to estimate the effects of the choice of matrix-element generator. These \ttbar\ samples are described in Ref.~\cite{hdamp}.

Additional \ttbar\ samples are generated for comparisons with unfolded data as follows. The predictions of the ME generators \pow\ and \mcnlo\ are interfaced to \hws~\cite{herwig, Herwig7_2} and \pye. In all \pow\ and \mcnlo\ samples mentioned above, the first emission is calculated from the leading-order real emission term, and further additional jets are simulated from parton showering, which is affected by significant theoretical uncertainties. Improved precision is expected from using \sherpa\ v2.2~\cite{Sherpa}, which models the inclusive and the one-additional-jet process using an NLO matrix element and up to four additional jets at leading-order (LO) accuracy using the ME + PS@NLO prescription~\cite{Hoeche:2012yf}. The sample used to compare to particle-level results presented here is generated with the central scale set to $\mu^2 = m_t ^2 + 0.5 \times (p_{\mathrm {T},t}^2 + p_{{\mathrm {T},\overline{t}}}^2)$, where $p_{\mathrm {T},t}$ and $p_{\mathrm {T},\overline{t}}$ refer to the \pt\ of the top and antitop quark, respectively, and with the matching scale set to 30~\gev. Furthermore, the NNPDF 3.0 PDF\,\cite{Ball:2014uwa} at next-to-next-to-leading order (NNLO) is used.

All $t\bar{t}$ samples are normalised to the cross-section calculated with the Top++2.0 program to NNLO in perturbative QCD, including soft-gluon resummation to NNLL~\cite{Czakon:2011xx}, assuming a top-quark mass of 172.5~\gev. 

Background processes are simulated using a variety of MC generators, as described below. Details of the background estimation are described in Section~\ref{sec:selection}. Single top-quark production in association with a $W$ boson ($Wt$) is simulated using \pow-Box v1+\py6 with the same parameters and PDF sets as those used for the nominal \ttbar\ sample and is normalised to the approximate NNLO cross-section ($71.7\pm 3.8$ pb) described in Ref.~\cite{Kidonakis:2010ux}. At NLO, part of the final state of $Wt$ production is identical to the final state of \ttbar\ production. The ``diagram removal'' (DR) generation scheme~\cite{diagram_subtraction} is used to remove this part of the phase space from the background calculation. 
A sample generated using an alternative ``diagram subtraction'' (DS) method~\cite{diagram_subtraction} is used to evaluate systematic uncertainties. Both samples are normalised to the generator cross-section.

The majority of backgrounds with at least one misidentified lepton in the selected sample arise from \ttbar production in which only one of the top quarks
decays semileptonically, which is simulated in the same way as the \ttbar production in which both top quarks decay leptonically.

\sherpa\ v2.1, interfaced to the CT10 PDF set, is used to model Drell--Yan production, specifically $Z/\gamma^*\rightarrow\tau^+\tau^-$. For this process, \sherpa\ calculates matrix elements at NLO for up to two partons and at LO for up to four partons using the OpenLoops~\cite{openloops} and Comix~\cite{comix} matrix-element generators. The matrix elements are merged with the \sherpa\ PS~\cite{Schumann:2007mg} using the ME + PS@NLO prescription~\cite{sherpanlo}. The total cross-section is normalised to NNLO predictions calculated using the FEWZ program~\cite{Anastasiou:2003ds} with the MSTW2008NNLO PDF \cite{Martin:2009iq}. \sherpa\ v2.1 with the CT10 PDF set is also used to simulate electroweak diboson production~\cite{ATL-PHYS-PUB-2016-002} ($WW$, $WZ$, $ZZ$), where both bosons decay leptonically. For diboson production, \sherpa\ v2.1 calculates matrix elements at NLO for zero additional partons, at LO for one to three additional partons (with the exception of $ZZ$ production, for which the one additional parton is also NLO), and using PS for all parton multiplicities of four or more. 

The ATLAS detector response is simulated~\cite{fullsim} using \GEANT4~\cite{Agostinelli:2002hh}. A ``fast simulation''~\cite{ATLAS:1300517}, utilising parameterised showers in the calorimeter, is used in the samples chosen to estimate \ttbar\ modelling uncertainties. Additional $ pp$ interactions are generated using \py8.186~\cite{pythia8} with tune A2 and overlaid with signal and background processes in order to simulate the effect of pile-up. The MC simulations are reweighted to match the distribution of the average number of interactions per bunch crossing that are observed in data, referred to as ``pile-up reweighting''. Corrections are applied to the MC simulation in order to improve agreement with data for the efficiencies of  reconstructed objects. The same reconstruction algorithms and analysis procedures are then applied to both data and MC simulation.

\label{sec:data_mc}
\label{sec:simulation}

\section{Object reconstruction}
This analysis selects reconstructed electrons, muons and jets. Electron candidates are identified by matching an inner-detector track to an isolated energy deposit in the electromagnetic calorimeter, within the fiducial region of transverse momentum $\pt>25$~\gev\ and pseudorapidity $|\eta|<2.47$. Electron candidates are excluded if the energy cluster is within the transition region between the barrel and the endcap of the electromagnetic calorimeter, $1.37 < |\eta| < 1.52$, and if they are also reconstructed as photons. Electrons are selected using a multivariate algorithm and are required to satisfy a likelihood-based quality criterion, in order to provide high efficiency and good rejection of fake and non-prompt electrons~\cite{Aad:2014fxa,ATLAS-CONF-2014-032}. Electron candidates must have tracks that pass the requirements of transverse impact parameter significance\footnote{The transverse impact parameter significance
is defined as $d_0^\text{sig} = d_0 / \sigma_{d_0}$, where $\sigma_{d_0}$ is the uncertainty in the transverse impact parameter $d_0$.} $|d_0^\text{sig}|<5$ and longitudinal impact parameter $|z_0 \sin\theta| < 0.5$~mm. Electrons must also pass isolation requirements based on inner-detector tracks and topological energy clusters varying as a function of $\eta$ and \pt. The track isolation cone size is given by the smaller of $\Delta R = 10$ \gev /\pt\ and $\Delta R = 0.2$, i.e. a cone which increases in size at lower \pt values, up to a maximum of 0.2. These requirements result in a 95\% efficiency of the isolation cuts for electrons from $Z\rightarrow e^+e^-$ decays with \pt\ of 25~\gev\ and 99\% for electrons with \pt\ above 60~\gev; when estimated in simulated \ttbar\ events, this efficiency is smaller by a few percent, due to the increased jet activity. Electrons that share a track with a muon are discarded. Double counting of electron energy deposits as jets is prevented by removing the closest jet with an angular
 distance  $\Delta R < 0.2$ from a reconstructed
 electron. Following this, the electron is discarded if a jet exists within $\Delta R < 0.4$ of the electron, to ensure sufficient separation from nearby jet activity.

Muon candidates are identified from a track in the inner detector matching a track in the muon spectrometer; the combined track is required to have $\pt > 25$~\gev\ and $|\eta| < 2.5$~\cite{Aad:2016jkr}. The tracks of muon candidates are required to have a transverse impact parameter significance $|d_0^\text{sig}|<3$ and a longitudinal impact parameter below 
0.5~mm. Muons are required to meet quality criteria and the same isolation requirement as applied to electrons, to obtain the same isolation efficiency performance as for electrons. These requirements reduce the contributions from fake and non-prompt muons. Muons may leave energy deposits in the calorimeter that could be misidentified as a jet, so jets with fewer than three associated tracks are removed if they are within $\Delta R < 0.4$
of a muon. Muons are discarded if they are separated from the nearest jet by $\Delta R < 0.4$, to reduce the background from muons originating in heavy-flavour decays inside jets.

Jets are reconstructed with the anti-$k_t$ algorithm~\cite{Cacciari:2005hq,Cacciari:2008gp}, using a radius parameter of $R = 0.4$, from topological clusters of energy deposits in the calorimeters. Jets are accepted within the range $\pt > 25$~\gev\ and $|\eta| < 2.5$, and are calibrated using simulation with corrections derived from data~\cite{ATL-PHYS-PUB-2015-015}. Jets likely to originate from pile-up are suppressed using a multivariate jet-vertex-tagger (JVT)~\cite{Aad:2015ina} for candidates with $\pt < 60$~\gev\ and $|\eta| < 2.4$. Jets containing $b$-hadrons are $b$-tagged using a multivariate discriminant~\cite{ATL-PHYS-PUB-2015-022}, which uses track impact parameters, track invariant mass, track multiplicity and secondary vertex information to discriminate those jets from light quark or gluon jets (``light jets''). The average $b$-tagging efficiency is 77\% for $b$-jets in simulated dileptonic \ttbar events with a purity of 95\%. The tagging algorithm gives a rejection factor of about 130 against light jets and about 4.5 against jets originating from charm quarks (``charm jets'').

\label{sec:object}

\section{Event selection and background estimates}
Signal events are selected by requiring exactly one electron and one muon of opposite electric charge (``opposite sign''), and at least two $b$-tagged jets. With this selection, almost all of the selected events are \ttbar\ events. The other processes that pass the signal selection are events with single top quarks ($Wt$), \ttbar\ events in the single-lepton decay channel with a misidentified (fake) lepton, $Z/\gamma^{*}\rightarrow\tau^{+}\tau^{-}(\rightarrow e\mu)$ and diboson events. Other backgrounds, including processes with two misidentified leptons, are negligible for the event selections used in this analysis.

Additional jets are defined as those produced in addition to the two highest-\pt\ $b$-tagged jets. They are identified as jets above \pt\ thresholds of 25~\gev, 40~\gev, 60~\gev\ and 80~\gev, independent of the jet flavour. In very rare cases, $b$-jets may also be produced in addition to the top-quark pair, for example through splitting of a very high momentum gluon, or through the decay of a Higgs boson into a bottom--antibottom pair, leading to events with more than two $b$-tagged jets. In this case, the two selected $b$-tagged jets with the highest \pt\ are assumed to originate from $t\bar{t}$ decay, and the others are considered as additional jets. This procedure ignores that occasionally a $b$-jet which is not the decay product of a top quark might have higher \pt\ than those from the top-quark decays. This is a negligible effect within the uncertainties of this measurement.

The single-top background is estimated from simulation, as described in Section~\ref{sec:simulation}. The background from \ttbar\ events in the lepton+jets channel with a fake lepton is estimated from a combination of data and simulation, as in Ref.\,\cite{Aaboud:2016pbd}. This method uses the observation that samples with a same-sign \emu\ pair and two $b$-tagged jets are dominated by events with a misidentified lepton, with a rate comparable to those in the opposite-sign sample. The contributions of events with misidentified leptons are therefore estimated as same-sign event counts in data, after subtraction of predicted prompt same-sign contributions multiplied by the ratio of opposite-sign to same-sign fake leptons, as predicted from the nominal \ttbar sample.

The backgrounds from $Z/\gamma^{*}\rightarrow\tau^{+}\tau^{-}$ and from diboson events are estimated from simulation and are below 1\%. The normalisation for the $Z/\gamma^{*}\rightarrow\tau^{+}\tau^{-}$ contribution is estimated from events with $Z/\gamma^{*}\rightarrow e^+ e^- $ or \mumu and two \bjet s within the acceptance of this analysis. The Monte Carlo prediction is scaled by $1.37\pm 0.30$ to fit the observed rate. 

\begin{figure}
\centering
\begin{tabular}{cc}
\subfloat[\label{fig:emu_electron_pt_1}]{\includegraphics[width=0.4\textwidth]{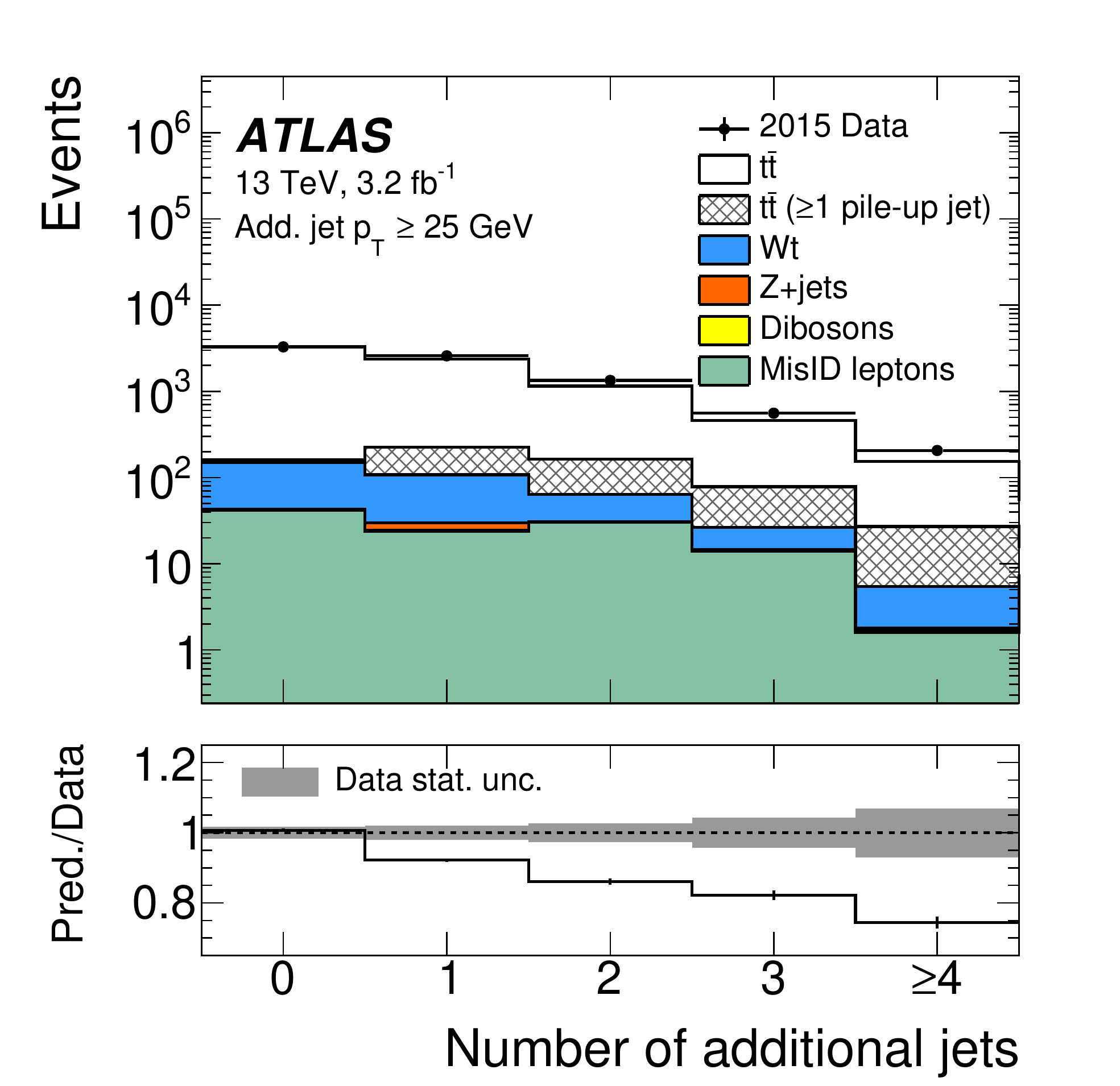}}&
\subfloat[\label{fig:emu_electron_pt_2}]{\includegraphics[width=0.4\textwidth]{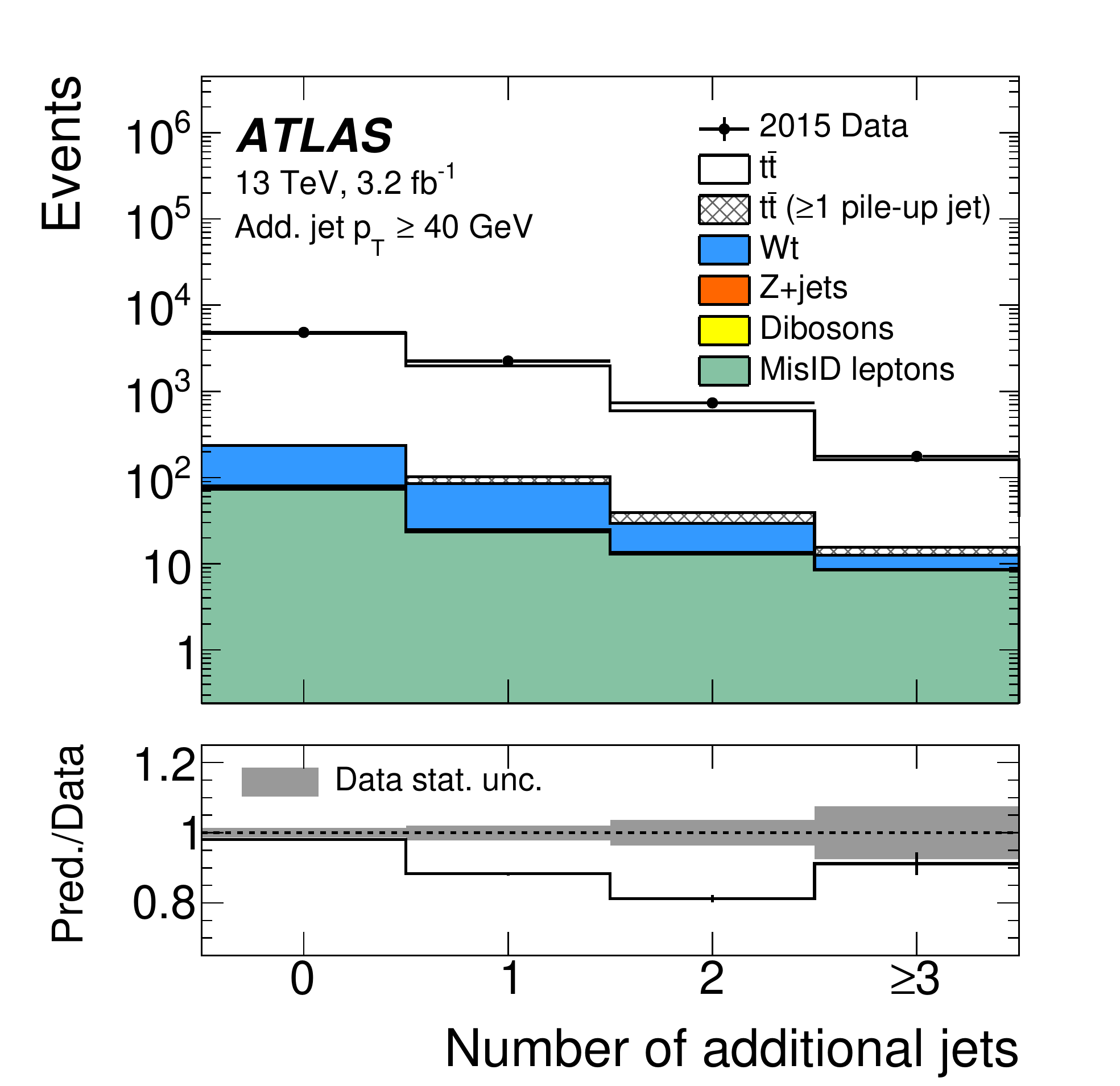}}\\
\subfloat[\label{fig:emu_electron_pt_3}]{\includegraphics[width=0.4\textwidth]{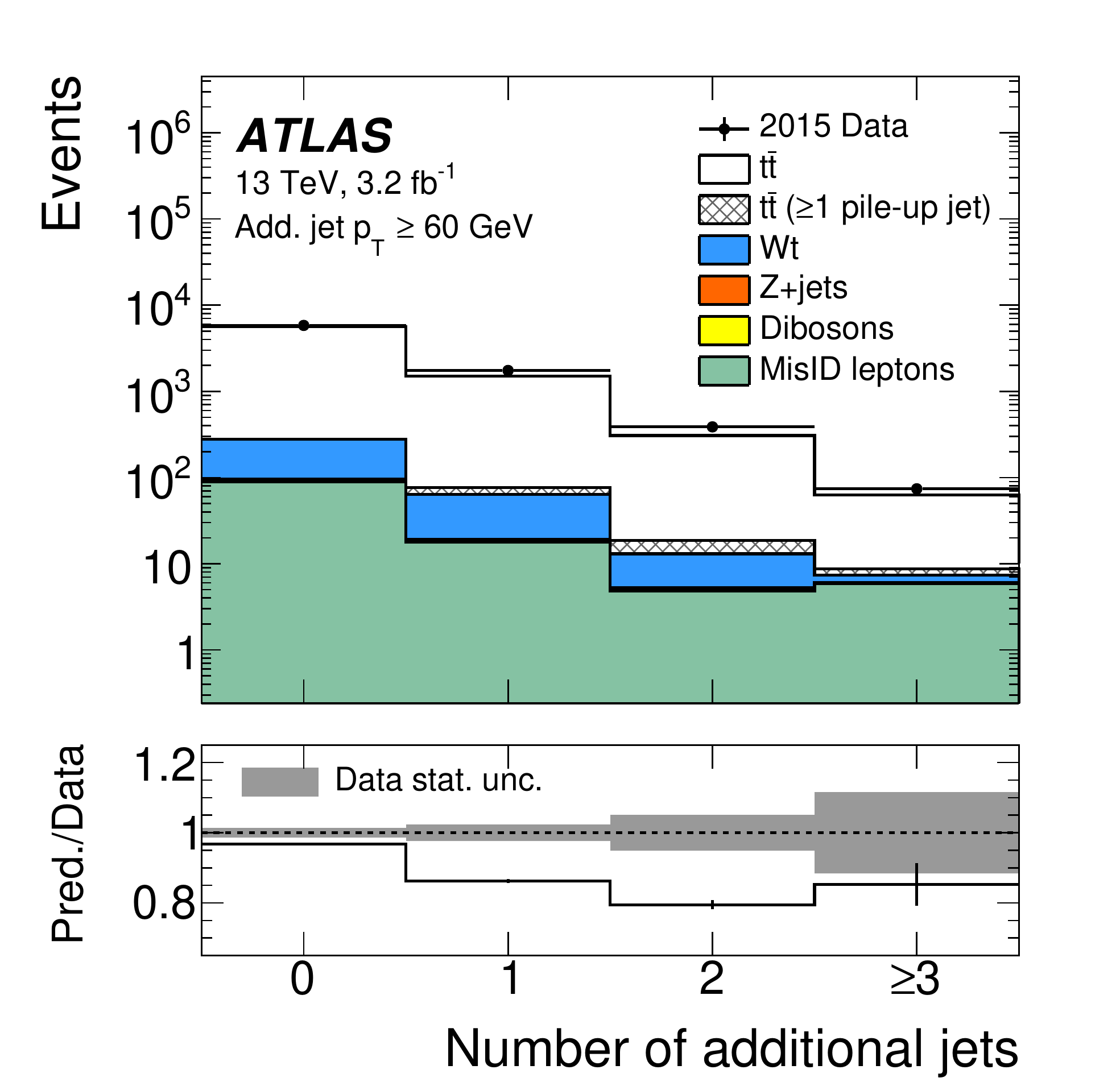}}&
\subfloat[\label{fig:emu_electron_pt_4}]{\includegraphics[width=0.4\textwidth]{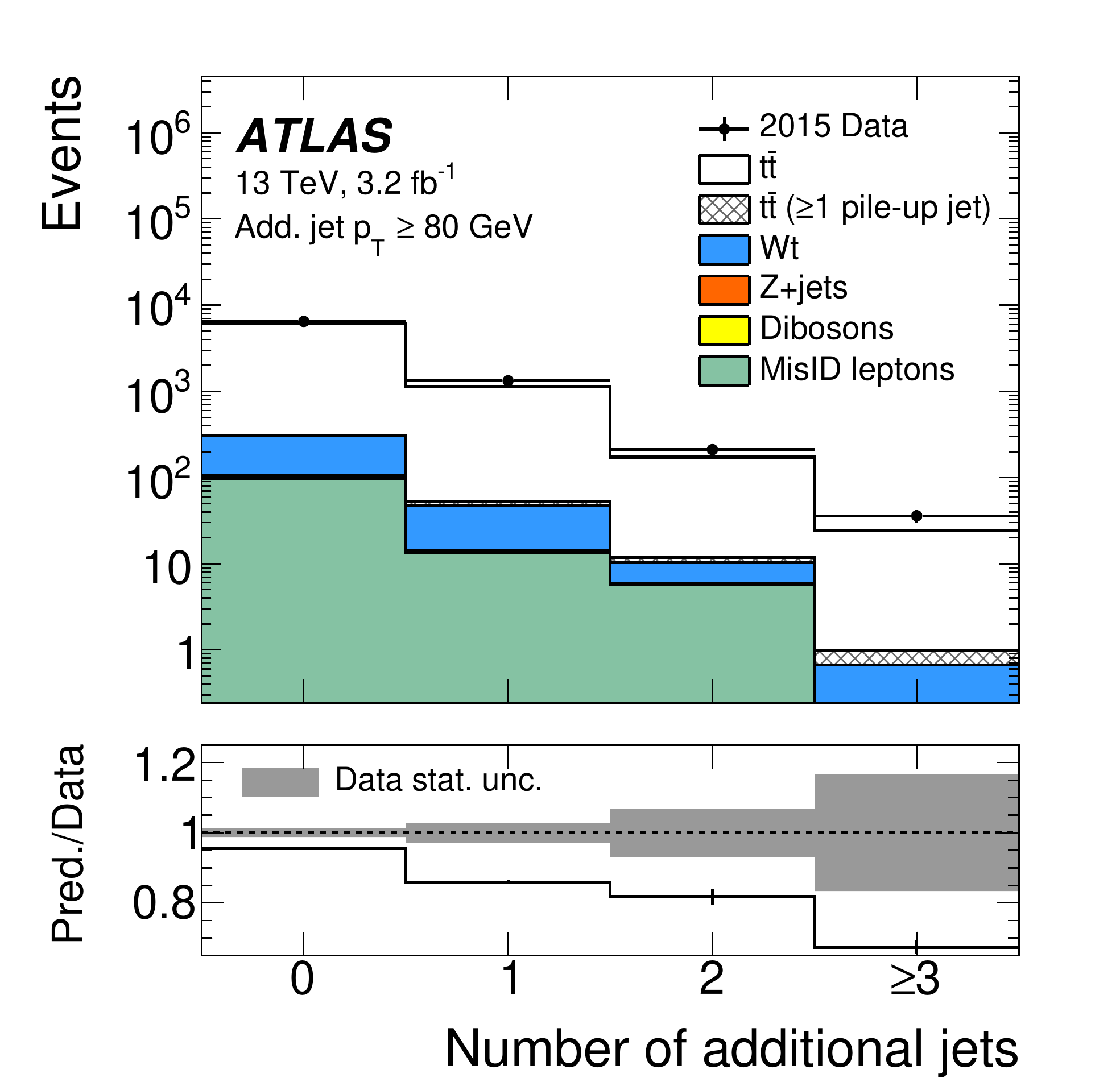}}\\
\end{tabular}
\caption{Multiplicity of additional jets with (a) \pt $>$ 25\,\gev, (b) \pt $>$ 40\,\gev, (c) \pt $>$ 60\,\gev, and (d) \pt $>$ 80\,\gev\ for selected events at reconstruction level in data and simulation. Simulated signal events with at least one additional jet identified as pile-up are indicated in grey. The contribution of pile-up jets to the backgrounds is negligible. The lower panel shows the ratio of the total prediction to the data (solid line), the grey band represents the statistical uncertainty of the measurement, and the error bars on the solid line show the statistical uncertainty in the signal MC sample.}
\label{fig:reconjet}
\end{figure}

\begin{figure}
\centering
\begin{tabular}{cc}
\subfloat[\label{fig:emu_electron_pt_5}]{\includegraphics[width=0.4\textwidth]{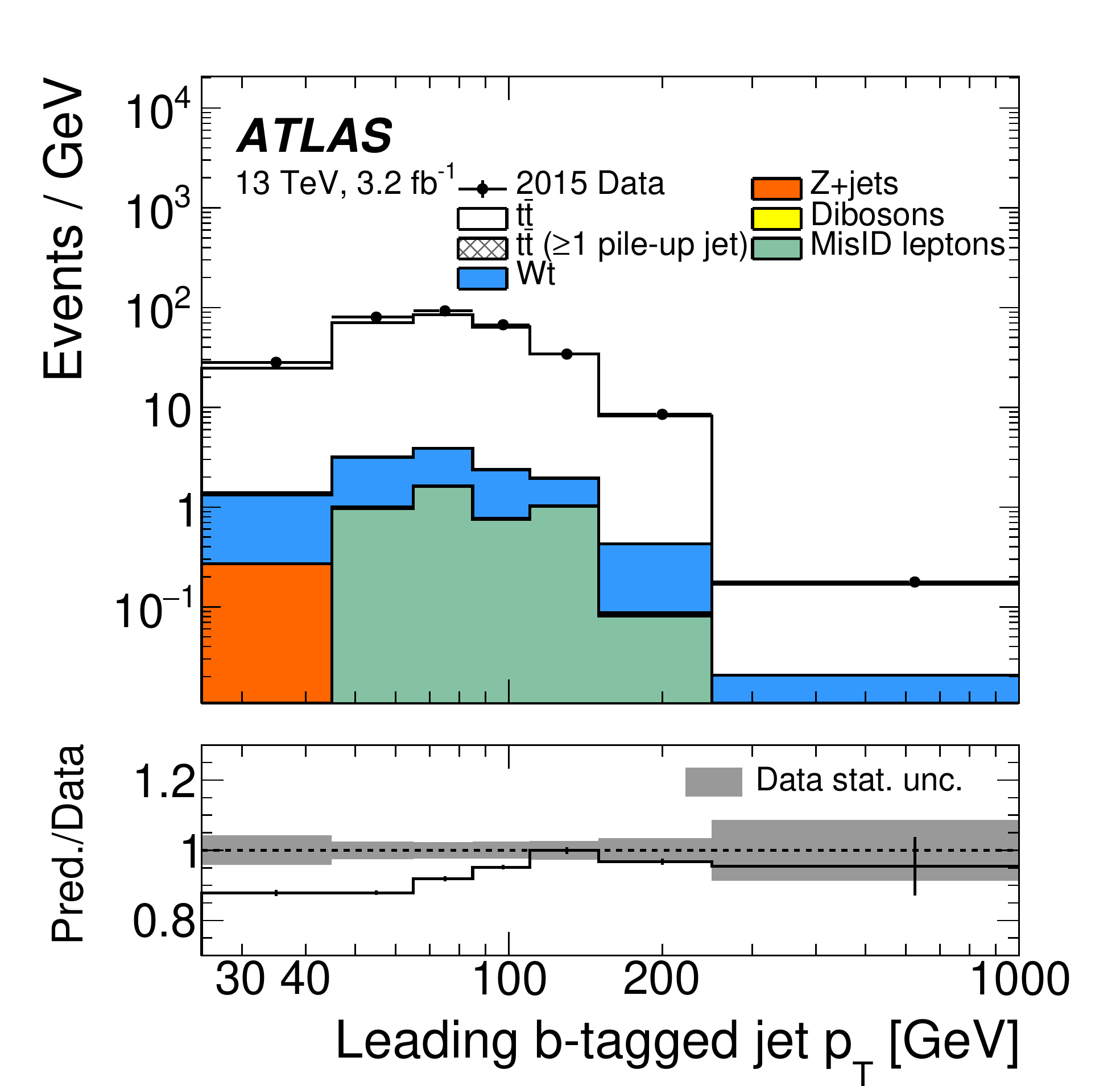}} &
\subfloat[\label{fig:emu_electron_pt_6}]{\includegraphics[width=0.4\textwidth]{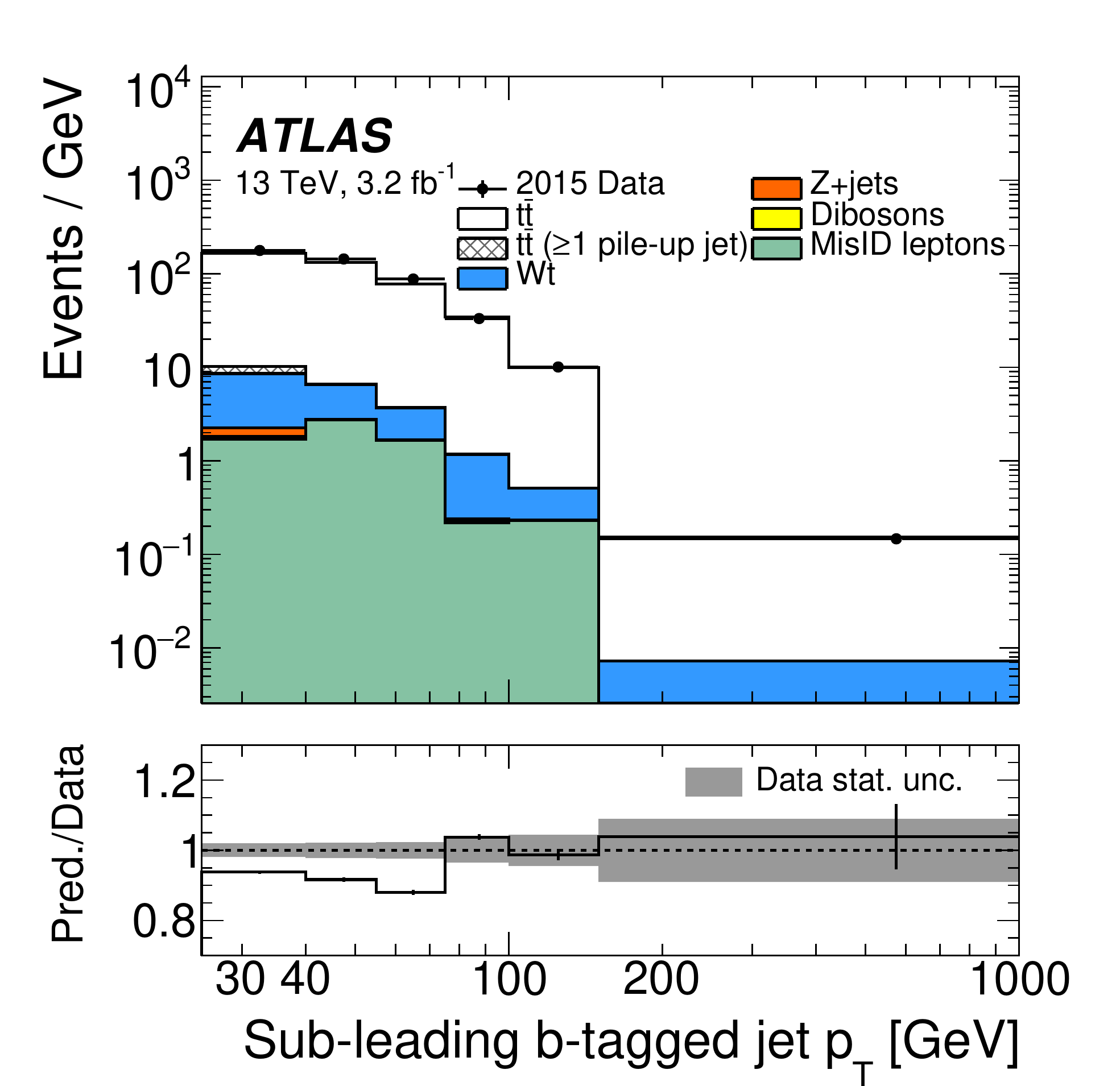}} 
\end{tabular}
\begin{tabular}{c}
\subfloat[\label{fig:emu_electron_pt_7}]{\includegraphics[width=0.4\textwidth]{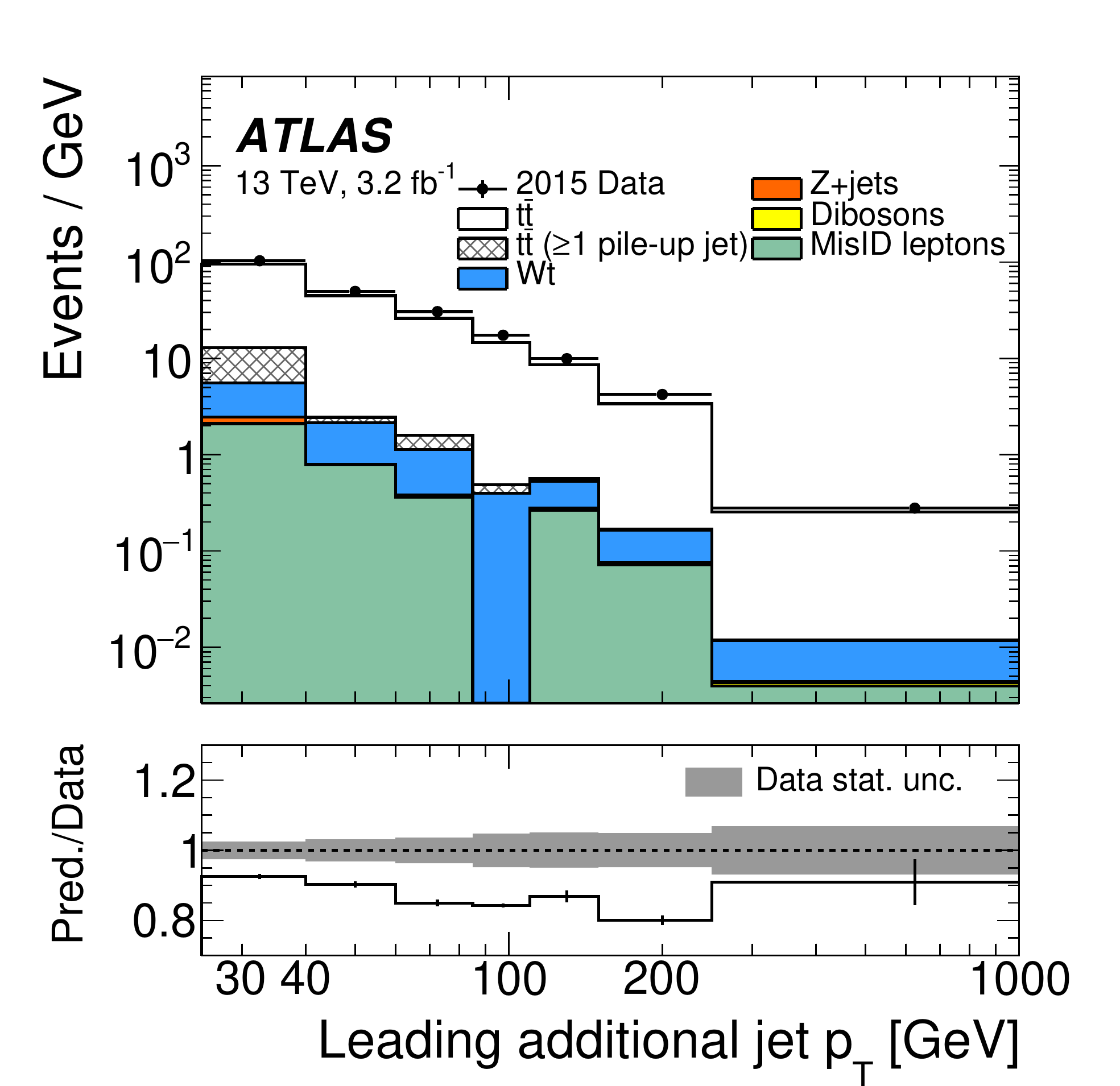}}\\
\end{tabular}
\caption{ (a) Leading \btagjet\ \pt, (b) sub-leading \btagjet\ \pt, and (c) leading additional-jet \pt\ for selected events at reconstruction level. The last bin includes overflows. Jets identified as pile-up in the \ttbar\ signal sample are indicated in grey. The contribution of pile-up jets to the backgrounds is negligible. The lower panel shows the ratio of the total prediction to the data (solid line), the grey band represents the statistical uncertainty of the measurement, and the error bars on the solid line shows the statistical uncertainty in the signal MC sample.}
\label{fig:recopt}
\end{figure}

After the event selection, only about 4.5\% of the events are background, as listed in Table~\ref{tab:yield_total}. The background is dominated by single top production (3.1\%) and fake leptons (1.6\%). The event yields and the relative background contributions vary with jet multiplicity and jet \pt\ as shown in Figures~\ref{fig:reconjet} and ~\ref{fig:recopt}, respectively. The single-top background dominates across all jet \pt\ values and at low additional jet multiplicities. At high jet multiplicities ($\geq 3$ additional jets) the fake-lepton background exceeds the number of single-top events. While the number of events observed in the 0-jet bin agrees with the prediction within the uncertainties, the data exceed the predictions increasingly with jet multiplicity, reaching a 25\% deviation for events with at least four additional jets above 25~\gev. 

The table and figures also list the contribution of \ttbar\ events with at least one additional jet identified as originating from pile-up (pile-up jets). These are signal events, but a few pile-up jets are still in the sample after object and event selection, as the background suppression of the JVT cut is very high but not 100\%. Due to the presence of at least one jet that does not originate from the hard interaction, these events may appear in the wrong jet
multiplicity bin. In the jet \pt\ spectra, pile-up jets contribute at low additional-jet \pt\ as the pile-up jets are generally softer than the jets in \ttbar\ events. For the same reason, pile-up jets only contribute significantly to the jet multiplicity distributions with the 25\,\gev\ threshold. In most of the events with remaining pile-up jets, only one of the additional jets is caused by pile-up. Any remaining pile-up jets can be identified in the simulation, but not in data. Therefore the data are corrected for pile-up jets in the unfolding procedure, as described later. 
\begin{table}[ht!]
\centering
\setlength\tabcolsep{1.5pt}
\begin{tabular}{|lrll|}
\hline
Process                                          & \multicolumn{3}{l|}{Yield}                   \\ \hline
\multicolumn{1}{|l|}{Single top ($Wt$)}          & 236  & $\pm 2$ (stat.)   & $\pm 46$ (syst.)  \\
\multicolumn{1}{|l|}{Fake leptons}               & 117  & $\pm 22$ (stat.)  & $\pm 120$ (syst.) \\
\multicolumn{1}{|l|}{$Z$+jets}                   & 6    & $\pm 3$ (stat.)   & $\pm 1$ (syst.)   \\
\multicolumn{1}{|l|}{Dibosons}                   & 3.1  & $\pm 0.4$ (stat.) & $\pm 1.5$ (syst.) \\ \hline
\multicolumn{1}{|l|}{Total background}           & 362  & $\pm 22$ (stat.)  & $\pm 130$ (syst.) \\ \hline
\multicolumn{1}{|l|}{$tt$ ($\geq 1$ pile-up jet} & 310  & $\pm 2$ (stat.)   & $\pm 88$ (syst.)  \\
\multicolumn{1}{|l|}{$tt$ (no pile-up jets)}     & 6850 & $\pm 11$ (stat.)  & $\pm 940$ (syst.) \\ \hline
\multicolumn{1}{|l|}{\textbf{Expected}}                   & 7520 & $\pm 25$ (stat.)  & $\pm 950$ (syst.) \\
\multicolumn{1}{|l|}{\textbf{Observed}}                   & 8050 &                   &                   \\ \hline
\end{tabular}
\caption{Yields of data and MC events fulfilling the selection criteria.
\label{tab:yield_total}}
\end{table}

\label{sec:selection}
\clearpage

\section{Sources of systematic uncertainty}
The systematic uncertainties of the reconstructed objects, in the signal modelling and in the background estimates, are evaluated as described in the  following.

The jet energy scale (JES) uncertainty is evaluated by varying 19 uncertainty parameters derived from {\textit in situ} analyses at $\sqrt{s} = 8$~\tev\ and extrapolated to data at $\sqrt{s} = 13$~\tev~\cite{jesrun2}. The JES uncertainty is 5.5\% for jets with \pt\ of 25~\gev\ and quickly decreases with increasing jet \pt, falling to below 2\% for jets above 80~\gev. The uncertainty in the jet energy resolution (JER) is calculated by extrapolating the uncertainties derived at $\sqrt{s} = 8$~\tev\ to $\sqrt{s} = 13$~\tev~\cite{jesrun2}. The uncertainty in JER is at most $3.5\%$~at \pt\ of 25~\gev, quickly decreasing with increasing jet \pt\ to below $2\%$~for jets above 50~\gev.

Uncertainties on the efficiency for tagging b-jets were determined using the methods described in Ref.~\cite{Aad:2015ydr} applied to dileptonic ttbar events in $\sqrt{s} =13$~\tev\ data.
The uncertainties on mistagging of charm and light jets  were determined using $\sqrt{s}=8$~\tev\ data as described in Refs.~\cite{ATLAS-CONF-2014-004} and  \cite{ATLAS-CONF-2014-046}.  Additional uncertainties are assigned to take into 
account the presence of the new IBL detector and the extrapolation to  $\sqrt{s}=13$~\tev\  \cite{ATL-PHYS-PUB-2015-022}.

 The lepton-related uncertainties are assessed mostly using $Z\rightarrow\mumu$ and $Z\rightarrow e^+e^-$ decays measured in $\sqrt{s}=13$~\tev\ data. The differences between the topologies of $Z$ and \ttbar\ pair production events are expected not to be significant for the estimation of uncertainties. 

The uncertainty associated with the amount of QCD initial- and final-state radiation is evaluated as the difference between the baseline MC sample and the corresponding RadHi and RadLo samples described in Section\,\ref{sec:data_mc}. The uncertainty due to  the choice of parton-shower and hadronisation algorithms in the signal modelling is assessed by comparing the baseline MC sample (\powpy) with \powhw. The uncertainty due to the use of a specific NLO MC sample with its particular matching algorithm is derived from the comparison of \powhw\ to the \mcnlohw\ sample. 

The uncertainty due to the particular PDF used for the signal model prediction is evaluated by taking the standard deviation of variations from 100 eigenvectors of the recommended Run-2 PDF4LHC\,\cite{pdf4lhc} set and adding them in quadrature with the difference between the central predictions from CT10 and CT14~\cite{ct14}.

The uncertainty in the single top-quark background is evaluated based on the 5.3\% error in the approximate NNLO cross-section prediction and by comparing samples with diagram removal and diagram subtraction schemes, as described in Section\,\ref{sec:data_mc}. The uncertainty in the background from fake leptons is estimated to be 100\% from the statistical uncertainty of the same-sign event counts in data and an interpolation error using the envelope of the differences of individual subcomponents (such as photon-conversion, heavy-flavour decay leptons, for example) of misidentified lepton background between the same-sign and the opposite-sign sample.

For $Z$+jets backgrounds, the scale factor derived in the $e^+e^-$ and $\mumu$ channels and used to reweight the signal-region distribution is varied by 22\%, corresponding to the difference in the scale factors derived in subsamples with and without an additional jet. This value covers the variations of the correction factor derived from subsets of events with different jet multiplicities. No theoretical uncertainty is applied to the $Z$+jets background normalisation as this is scaled to data.

The uncertainty in the amount of pile-up is estimated by changing the nominal MC reweighting factors to vary the number of interactions per bunch crossing in data up and down by 10\%. Two methods were used to estimate the amount of  interactions per bunch crossing. The first method calculated the number of interactions  using the instantaneous luminosity and the inelastic proton-proton cross section \cite{pileup, inelastic}. The results of the calculation were compared to results from a data-driven method based on the number of reconstructed vertices. The difference between the correlation of the two methods in data and MC is taken as the uncertainty.

The uncertainty due to the 2--3\% loss of hard-scatter jets due to the JVT cut is estimated using $Z$+jet events. The uncertainty in the efficiency of the JVT cut to reduce pile-up jets is estimated by using a sideband method. The JVT cut is inverted in simulation to estimate the number of pile-up jets and derive a scale factor to describe the number of pile-up jets in data. This factor is then used to scale the predicted number of pile-up jets in the signal region (with the JVT cut applied). Scale factors are also derived using the samples with increased and decreased pile-up mentioned above, and the larger of two variations is taken as systematics.

\label{sec:systematics}

\section{Definition of the fiducial phase space}
For the measurement of the jet multiplicity, the jet \pt\ spectra and the gap fractions, the data are corrected to particle level by comparing to events from MC generators in the fiducial volume described below. The fiducial volume, i.e., the object definitions and the kinematic phase space at particle level, is designed to match the reconstruction level as closely as possible and follow closely the definitions in Refs.\,\cite{TOPQ-2012-03, TOPQ-2015-04}. Leptons and jets are defined using particles with a mean lifetime greater than $0.3 \times 10^{-10}$ s, directly produced in $pp$ interactions or from subsequent decays of particles with a shorter lifetime. Leptons from $W$ boson decays ($e$, $\mu, \nu_e, \nu_{\mu}, \nu_{\tau})$ are identified as such by requiring that they are not hadron decay products. Electron and muon four-momenta are calculated after the addition of photon four-momenta within a cone of $\Delta R =0.1$ around their original directions. 

Jets are defined using the anti-$k_t$ algorithm with a radius parameter of 0.4. All particles are considered for jet clustering, except for leptons from $W$ decays as defined above (i.e., neutrinos from hadron decays are included in jets) and any photons associated with the selected electrons or muons. Jets initiated by $b$-quarks are identified as such, i.e., identified as $b$-jets if a hadron with \pt $>5$~\gev\ containing a $b$-quark is associated with the jet through a ghost-matching technique as described in Ref. \cite{Cacciari:2008gn}.

The cross-section is defined using events with exactly one electron and one muon  with opposite-sign directly from $W$ boson decays,  i.e. excluding electrons and muons from decay of the $\tau$ leptons. In addition, at least two \bjet s each with $\pt > 25$~\gev\ and $| \eta |<2.5$ are required. Following the reconstructed object selection, events with jet--electron pairs or jet--muon pairs with $\Delta R < 0.4$ are excluded. Additional jets are considered within $| \eta |<2.5$ for \pt\ thresholds of 25\,\gev\ or higher, independently of their flavour.
\label{sec:fiducial}

\section{Measurement of jet multiplicities and \pt\ spectra}

The multiplicities of additional reconstructed jets with different \pt\ thresholds are corrected to particle level within the fiducial volume as defined above. Even though the kinematic range of the measurement is chosen to be the same for particle-level and reconstruction-level objects, corrections are necessary due to the efficiencies and detector resolutions that cause differences between reconstruction-level and particle-level jet distributions. Examples include events in which one or more particle-level jets do not pass the \pt\ threshold for reconstruction-level jets and when the selection efficiency for inclusive $t\bar{t}$ events changes as a function of jet multiplicity. Furthermore, additional reconstructed jets without a corresponding particle-level jet may appear due to pile-up, or if a jet migrates into the fiducial volume due to an upward fluctuation caused by the \pt\ resolution, or if a single particle-level jet is reconstructed as two separate jets. These effects lead to migrations between bins and are taken into account within an iterative Bayesian unfolding \cite{Bayes}.

The reconstructed jet multiplicity measurements are corrected separately for each additional-jet \pt\ threshold according to
\begin{equation}
N^i_{\mathrm {unfold}} =\frac{1}{ f^i_{\mathrm {eff}}} \cdot \sum_j (M^{-1})^{\mathrm {part,}i}_{\mathrm {reco,}j} \cdot f^j_{\mathrm {accept}} (N_{\mathrm {data}}^{j}- N_{\mathrm {bg}}^j),
\label{eqn:unfold}
\end{equation}
where $N^i_{\mathrm {unfold}}$ is the total number of fully corrected particle-level events with particle-level jet multiplicity $i$. The term $f^i_{\mathrm{eff}}$ represents the efficiency to reconstruct an event with $i$ additional jets, defined as the ratio of events with $i$ particle-level jets that fulfil both the fiducial volume selection at particle-level and the reconstruction-level selection, $N^i_{\mathrm {reco \wedge part}}$, to the number of events that fulfil the particle-level selection, $N^i_{\mathrm {part}}$:
\begin{equation}
f^i_{\mathrm {eff}} = \frac{N^i_{\mathrm {reco \wedge part}}}{N^i_{\mathrm {part}}}.
\label{eqn:feff}
\end{equation}
The resulting ratio $f^i_{\mathrm {eff}} $ is approximately 0.33 and has very small dependence on the jet multiplicity. The analysis of different \ttbar\ MC samples results in values of $f^i_{\mathrm {eff}}$ which vary by up to 10\%. The variations of $ f^i_{\mathrm {eff} }$ between different \pt\ thresholds are less than 2\%. The function $ f^j_{\mathrm {accept}}$ is the probability of an event fulfilling the reconstruction-level selection and with $j$ reconstructed jets, $N^{j}_{\mathrm {reco}}$, to also be within the particle-level acceptance defined in Section\,\ref{sec:fiducial}: 
\begin{equation}
f^j_{\mathrm {accept}} = \frac{N^{j}_{\mathrm {reco \wedge part}}}{N^{j}_{\mathrm {reco}}}.
\label{eqn:facc}
\end{equation}
The variable $N^{j}_{\mathrm {data}}$ is the number of events in data with $j$ reconstructed jets and $N^j_{\mathrm {bg}}$ is the number of background events, as evaluated in Section\,\ref{sec:selection}. The resulting $f^j_{\mathrm {accept}}$ decreases from around 0.85 for events without additional jets to about 0.76 for the highest jet multiplicities. The MC predictions of $f^j_{\mathrm {accept}}$ agree within 1\% for events without any additional jets and within 5\% at high jet multiplicities. Only \mcnlohw\ predicts a smaller change as a function of the number of jets.

\begin{figure}
\centering
\begin{tabular}{cc}
\subfloat[\label{fig:responsenjet25}]{\includegraphics[width=0.4\textwidth]{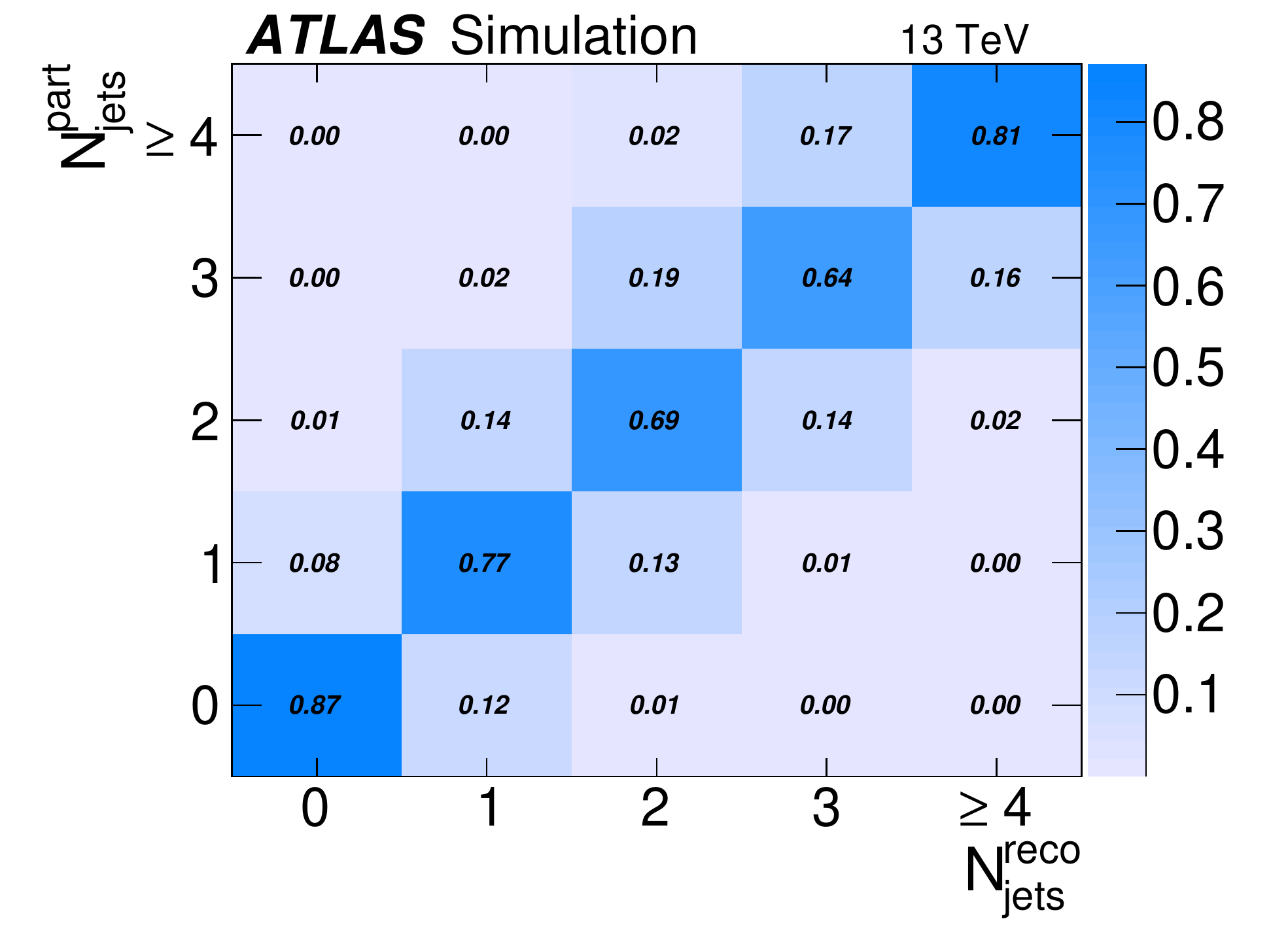}}&
\subfloat[\label{fig:responsenjet40}]{\includegraphics[width=0.4\textwidth]{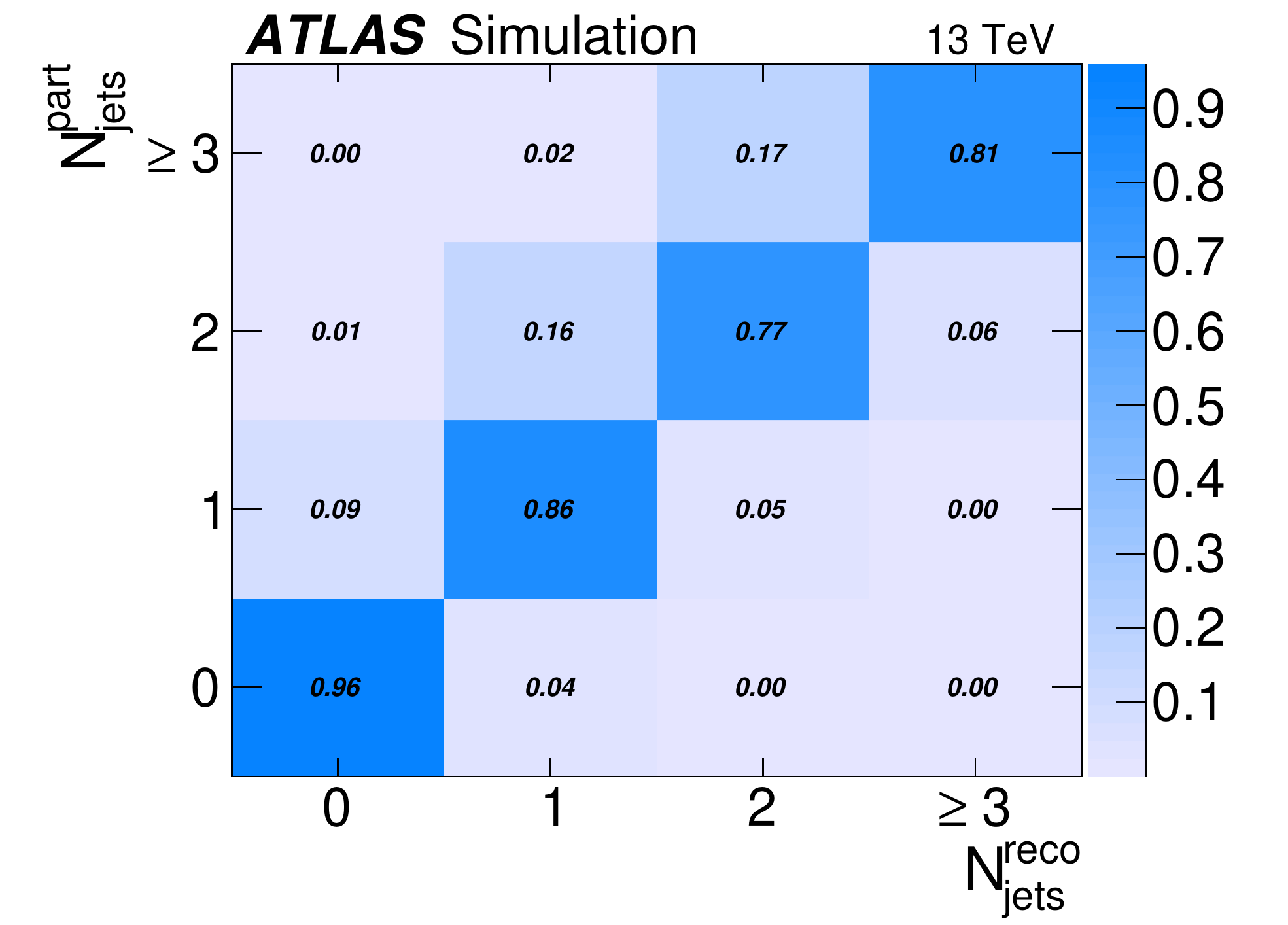}}\\
\subfloat[\label{fig:responseladdpt}]{\includegraphics[width=0.4\textwidth]{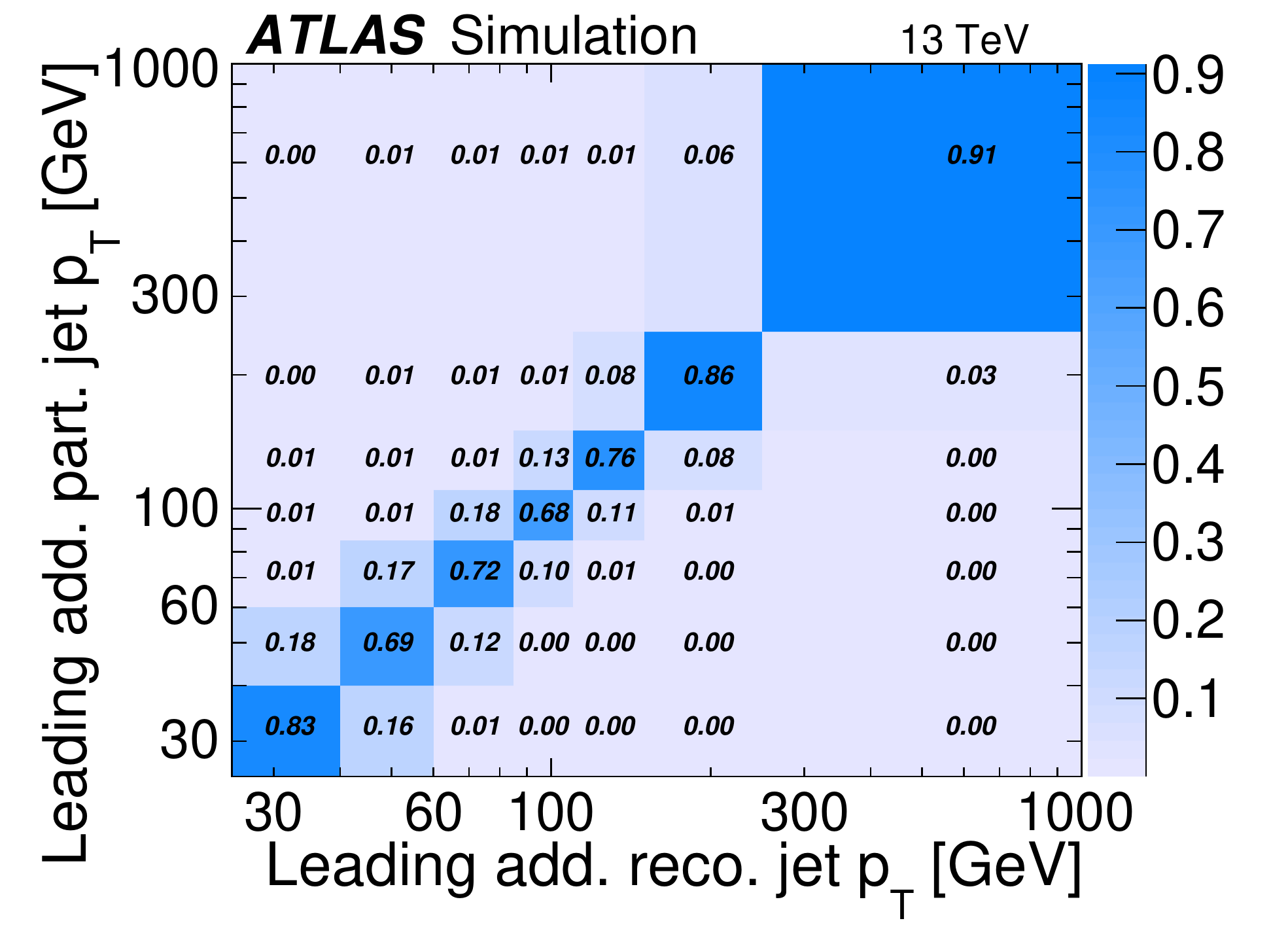}}&
\subfloat[\label{fig:responselbjetpt}]{\includegraphics[width=0.4\textwidth]{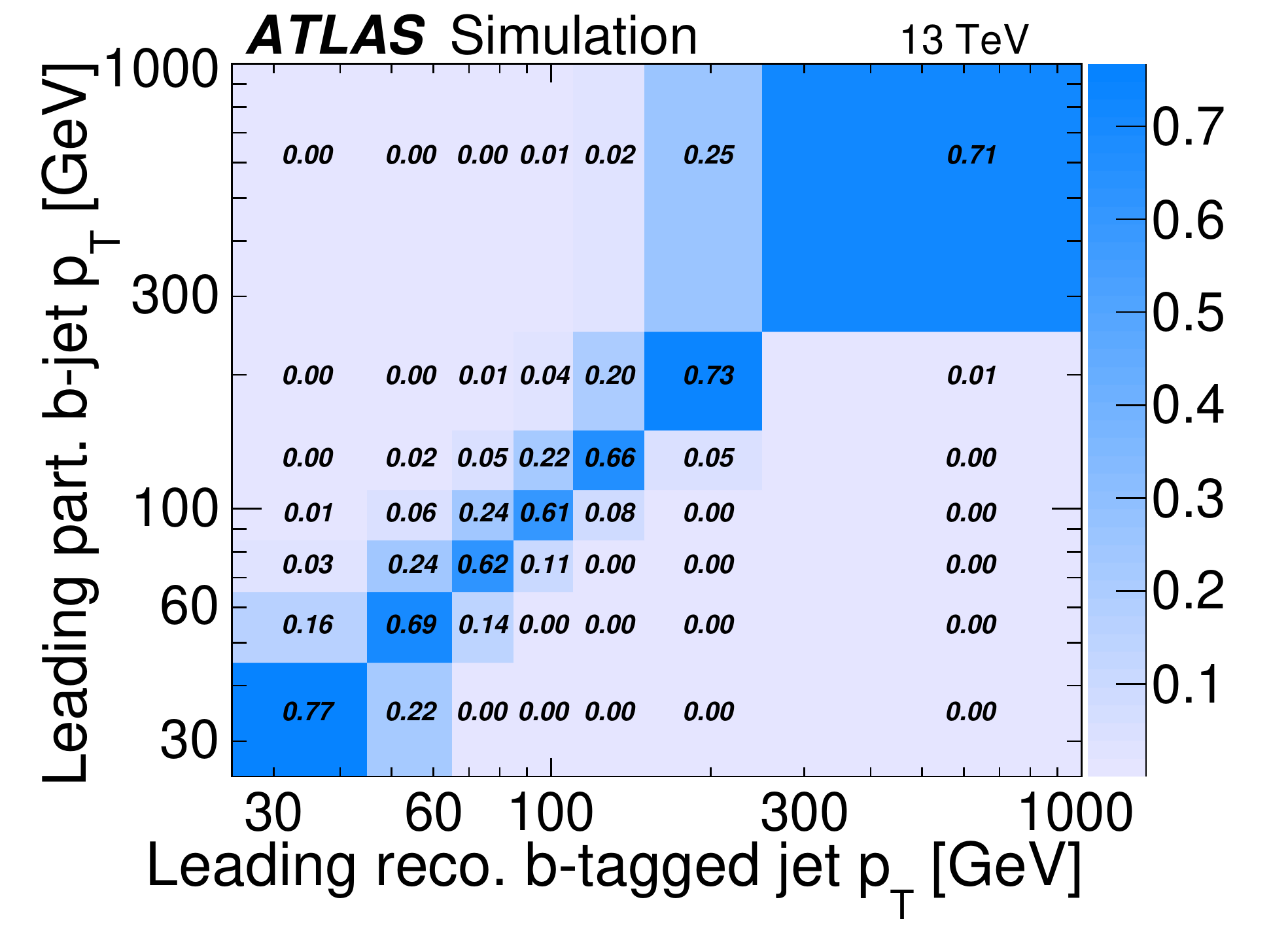}}\\
\end{tabular}
\caption{Unfolding response matrices to match distributions (jet multiplicity, jet \pt) at reconstruction level to particle-level distributions in the fiducial phase space. Only events that fulfil the reconstruction- (particle-) level selection are included. Matrices to unfold (a) jet multiplicity for additional jets with \pt $> 25$~\gev, (b) jet multiplicity for additional jets with \pt $>40$~\gev, (c) jet \pt\ of the leading additional jet, and (d) jet \pt\ of the leading \bjet.}
\label{fig:matrix}
\end{figure}

The response matrix $M^{\mathrm {part,}i}_{\mathrm {reco,} j}$ represents the probability $P(N^j_{\mathrm {reco}} | N^i_{\mathrm {part}})$ of finding an event with true particle-level jet multiplicity $i$ with a reconstructed jet multiplicity $j$. As shown in Figure~\ref{fig:matrix}, at the higher jet \pt\ thresholds, at least 77\% of the events have the same jet multiplicity at particle level and at reconstruction level. At the 25~\gev\ threshold, the agreement still exceeds 64\%. The worse agreement can be explained in part by the presence of pile-up jets, which leads to events with more reconstructed than particle-level jets. There are almost no events with a difference of more than one jet between particle and reconstruction-level multiplicity.
 
As part of the Bayesian unfolding using Equation~(\ref{eqn:unfold}), $M^{\mathrm {part,}i}_{\mathrm {reco,}j}$ is calculated iteratively, i.e., the result of the first iteration is used as the reconstruction-level jet multiplicity for the following one. The corrected spectra are found to converge after four iterations of the Bayesian unfolding algorithm. 
 
 The unfolded additional-jet multiplicity distributions are normalised after the last iteration according to 
 \begin{equation}
 \frac{1}{\sigma} \frac{{\textrm d} \sigma}{{\textrm d}N^{i}} = \frac{N^{i}_{{\mathrm {unfold}}}}{ \sum_{i} N^{i}_{{\mathrm{ unfold}}}}, 
\end{equation}
where $N^{i}_{\mathrm {unfold}}$, as defined in Equation~(\ref{eqn:unfold}), corresponds to the number of events with $i$ jets after full unfolding and $\sigma$ 
is the measured \ttbar\ production cross section in the fiducial volume. 

A potential bias of the unfolded results due to data statistics and the unfolding procedure is investigated using pseudo-experiments by performing Gaussian sampling of the reconstruction-level distributions with statistical power equivalent to that present in data. The size of the bias, defined as the relative difference between the unfolded and predicted particle-level distributions, is found to be within the statistical uncertainty of the data. To check the size of a potential bias of the unfolding due to the relation between reconstructed and particle level distributions, the particle-level distributions are reweighted to alternative MC samples. Pseudo-experiments are performed based on the resulting alternative spectrum at reconstruction level. The pseudo-experiments are unfolded using the original correction procedure. The relative difference between the unfolded particle-level distribution and the predicted particle-level distribution from the alternative MC sample is found to be well within the modelling uncertainty. In addition, it is ensured that differences between the nominal and alternative particle-level distributions are at least as large as the difference between data and the predicted reconstruction-level distributions.

The effect of the uncertainties listed in Section\,\ref{sec:systematics} on the unfolded multiplicity and jet spectra is evaluated as follows. The uncertainties due to detector-related effects, such as JES, JER and $b$-tagging and data statistics, are propagated through the unfolding by varying the reconstructed objects for each uncertainty component by $\pm 1\sigma$. The modified spectrum is then used as $N_{\mathrm {data}}^{j}$ in Equation~(\ref{eqn:unfold}) for the iterative unfolding and the difference on the particle-level distribution is taken as the systematic uncertainty. 

The uncertainties due to the MC modelling of the QCD initial- and final-state radiation (ISR/FSR) and the parton-shower uncertainty are evaluated by replacing the data with the corresponding alternative MC sample and using the response matrix and the correction factors from the baseline \ttbar\ MC sample for unfolding. The result is compared to the particle-level distribution of the alternative MC sample and the difference is taken as a systematic uncertainty. The uncertainties due to the MC modelling of the NLO matrix element and the matching algorithm are estimated in a similar way by replacing the data with the \mcnlohw\ sample but using the response matrix and correction factors from \powhw. The resulting uncertainties are symmetrised for each component.

To unfold the leading and sub-leading \bjet\ \pt\ and the leading additional-jet \pt, the same ansatz is used as for the jet multiplicity measurement, but with the jet \pt\ instead of the jet multiplicity in the matrix, the acceptance and the efficiency formula. The binning is chosen to limit the migration, such that most events have reconstruction-level jet \pt\ in the same region as the particle-level jet \pt, and to limit the uncertainty due to data statistics. The efficiency correction $f^i_{\mathrm {eff}} $ for the \bjet s has a significant \pt\ dependence: it is around 0.2 for the lowest \pt\ bin and reaches approximately 0.35 at \pt\ of ~80~\gev. The efficiency for the additional jet varies only slightly between 0.28 and 0.31. The acceptance correction is between 0.8 and 0.9 for all jets and almost independent of \pt, except at very low \pt, at which it decreases significantly, to 0.56 for the leading additional jet. The unfolding response matrix presented in Figure\,\ref{fig:matrix} shows that more than 60\% of the jets are in the same \pt\ bin at particle and reconstruction level. 

The spectra are normalised after the last iteration similarly to those in the jet multiplicity measurement:
 \begin{equation}
 \frac{1}{\sigma} \frac{{\textrm d} \sigma}{{\textrm d} \pt^i} = \frac{N^{i}_{\pt,{\mathrm{ unfold}}}}{ \Delta \pt^i \sum_{i} N^{i}_{ \pt,{\mathrm{unfold}}}}, 
\end{equation}
where $N^{i}_{ \pt,{\mathrm{unfold}}}$, as defined in Equation~(\ref{eqn:unfold}), corresponds to the number of events with the jet \pt\ in bin $i$ after full unfolding.

The measurement of the 
jet \pt\ spectra is as stable as the jet multiplicity measurements and the biases are small.

\label{sec:jets}

\subsection{Jet multiplicity  results}

\begin{figure}[htbp]
\centering
\begin{tabular}{cc}
\subfloat[\label{fig:3a_1}] {\includegraphics[width=0.42\textwidth]{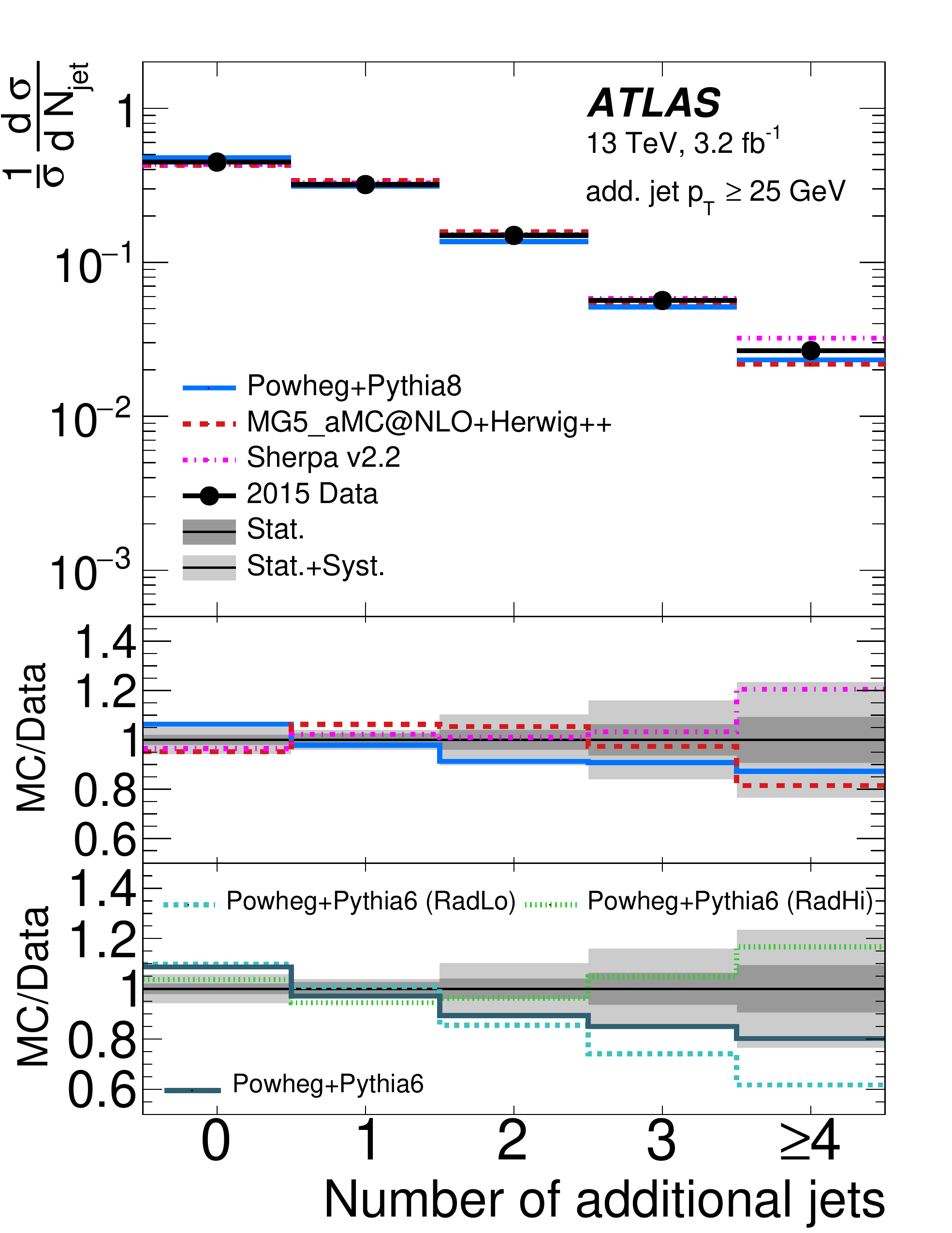}} &
\subfloat[\label{fig:3b_1}] {\includegraphics[width=0.42\textwidth]{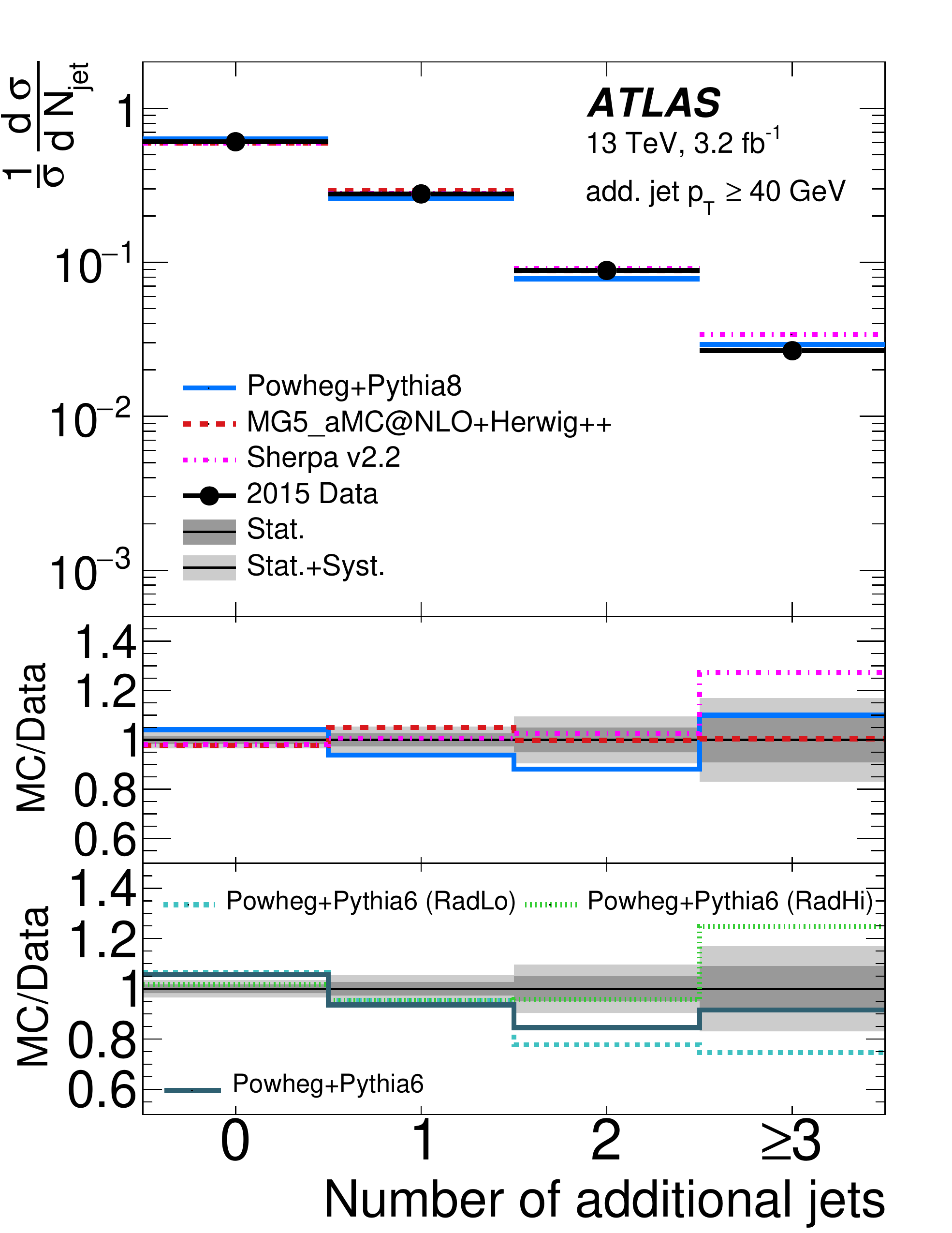}}\\
\subfloat[\label{fig:3ci}] {\includegraphics[width=0.42\textwidth]{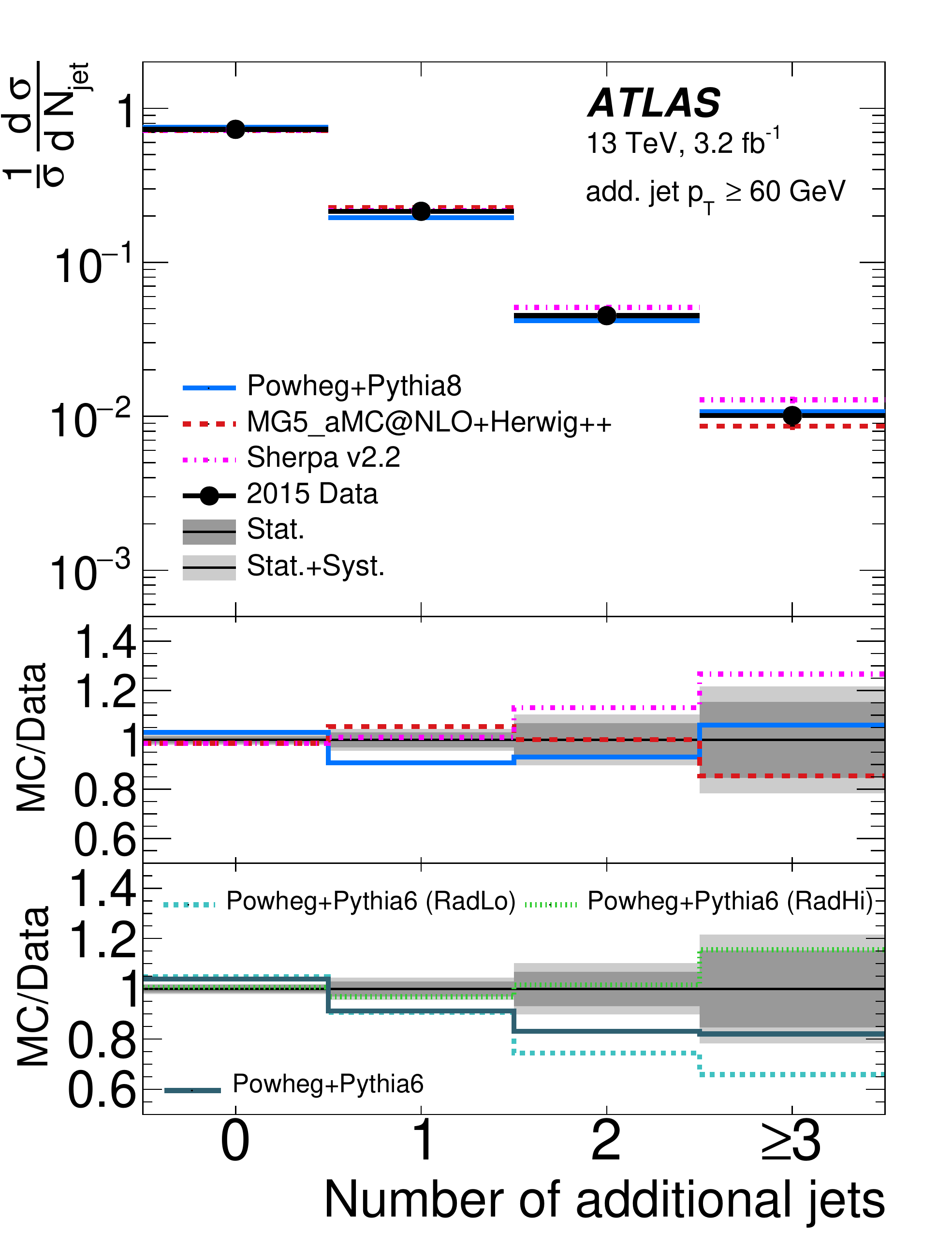} }&
\subfloat[\label{fig:3dl}] {\includegraphics[width=0.42\textwidth]{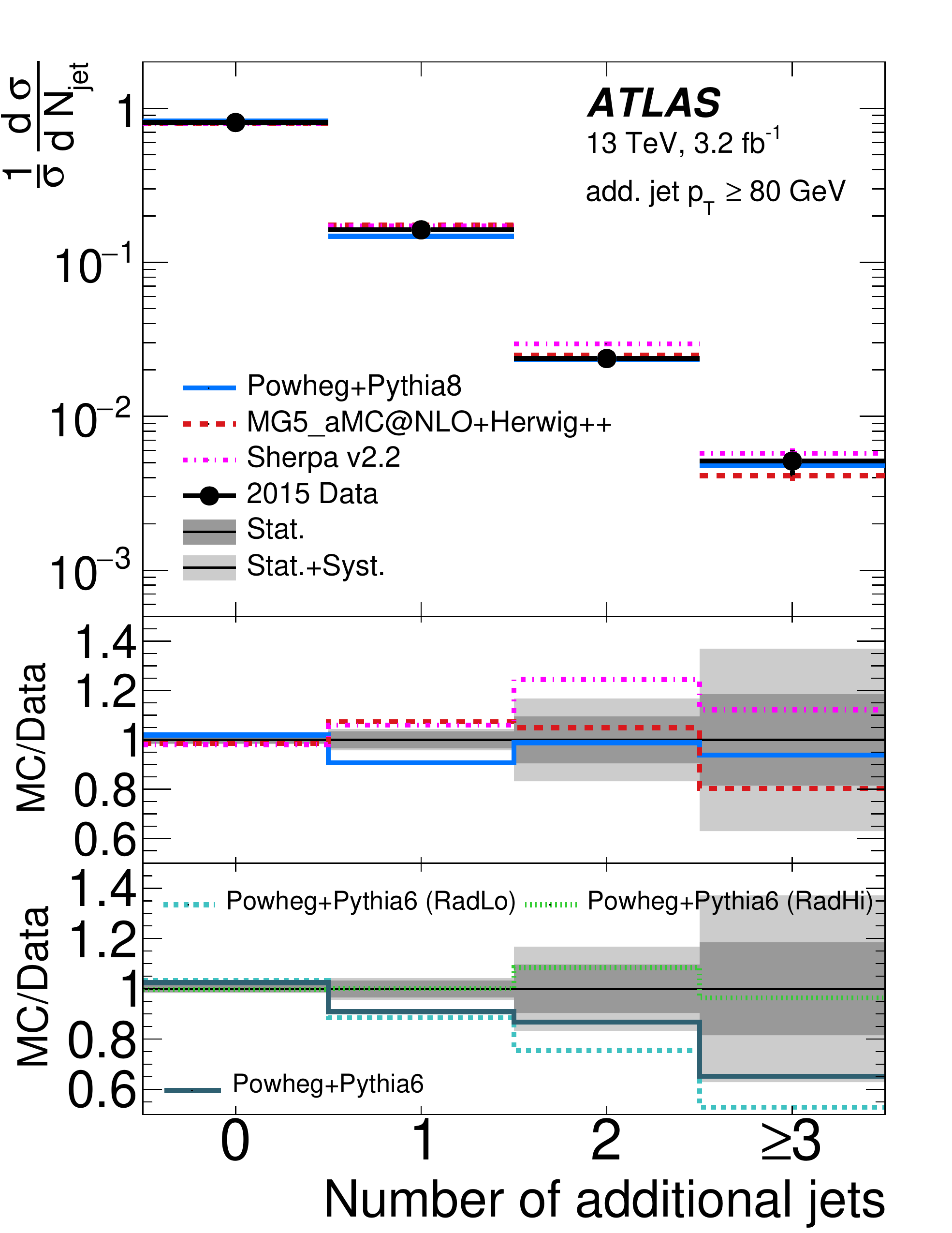}}\\
\end{tabular}
\caption{Unfolded jet multiplicity distribution for different \pt\ thresholds of the additional jets, for (a) additional jet \pt$>25$~\gev, (b) additional jet \pt$>40$~\gev, (c) additional jet \pt$>60$~\gev, and (d) additional jet \pt$>80$~\gev. Comparison to different MC predictions is shown for these distribution in first panel. \ The middle and bottom panels show the ratios of different MC predictions of the normalised cross-section to the measurement and the ratios of \powpy\ predictions with variation of the QCD radiation to the measurement, respectively. The shaded regions show the statistical uncertainty (dark grey) and total uncertainty (light grey).    
 \label{fig:resultmult}}
\end{figure}

The unfolded normalised cross-sections are shown in Figure~\ref{fig:resultmult} and are compared to different MC predictions. Events with up to three additional jets with \pt\ above 25~\gev\ are measured exclusively (four jets inclusively) and up to two additional jets exclusively (three inclusively) for the higher \pt\ thresholds. Tables~\ref{tab:uncnjets25} to \ref{tab:uncnjets80} list the detailed composition of the uncertainties for 25~\gev\ to 80~\gev. The jet multiplicity distributions are measured with an uncertainty of 4--5\% for one additional jet, about 10\% for two additional jets, and around 20\% for the highest jet multiplicity bin, except for the 80~\gev\ threshold where the statistical uncertainty is larger for higher jet multiplicity bins. Systematic uncertainties dominate in all the measurements. In almost all bins for all \pt\ thresholds, the JES uncertainty dominates, followed by the modelling uncertainty.

\begin{figure}[ht]
\centering
\begin{tabular}{cc}
\subfloat[\label{fig:2a}] {\includegraphics[width=0.42\textwidth]{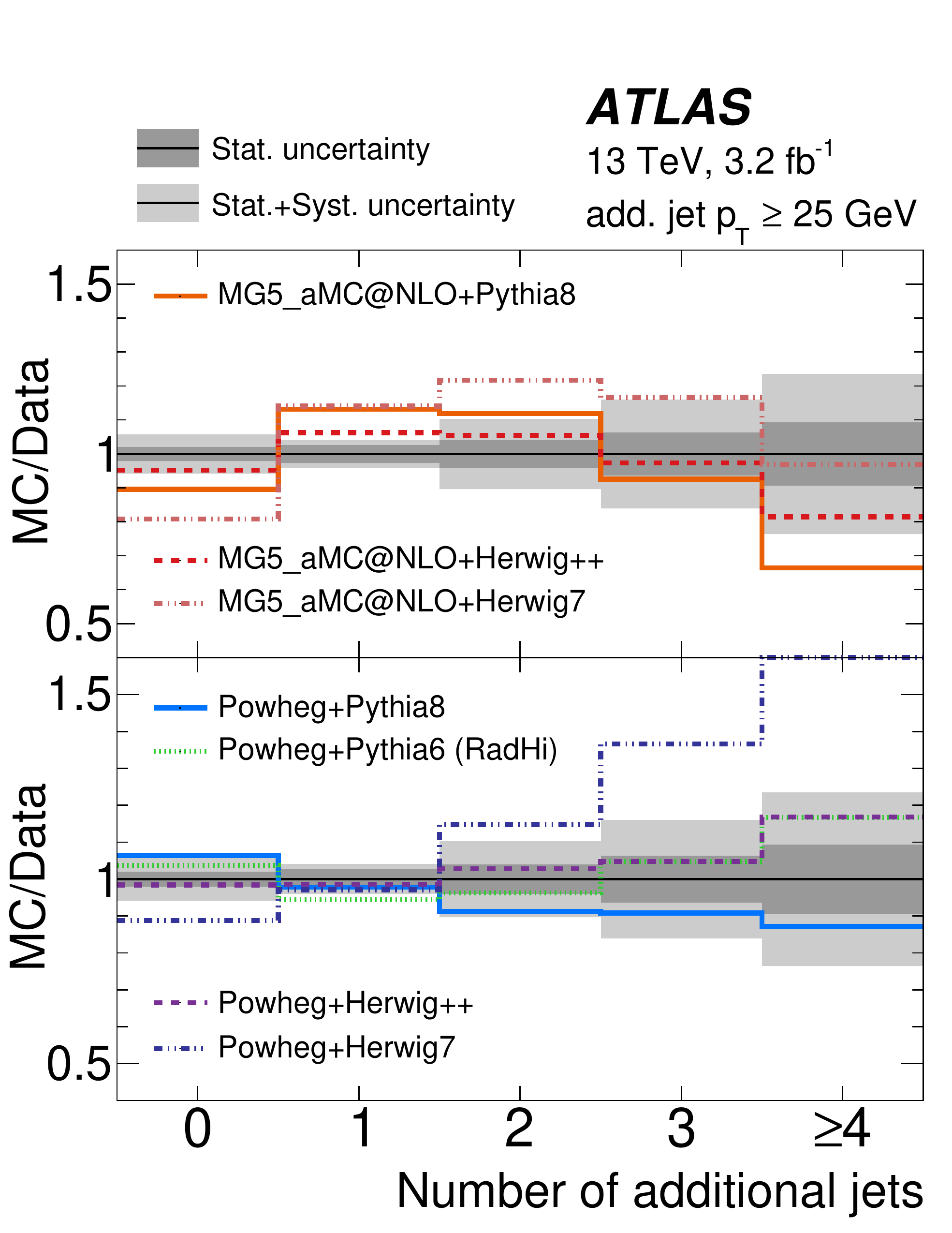}}&
\subfloat[\label{fig:2c}] {\includegraphics[width=0.42\textwidth]{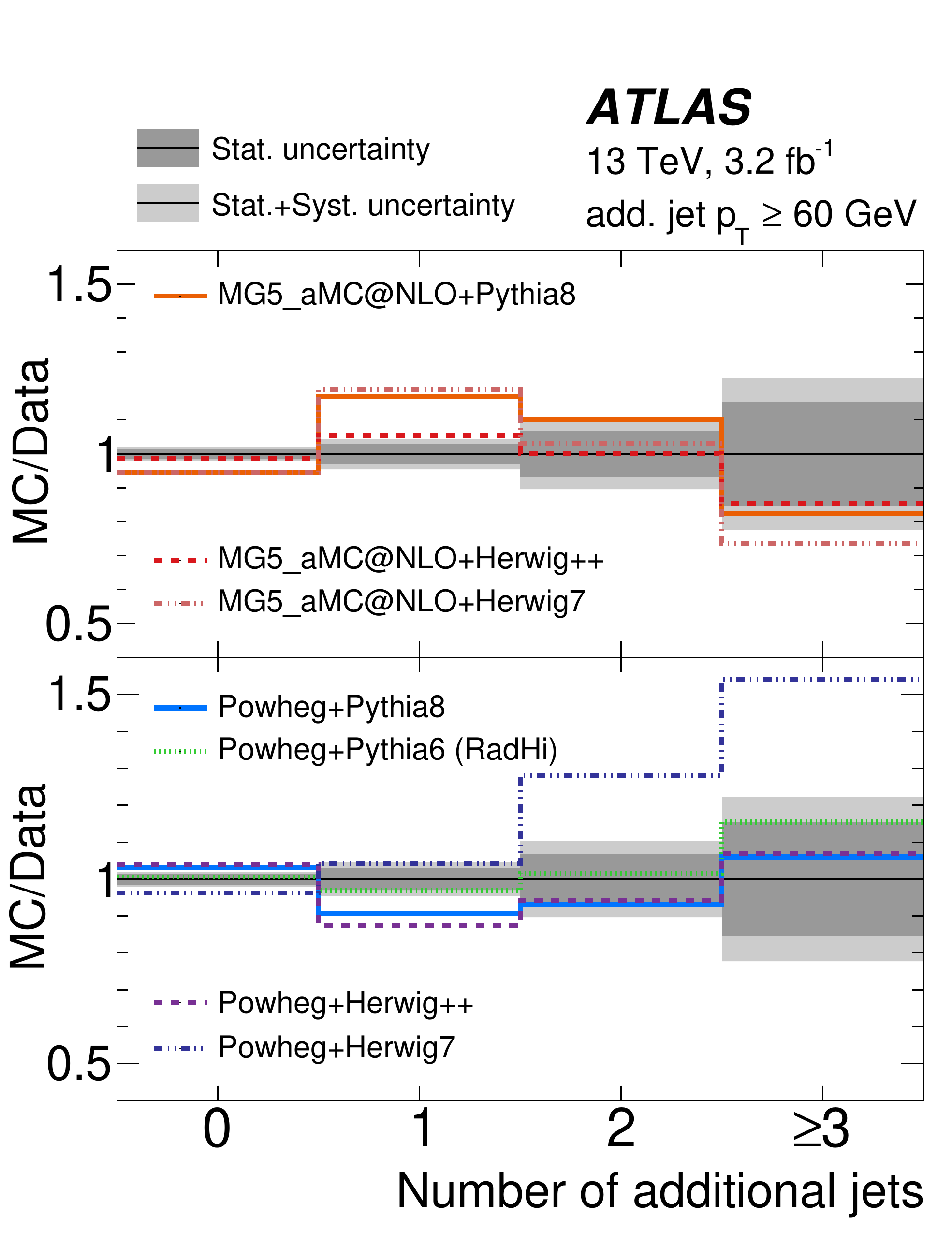}} \\
\end{tabular}
\caption{Ratios of jet multiplicity distribution for different \pt\ thresholds of the additional jets predicted by various MC generators to the unfolded data, for (a) additional jet \pt$>25$~\gev, (b) additional jet \pt$>60$~\gev. The shaded regions show the statistical uncertainty (dark grey) and total uncertainty (light grey). 
\label{fig:multratios}}
\end{figure}

The data are compared to \pow\ and \mcnlo\ matched with different shower generators, namely \pye, \hw, and \hws\ and to \sherpa, as shown in Figures\,\ref{fig:resultmult} and \ref{fig:multratios}. Most predictions are within uncertainties and only slight deviations are visible except for \powhws, which deviates significantly from the data for all \pt\ thresholds. The \mcnlo\ predictions agree within 5--10\% regardless of which parton shower is used (except \hws),  and the \pow\ predictions vary slightly more. The variations are larger when using different matrix elements but the same parton shower.

The unfolded data are compared with different MC predictions using $\chi^2$ tests. Full covariance matrices are produced from the unfolding taking into account statistical and all systematic uncertainties. The correlation of the measurement bins is similar for all jet \pt thresholds:     
strong anti-correlations exist between events with no additional jet and events with any number of additional jets.
Positive correlations exist between the bins with one and two additional jets.
\ The $\chi^2$ is determined using:
\begin{eqnarray}
\chi^{2} = S^{\mathrm{T}}_{n-1} {\textrm Cov}^{-1}_{n-1} S_{n-1}
\end{eqnarray}

where $S_{n-1}$ is a column vector representing the difference between the unfolded data and the MC generator predictions of the normalised cross-section for one less than the total number of bins in the distribution, and ${\textrm Cov}_{n-1}$ is a matrix with $n-1$ rows and the respective $n-1$ columns of the full covariance matrix. The full covariance matrix is singular and non-invertible, as it is evaluated using normalised distributions. The $p$-values are determined using the $\chi^2$ and $n-1$ degrees of freedom. Table~\ref{tab:chi2MeasPrednjets} shows the $\chi^2$ and $p$-values.

A statistical comparison taking into account the bin correlations indicates that the agreement with data is slightly better for \mcnlohw, as shown in Table\,\ref{tab:chi2MeasPrednjets}. The ratio of the data to predictions of \powpy\ with different levels of QCD radiation both in the matrix-element calculation and in the parton shower is also shown. \powpy\ (RadLo) does not describe the data well. The central prediction of \powpy\ yields fewer jets than in data; however, the predictions are still within the experimental uncertainties. \powpy\ (RadHi) describes the data most consistently, which is also confirmed by high $p$-values for all \pt\ thresholds. The \powpy\ (RadLo) sample has $p$-values around 0.5 and the central sample mostly between 0.8 and 0.9.

\begin{table}[ht!]
\begin{center}
\sisetup{retain-explicit-plus}
\sisetup{round-mode = places}
\begin{tabular}{|l|r|r|r|r|r|}
\hline
\multicolumn{6}{|c|}{Relative uncertainty in [\%] in additional jets multiplicity}\\
Sources    &  0   & 1   & 2  & 3  & $\ge$ 4  \\
\hline
\hline

Data statistics & \num[round-precision=1]{2.1} & \num[round-precision=1]{2.7} & \num[round-precision=1]{4.0} & \num[round-precision=1]{6.0} & \num[round-precision=1]{9.0} \\
JES/JER   & \num[round-precision=1]{5.0} & \num[round-precision=1]{1.8} & \num[round-precision=1]{7.0} & \num[round-precision=1]{12.0} & \num[round-precision=1]{16.0} \\ 
$b$-tagging  & \num[round-precision=1]{0.5} & \num[round-precision=1]{0.2} & \num[round-precision=1]{0.7} & \num[round-precision=1]{1.4} & \num[round-precision=1]{2.0} \\
ISR/FSR modelling & \num[round-precision=1]{0.4} & \num[round-precision=1]{0.5} & \num[round-precision=1]{2.2} & \num[round-precision=1]{3.8} & \num[round-precision=1]{6.0} \\
Signal modelling & \num[round-precision=1]{1.9} & \num[round-precision=1]{2.0} & \num[round-precision=1]{5.6} & \num[round-precision=1]{6.0} & \num[round-precision=1]{11.0} \\ 
Other  & \num[round-precision=1]{1.4} & \num[round-precision=1]{0.9} & \num[round-precision=1]{2.5} & \num[round-precision=1]{3.3} & \num[round-precision=1]{5.0} \\
\hline
\textbf{Total} & \textbf{\num[round-precision=1]{6.0}} & \textbf{\num[round-precision=1]{4.0}} & \textbf{\num[round-precision=1]{10.0}} & \textbf{\num[round-precision=1]{16.0}} & \textbf{\num[round-precision=1]{24.0}} \\ 
\hline
\end{tabular}  \caption{Summary of relative uncertainties in [\%] for the jet multiplicity measurement using a jet \pt\ threshold of 25~\gev. "Signal modelling" sources  of systematic uncertainty includes the hadronisation, parton shower and NLO modelling uncertainties. "Other" sources of systematic uncertainty refers to lepton and jet selection efficiencies, background (including pile-up jets) estimations, and the PDF. \label{tab:uncnjets25}}
\end{center}
\end{table}

\begin{table}[ht!]
\sisetup{retain-explicit-plus}
\sisetup{round-mode = places}
\begin{center}
\begin{tabular}{|l|r|r|r|r|}

\hline
             \multicolumn{5}{|c|}{Relative uncertainty in [\%] in additional jets multiplicity}\\
       Sources    &  0   & 1   & 2  & $\ge$ 3  \\
\hline
\hline

Data statistics & \num[round-precision=1]{1.7} & \num[round-precision=1]{2.7} & \num[round-precision=1]{5.0} & \num[round-precision=1]{9.0} \\
JES/JER   & \num[round-precision=1]{2.0} & \num[round-precision=1]{2.5} & \num[round-precision=1]{6.0} & \num[round-precision=1]{9.0} \\ 
$b$-tagging  & \num[round-precision=1]{0.3} & \num[round-precision=1]{0.4} & \num[round-precision=1]{1.1} & \num[round-precision=1]{1.8} \\
ISR/FSR modelling & \num[round-precision=1]{0.2} & \num[round-precision=1]{0.4} & \num[round-precision=1]{3.0} & \num[round-precision=1]{6.0} \\
Signal modelling & \num[round-precision=1]{2.0} & \num[round-precision=1]{3.7} & \num[round-precision=1]{4.4} & \num[round-precision=1]{9.0} \\ 
Other & \num[round-precision=1]{0.7} & \num[round-precision=1]{0.8} & \num[round-precision=1]{1.5} & \num[round-precision=1]{4.1} \\ 
\hline
\textbf{Total} & \textbf{\num[round-precision=1]{3.4}} & \textbf{\num[round-precision=1]{5.0}} & \textbf{\num[round-precision=1]{10.0}} & \textbf{\num[round-precision=1]{17.0}} \\
\hline
\end{tabular}
\caption{Summary of relative uncertainties in [\%] for the jet multiplicity measurement using a jet \pt\ threshold of 40~\gev. "Signal modelling" sources  of systematic uncertainty includes the hadronisation, parton shower and NLO modelling uncertainties.  "Other" sources of systematic uncertainty refer to lepton and jet selection efficiencies, background (including pile-up jets) estimations, and the PDF. \label{tab:uncnjets40}}
\end{center}
\end{table}

\begin{table}[ht!]
\begin{center}
\sisetup{retain-explicit-plus}
\sisetup{round-mode = places}
\begin{tabular}{|l|r|r|r|r|}
  \hline
  \multicolumn{5}{|c|}{Relative uncertainty in [\%] in additional jets multiplicity}\\
  Sources    &  0   & 1   & 2  & $\ge$ 3  \\
  \hline
  \hline

Data statistics & \num[round-precision=1]{1.5} & \num[round-precision=1]{3.0} & \num[round-precision=1]{7.0} & \num[round-precision=1]{15.0} \\
JES/JER   & \num[round-precision=1]{0.9} & \num[round-precision=1]{2.3} & \num[round-precision=1]{4.2} & \num[round-precision=1]{7.0} \\
$b$-tagging  & \num[round-precision=1]{0.2} & \num[round-precision=1]{0.6} & \num[round-precision=1]{1.2} & \num[round-precision=1]{2.0} \\ 
ISR/FSR modelling & \num[round-precision=1]{0.2} & \num[round-precision=1]{1.2} & \num[round-precision=1]{2.2} & \num[round-precision=1]{1.1} \\ 
Signal modelling  & \num[round-precision=1]{0.7} & \num[round-precision=1]{1.6} & \num[round-precision=1]{5.0} & \num[round-precision=1]{9.0} \\
Other  & \num[round-precision=1]{0.8} & \num[round-precision=1]{0.8} & \num[round-precision=1]{3.2} & \num[round-precision=1]{10.0} \\
\hline
\textbf{Total} & \textbf{\num[round-precision=1]{2.0}} & \textbf{\num[round-precision=1]{4.4}} & \textbf{\num[round-precision=1]{10.0}} & \textbf{\num[round-precision=1]{22.0}} \\
\hline
\end{tabular}  \caption{Summary of relative uncertainties in [\%] for the jet multiplicity measurement using a jet \pt\ threshold of 60~\gev. \ "Signal modelling" sources  of systematic uncertainty includes the hadronisation, parton shower and NLO modelling uncertainties. "Other" sources of systematic uncertainty refer to lepton and jet selection efficiencies, background (including pile-up jets) estimations, and the PDF. \label{tab:uncnjets60}}
\end{center}
\end{table}

\begin{table}[ht!]
\begin{center}
\sisetup{retain-explicit-plus}
\sisetup{round-mode = places}
\begin{tabular}{|l|r|r|r|r|}
  \hline
    \multicolumn{5}{|c|}{Relative uncertainty in [\%] in additional jets multiplicity}\\
  Sources    &  0   & 1   & 2  & $\ge$ 3  \\
\hline
\hline

Data statistics & \num[round-precision=1]{1.4} & \num[round-precision=1]{3.3} & \num[round-precision=1]{10.0} & \num[round-precision=1]{19.0} \\
JES/JER   & \num[round-precision=1]{0.4} & \num[round-precision=1]{1.8} & \num[round-precision=1]{5.0} & \num[round-precision=1]{6.0} \\ 
$b$-tagging  & \num[round-precision=1]{0.1} & \num[round-precision=1]{0.6} & \num[round-precision=1]{1.4} & \num[round-precision=1]{2.4} \\
ISR/FSR modelling & \num[round-precision=1]{0.1} & \num[round-precision=1]{1.3} & \num[round-precision=1]{6.0} & \num[round-precision=1]{4.5} \\
Signal modelling & \num[round-precision=1]{0.2} & \num[round-precision=1]{0.6} & \num[round-precision=1]{10.0} & \num[round-precision=1]{31.0} \\ 
Other  & \num[round-precision=1]{0.8} & \num[round-precision=1]{1.4} & \num[round-precision=1]{3.1} & \num[round-precision=1]{6.0} \\ 
\hline
\textbf{Total} & \textbf{\num[round-precision=1]{1.7}} & \textbf{\num[round-precision=1]{4.3}} & \textbf{\num[round-precision=1]{17.0}} & \textbf{\num[round-precision=1]{37.0}} \\
\hline
\end{tabular}  \caption{Summary of relative uncertainties in [\%] for the jet multiplicity measurement using a jet \pt\ threshold of 80~\gev. \ "Signal modelling" sources  of systematic uncertainty includes the hadronisation, parton shower and NLO modelling uncertainties. "Other" sources of systematic uncertainty refer to lepton and jet selection efficiencies, background (including pile-up jets) estimations, and the PDF. \label{tab:uncnjets80}}
\end{center}
\end{table}

\begin{table}[ht!]
\sisetup{retain-explicit-plus}
\sisetup{round-mode = places}
\begin{center}
\begin{tabular}{|l|r|r|r|r|r|r|r|}
    \hline
         &  \multicolumn{7}{c|}{Relative uncertainty in leading $b$-jet $p_{\textrm T}$ [GeV] in [\%]} \\
    Sources         &  25--45 & 45--65 & 65--85 & 85--110 & 110--150 & 150--250 & $>$ 250\\
\hline
\hline

Data statistics  & \num[round-precision=1]{11.0} & \num[round-precision=1]{5.0} & \num[round-precision=1]{4.3} & \num[round-precision=1]{4.2} & \num[round-precision=1]{4.4} & \num[round-precision=1]{5.0} & \num[round-precision=1]{12.0} \\
JES/JER  & \num[round-precision=1]{11.0} & \num[round-precision=1]{2.3} & \num[round-precision=1]{1.3} & \num[round-precision=1]{2.4} & \num[round-precision=1]{3.2} & \num[round-precision=1]{4.2} & \num[round-precision=1]{6.0} \\ 
$b$-tagging & \num[round-precision=1]{6.0} & \num[round-precision=1]{1.1} & \num[round-precision=1]{0.9} & \num[round-precision=1]{1.0} & \num[round-precision=1]{1.9} & \num[round-precision=1]{5.0} & \num[round-precision=1]{14.0} \\ 
ISR/FSR modelling  & \num[round-precision=1]{6.0} & \num[round-precision=1]{0.9} & \num[round-precision=1]{1.0} & \num[round-precision=1]{2.1} & \num[round-precision=1]{3.1} & \num[round-precision=1]{0.9} & \num[round-precision=1]{0.1} \\ 
Signal modelling & \num[round-precision=1]{9.0} & \num[round-precision=1]{2.0} & \num[round-precision=1]{5.0} & \num[round-precision=1]{6.0} & \num[round-precision=1]{2.1} & \num[round-precision=1]{0.4} & \num[round-precision=1]{15.0} \\
Other  & \num[round-precision=1]{4.4} & \num[round-precision=1]{3.0} & \num[round-precision=1]{1.4} & \num[round-precision=1]{1.7} & \num[round-precision=1]{3.0} & \num[round-precision=1]{2.2} & \num[round-precision=1]{10.0} \\ 
\hline
\textbf{Total} & \textbf{\num[round-precision=1]{20.0}} & \textbf{\num[round-precision=1]{7.0}} & \textbf{\num[round-precision=1]{7.0}} & \textbf{\num[round-precision=1]{8.0}} & \textbf{\num[round-precision=1]{8.0}} & \textbf{\num[round-precision=1]{8.0}} & \textbf{\num[round-precision=1]{26.0}}\\

\hline
\end{tabular}
\caption{Summary of  relative measurement uncertainties in [\%] for the leading $b$-jet $p_{\textrm T}$ distribution. \ "Signal modelling" sources  of systematic uncertainty includes the hadronisation, parton shower and NLO modelling uncertainties. "Other" sources of systematic uncertainty refers to lepton and jet selection efficiencies, background (including pile-up jets) estimations, and the PDF. \label{tab:uncleadbpt}}
\end{center}
\end{table}

\begin{table}[ht!]
\begin{center}
\sisetup{retain-explicit-plus}
\sisetup{round-mode = places}
\begin{tabular}{|l|r|r|r|r|r|r|}
    \hline
    &  \multicolumn{6} {c|} {Relative uncertainty in sub-leading $b$-jet $p_{\textrm T}$ [GeV] in [\%]} \\
 Sources    &  25--40 & 40--55 & 55--75 & 75--100 & 100--150 & $>$ 150 \\
\hline
\hline

Data statistics  & \num[round-precision=1]{4.0} & \num[round-precision=1]{4.2} & \num[round-precision=1]{3.9} & \num[round-precision=1]{6.0} & \num[round-precision=1]{7.0} & \num[round-precision=1]{11.0} \\ 
JES/JER  & \num[round-precision=1]{5.0} & \num[round-precision=1]{2.5} & \num[round-precision=1]{3.4} & \num[round-precision=1]{3.8} & \num[round-precision=1]{3.6} & \num[round-precision=1]{6.0} \\ 
$b$-tagging & \num[round-precision=1]{2.8} & \num[round-precision=1]{1.2} & \num[round-precision=1]{2.2} & \num[round-precision=1]{2.3} & \num[round-precision=1]{3.6} & \num[round-precision=1]{11.0} \\ 
ISR/FSR modelling  & \num[round-precision=1]{0.3} & \num[round-precision=1]{2.7} & \num[round-precision=1]{1.2} & \num[round-precision=1]{1.3} & \num[round-precision=1]{3.2} & \num[round-precision=1]{0.3} \\
Signal modelling & \num[round-precision=1]{6.0} & \num[round-precision=1]{1.9} & \num[round-precision=1]{6.0} & \num[round-precision=1]{8.0} & \num[round-precision=1]{6.0} & \num[round-precision=1]{5.0} \\
Other  & \num[round-precision=1]{1.4} & \num[round-precision=1]{1.8} & \num[round-precision=1]{1.9} & \num[round-precision=1]{3.4} & \num[round-precision=1]{3.1} & \num[round-precision=1]{3.9} \\ 
\hline
\textbf{Total}  & \textbf{\num[round-precision=1]{9.0}} & \textbf{\num[round-precision=1]{6.0}} & \textbf{\num[round-precision=1]{9.0}} & \textbf{\num[round-precision=1]{11.0}} & \textbf{\num[round-precision=1]{11.0}} & \textbf{\num[round-precision=1]{18.0}} \\
\hline
\end{tabular}
\caption{Summary of  relative measurement uncertainties in [\%] for the sub-leading $b$-jet $p_{\textrm T}$ distribution. \"Signal modelling" sources  of systematic uncertainty includes the hadronisation, parton shower and NLO modelling uncertainties. "Other" sources of systematic uncertainty refers to lepton and jet selection efficiencies, background (including pile-up jets) estimations, and the PDF. \label{tab:uncnleadbpt}}
\end{center}
\end{table}

\begin{table}[ht!]
\begin{center}
\sisetup{retain-explicit-plus}
\sisetup{round-mode = places}
\begin{tabular}{|l|r|r|r|r|r|r|r|}
    \hline
    &  \multicolumn{7}{c|}{Relative uncertainty in leading additional jet $p_{\textrm T}$ [GeV] in [\%]} \\
Sources  &  25--40 & 40--60 & 60--85 & 85--110 & 110--150 & 150--250 & $>250$\\
\hline
\hline

Data statistics  & \num[round-precision=1]{3.8} & \num[round-precision=1]{6.0} & \num[round-precision=1]{6.0} & \num[round-precision=1]{8.0} & \num[round-precision=1]{8.0} & \num[round-precision=1]{6.0} & \num[round-precision=1]{8.0} \\
JES/JER  & \num[round-precision=1]{2.9} & \num[round-precision=1]{3.3} & \num[round-precision=1]{2.1} & \num[round-precision=1]{2.7} & \num[round-precision=1]{3.8} & \num[round-precision=1]{3.8} & \num[round-precision=1]{4.2} \\
$b$-tagging & \num[round-precision=1]{0.3} & \num[round-precision=1]{0.2} & \num[round-precision=1]{0.6} & \num[round-precision=1]{0.4} & \num[round-precision=1]{0.6} & \num[round-precision=1]{0.4} & \num[round-precision=1]{1.3} \\ 
ISR/FSR modelling  & \num[round-precision=1]{0.6} & \num[round-precision=1]{1.6} & \num[round-precision=1]{1.4} & \num[round-precision=1]{0.7} & \num[round-precision=1]{2.4} & \num[round-precision=1]{4.0} & \num[round-precision=1]{2.1} \\
Signal modelling  & \num[round-precision=1]{2.5} & \num[round-precision=1]{4.0} & \num[round-precision=1]{3.6} & \num[round-precision=1]{10.0} & \num[round-precision=1]{8.0} & \num[round-precision=1]{8.0} & \num[round-precision=1]{4.0} \\
Other   & \num[round-precision=1]{1.5} & \num[round-precision=1]{2.8} & \num[round-precision=1]{1.8} & \num[round-precision=1]{3.4} & \num[round-precision=1]{2.4} & \num[round-precision=1]{1.6} & \num[round-precision=1]{1.8} \\ 
\hline
\textbf{Total}  & \textbf{\num[round-precision=1]{6.0}} & \textbf{\num[round-precision=1]{8.0}} & \textbf{\num[round-precision=1]{8.0}} & \textbf{\num[round-precision=1]{13.0}} & \textbf{\num[round-precision=1]{12.0}} & \textbf{\num[round-precision=1]{11.0}} & \textbf{\num[round-precision=1]{11.0}} \\ 
\hline
\end{tabular}
\caption{Summary of relative measurement uncertainties in [\%] for the leading additional jet $p_{\textrm T}$ distribution. "Signal modelling" sources  of systematic uncertainty includes the hadronisation, parton shower and NLO modelling uncertainties. "Other" sources of systematic uncertainty refers to lepton and jet selection efficiencies, background (including pile-up jets) estimations, and the PDF. \label{tab:uncleadaddpt}}
\end{center}
\end{table}

\subsection{Jet \pt\ spectra results}
\begin{figure}[htbp]
\centering
\begin{tabular}{cc}
\subfloat[\label{fig:3a_2}] {\includegraphics[width=0.42\textwidth]{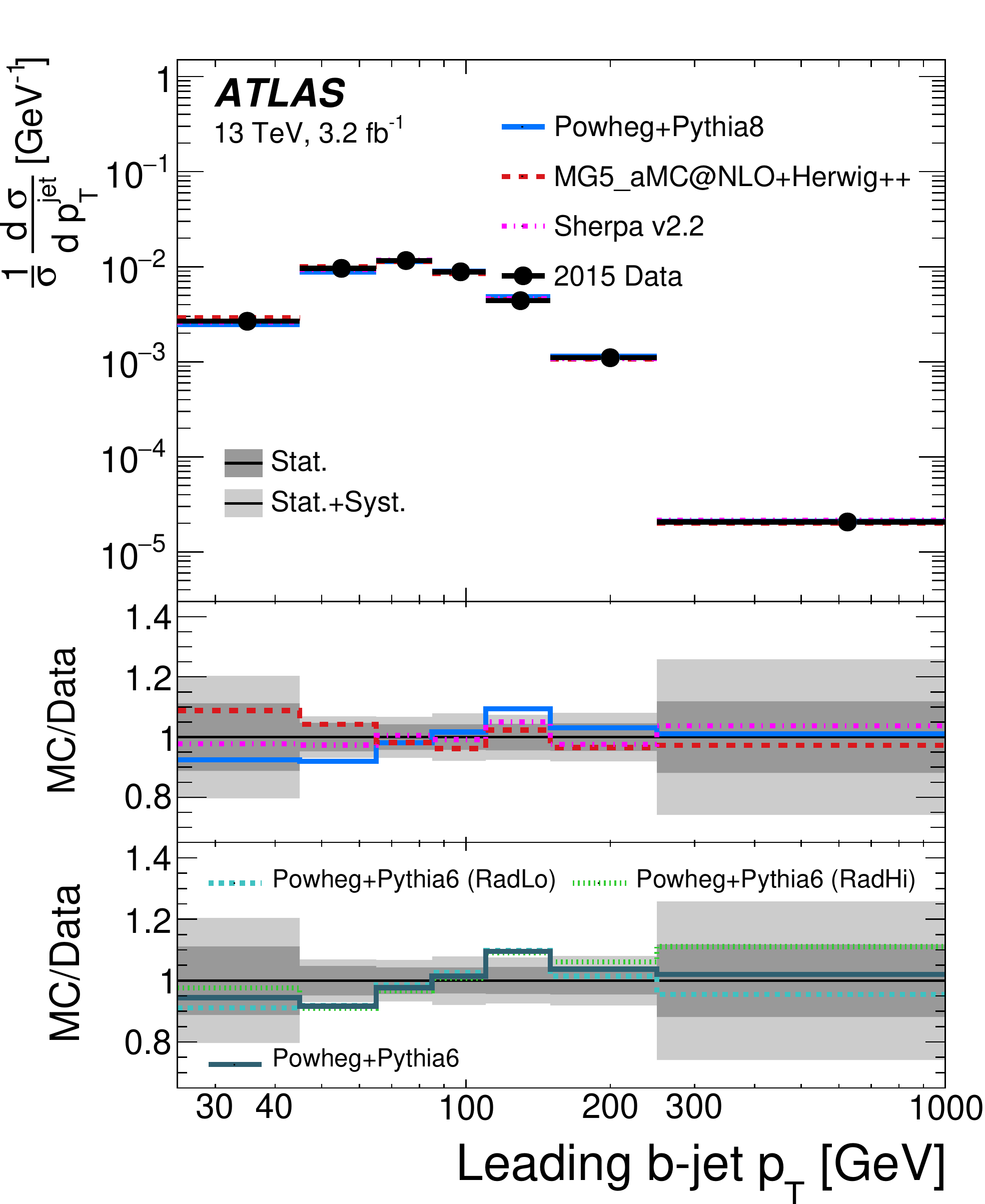}} &
\subfloat[\label{fig:3b_2}] {\includegraphics[width=0.42\textwidth]{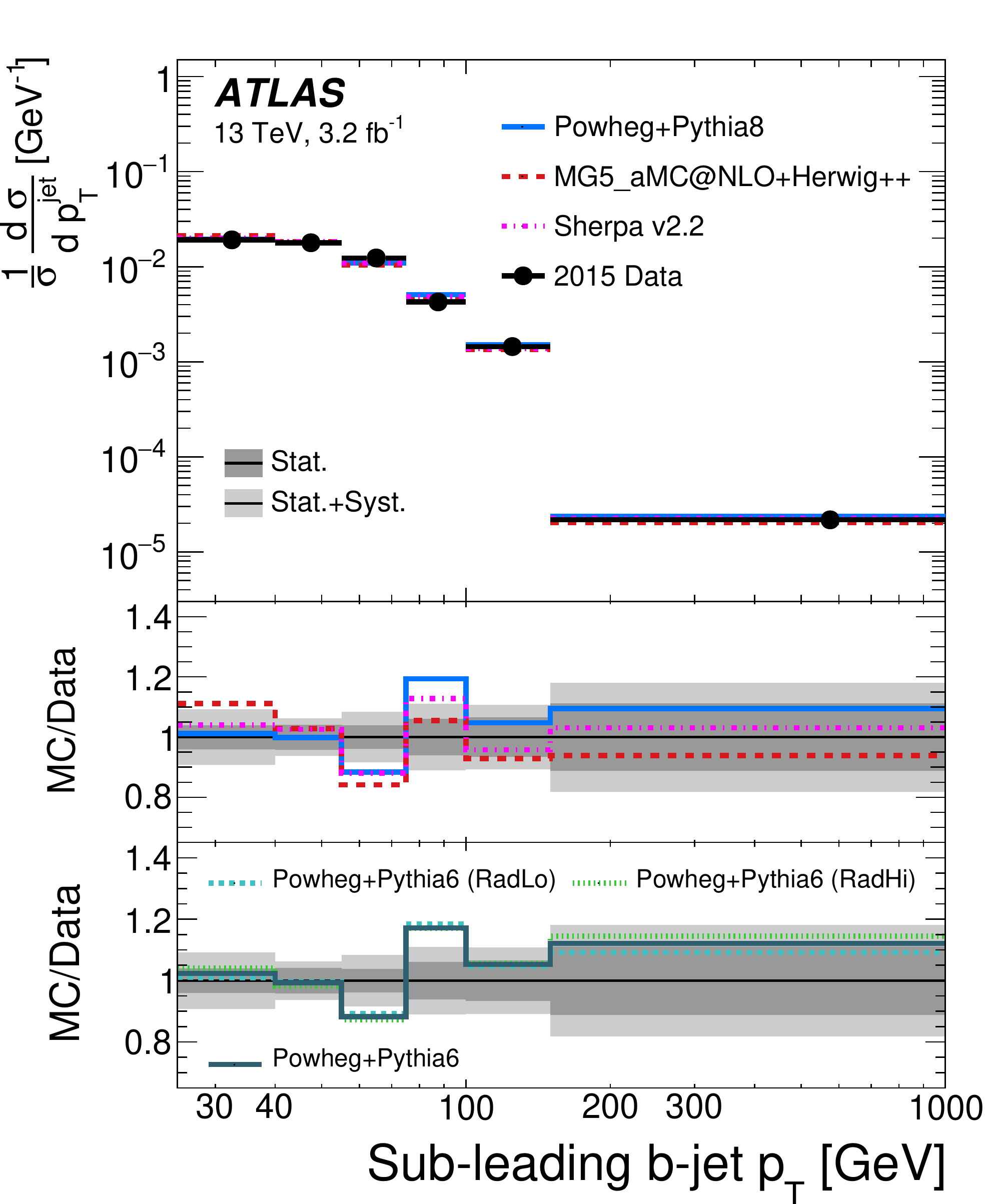}}
\end{tabular}
\begin{tabular}{c}
\subfloat[\label{fig:3ci_2}] {\includegraphics[width=0.42\textwidth]{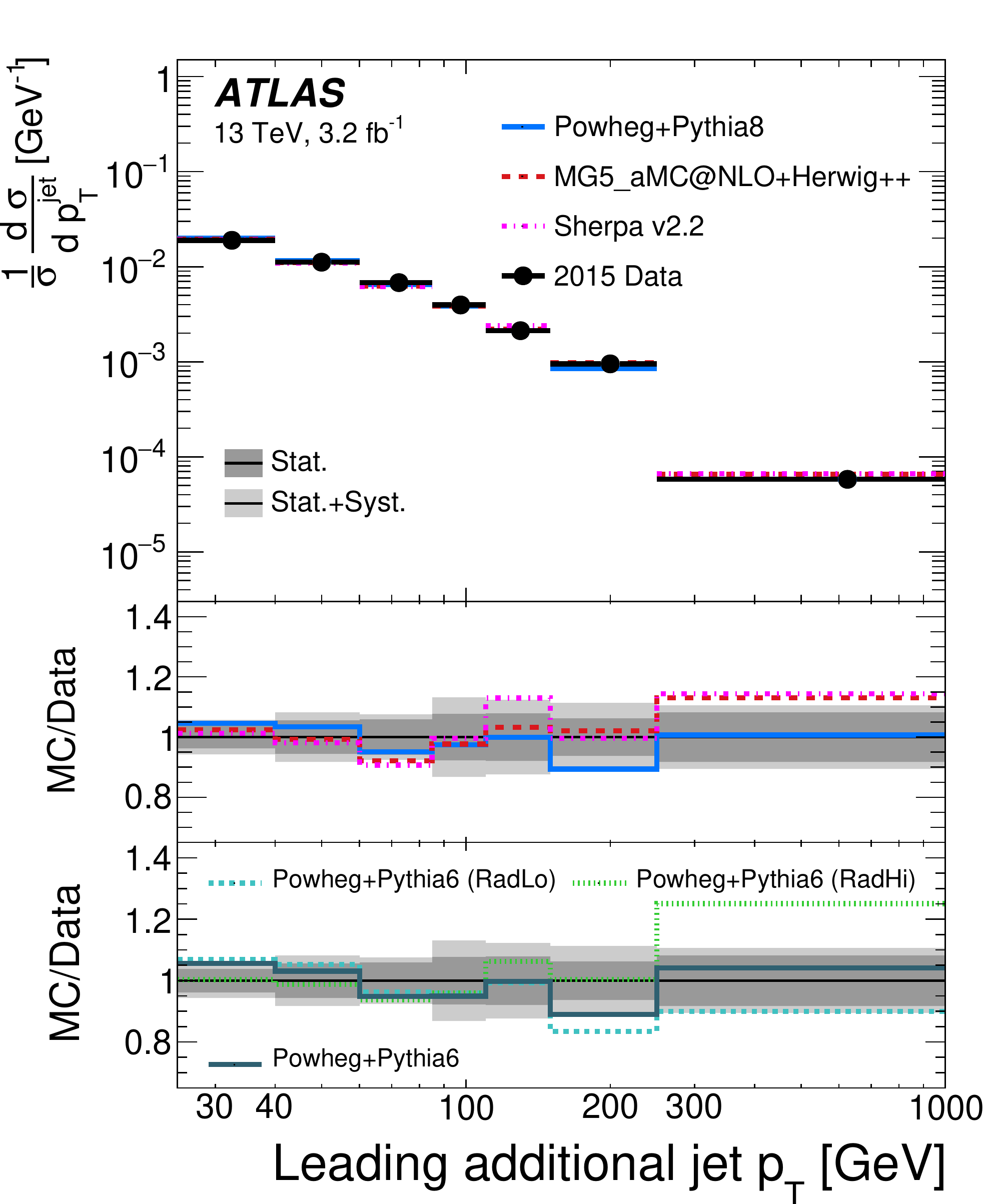} }
\end{tabular}
\caption{Unfolded jet \pt\ distribution for (a) leading $b$-jet, (b) sub-leading $b$-jet and (c) leading additional jet. Comparison to different MC predictions is shown for these distribution in first panel. \ The middle and bottom panels show the ratios of different MC predictions of the normalised cross-section to the measurement and the ratios of \powpy\ predictions with variation of the QCD radiation to the measurement, respectively. The shaded regions show the statistical uncertainty (dark grey) and total uncertainty (light grey).
\label{fig:resultpt}}
\end{figure}

The particle-level normalised cross-sections differential in jet \pt\ are shown in Figure\,\ref{fig:resultpt} and are  compared to different MC predictions. The total uncertainty in the \pt\ measurements is 5--11\%, although higher at some edges of the phase space. The uncertainty is dominated by the statistical uncertainty in almost all bins. The systematic uncertainties are listed in Tables\,\ref{tab:uncleadbpt} to \ref{tab:uncleadaddpt}. JES/JER, NLO generator modelling and PS/hadronisation  are all significant and one of them is always the dominant source of systematic uncertainty. JES/JER is the main source of uncertainty in the lowest \pt\ bins of all measurements.

\begin{table}[ht!]
\begin{center}
\footnotesize
\begin{tabular}{|l|c|c|c|c|c|c|c|c|}
    \hline
 & \multicolumn{2}{c|}{\pt$ > 25$~\gev} & \multicolumn{2} {c|} {\pt$ > 40$~\gev} & \multicolumn{2}{c|}{\pt$ > 60$~\gev}  & \multicolumn{2} {c|}{\pt$ > 80$~\gev} \\
    \cline{2-9}
 Generator & $\chi^2$/NDF  & $p$-value & $\chi^2$/NDF  & $p$-value & $\chi^2$/NDF  & $p$-value & $\chi^2$/NDF  & $p$-value  \\
 \hline \hline 
\pps & 0.82/4 & 0.94 & 0.83/3 & 0.84 & 1.01/3 & 0.80 & 1.82/3 & 0.61 \\
\peight & 0.43/4 & 0.98 & 0.90/3 & 0.83 &  0.64/3 & 0.89 & 1.09/3 & 0.78 \\
\powhw &   0.51/4 & 0.97 &  0.88/3 & 0.83  & 1.46/3 & 0.69 &  2.58/3 & 0.46   \\
\powhws &  8.62/4 & 0.07 &  4.87/3 & 0.18  & 3.17/3 & 0.37 & 2.57/3 & 0.46   \\
\mcnlo+\pye & 5.51/4 & 0.24 &  3.10/3 & 0.38 & 2.25/3 & 0.52 & 2.20/3 & 0.53 \\
\mcnlohw & 1.28/4 &  0.86 &  0.49/3 & 0.92 & 0.34/3 & 0.95 & 0.40/3 & 0.94 \\
\mcnlohws & 3.14/4 & 0.54 &  4.31/3 & 0.23 & 3.57/3 & 0.31 & 2.87/3 & 0.41 \\
\sherpa\ v2.2 &  0.43/4 & 0.98 &   0.85/3 & 0.84  & 0.74/3 & 0.86 & 0.79/3 & 0.85 \\
\hline
\pps\ (RadHi) & 1.20/4 & 0.88 &  1.06/3 & 0.79 & 0.22/3 & 0.97 &   0.22/3 & 0.97 \\
\pps\ (RadLo) & 4.15/4 & 0.39 &  2.05/3 & 0.56 & 2.08/3 & 0.56 &   2.87/3 & 0.41 \\
\hline
\end{tabular}
\caption{Values of $\chi^2/\text{NDF}$ and $p$-values between the unfolded normalised cross-section and the predictions for additional-jet multiplicity measurements. The number of degrees of freedom is equal to the number of bins minus one.
\label{tab:chi2MeasPrednjets}}
\end{center}
\end{table}

\begin{table}[ht!]
\begin{center}
\footnotesize
\begin{tabular}{|l|c|c|c|c|c|c|}
  \hline
 & \multicolumn{2}{c|}{Leading $b$-jet \pt} & \multicolumn{2}{c|}{Sub-leading $b$-jet \pt} & \multicolumn{2}{c|}{Leading additional jet \pt} \\
    \cline{2-7}
 Generator & $\chi^2$/NDF  & $p$-value & $\chi^2$/NDF  & $p$-value & $\chi^2$/NDF  & $p$-value \\
 \hline \hline
\pps & 2.24/6 & 0.90 & 5.85/5 & 0.32 & 3.50/6 & 0.74 \\
\peight & 1.94/6 & 0.93  & 6.33/5 & 0.28 & 2.28/6 & 0.89 \\
\powhw & 1.95/6 & 0.92  & 6.91/5 & 0.23 & 18.5/6 & 0.01  \\
\powhws & 1.26/6 & 0.97  & 5.44/5 & 0.36 & 1.95/6 & 0.92  \\
\mcnlo+\pye & 1.99/6 & 0.92 & 6.76/5 & 0.24 &  10.5/6 & 0.10  \\
\mcnlohw & 2.03/6 & 0.92 & 6.94/5 & 0.23 & 2.97/6 & 0.81  \\
\mcnlohws & 1.32/6 & 0.97 & 4.80/5 & 0.44 &  2.31/6 & 0.89  \\
\sherpa v2.2 &  0.71/6 & 0.99 & 5.37/5 & 0.37 & 4.03/6 & 0.67  \\
\hline
\pps\ (RadHi) & 2.79/6 & 0.83  & 6.55/5 & 0.26 &  1.68/6 & 0.95  \\
\pps\ (RadLo) & 2.16/6 & 0.90  & 5.55/5 & 0.35 &  3.27/6 & 0.77  \\
\hline
\end{tabular}
\caption{Values of $\chi^2/\text{NDF}$  and $p$-values between the unfolded normalised cross-section and the predictions for the jet \pt\ measurements. \ The number of degrees of freedom is equal to one less than the number of bins in the distribution.
\label{tab:chi2MeasPredjetpt}}
\end{center}
\end{table}

\begin{figure}[htbp]
\centering
\begin{tabular}{cc}
\subfloat[\label{fig:3a_1b}] {\includegraphics[width=0.42\textwidth]{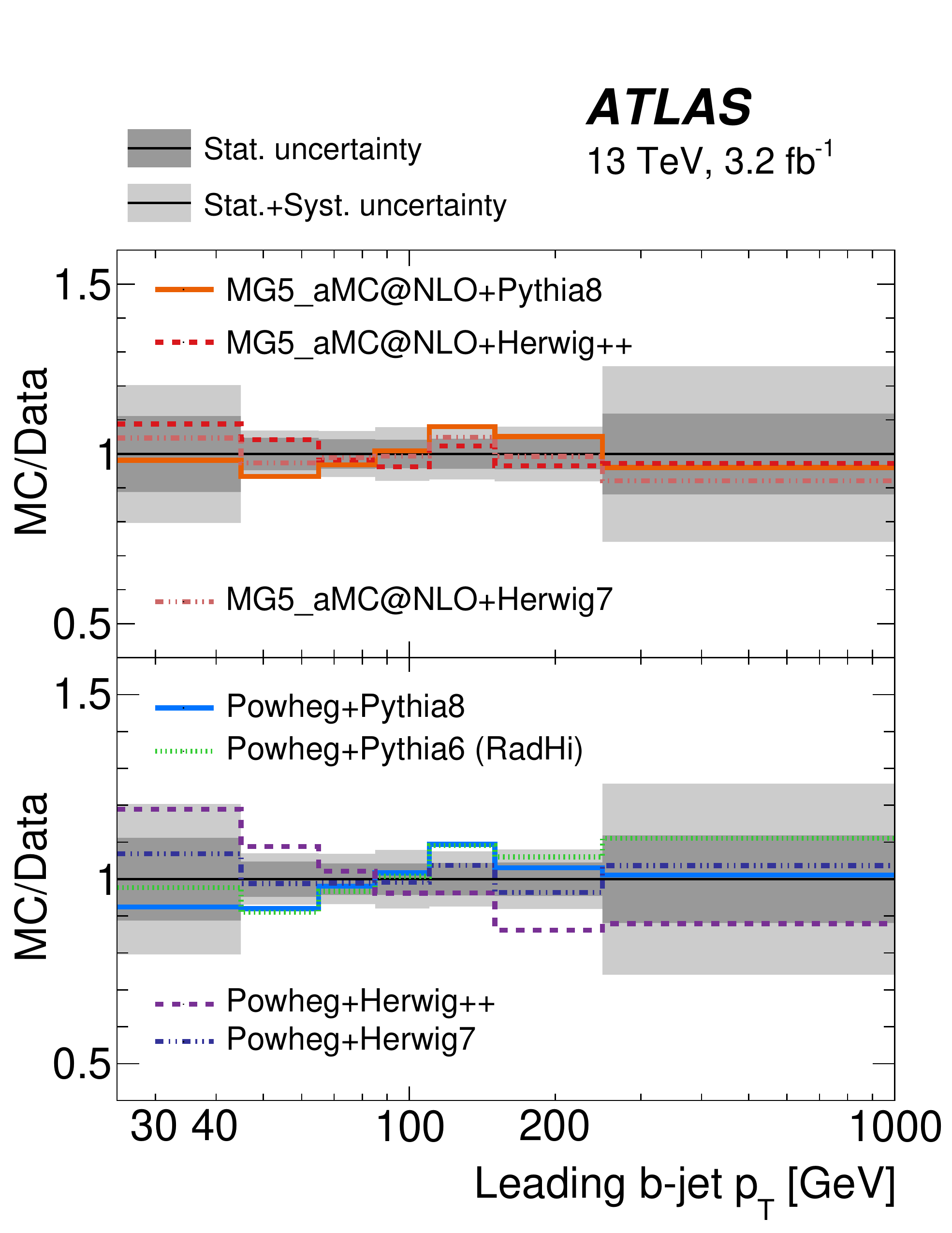}} &
\subfloat[\label{fig:3b_1b}] {\includegraphics[width=0.42\textwidth]{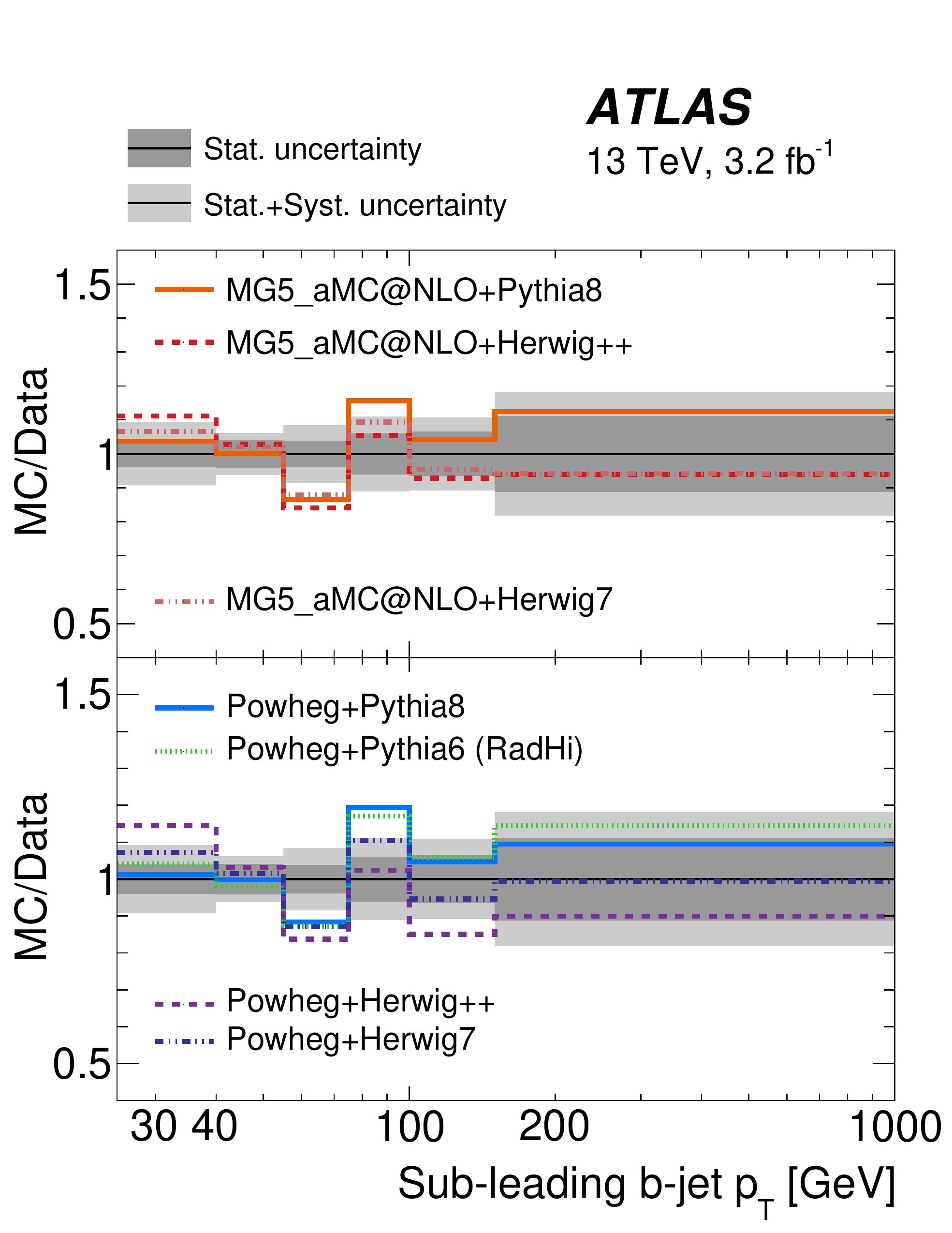}}
\end{tabular}
\begin{tabular}{c}
\subfloat[\label{fig:3ci_b}] {\includegraphics[width=0.42\textwidth]{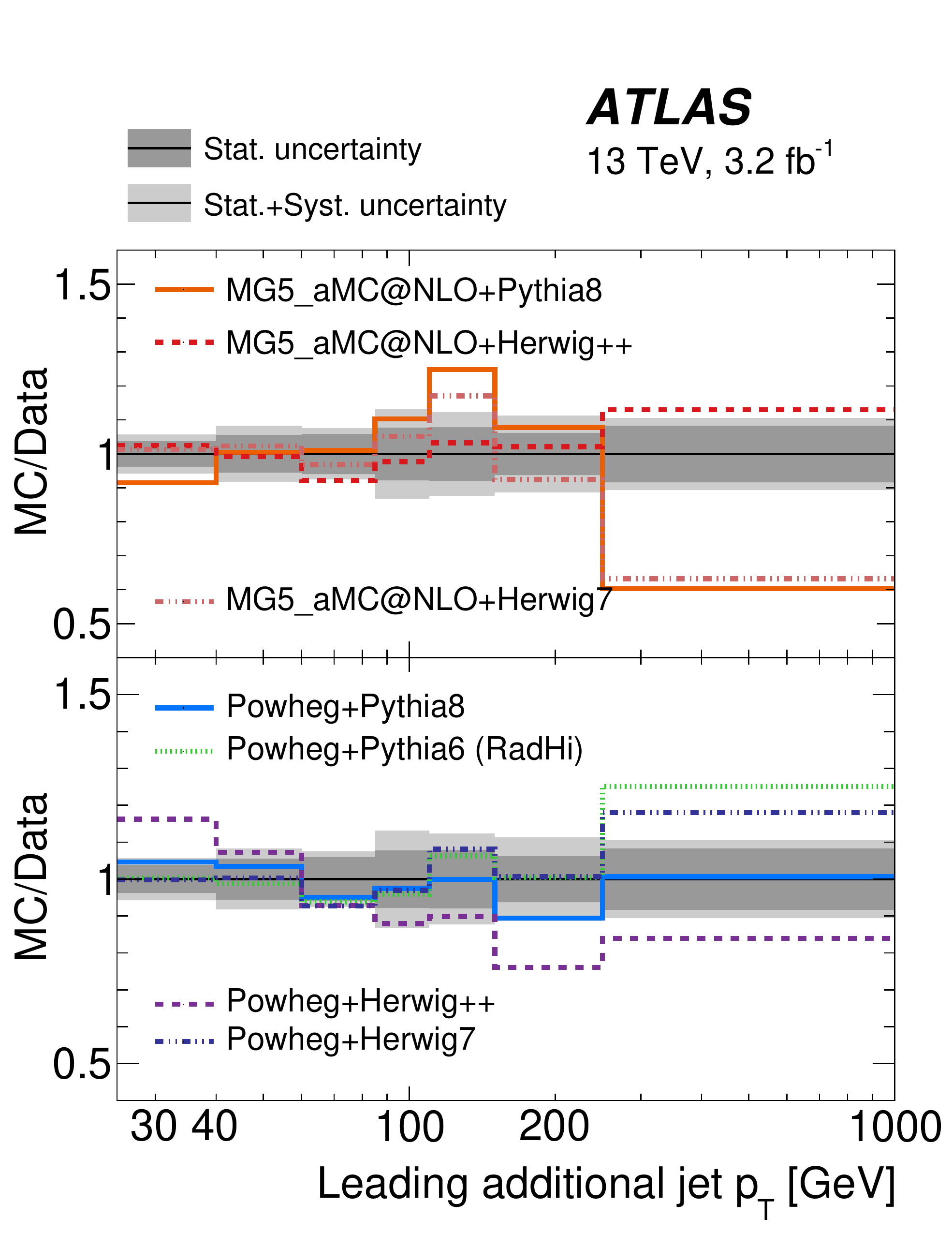} }
\end{tabular}
\caption{Ratios of jet \pt\ distribution for (a) leading $b$-jet, (b) sub-leading $b$-jet and (c) leading additional jet predicted by various MC generators to the unfolded data. \ The shaded regions show the statistical uncertainty (dark grey) and total uncertainty (light grey). 
\label{fig:ptratios}}
\end{figure}

The predictions agree with data for all jet \pt\ distributions as shown in Figure\,\ref{fig:resultpt} and  Figure\,\ref{fig:ptratios}, although the predictions of \powhw\ and \mcnlopye\ do not give a good description of the leading additional-jet \pt\ distribution, which is consistent with the jet multiplicity results. This is reflected by the statistical comparison as well (Table\,\ref{tab:chi2MeasPredjetpt}).

\clearpage

\clearpage

\section{Gap fraction measurements}
The jet activity is also studied by measuring the gap fraction $f_{\mathrm{gap}}$, defined as the fraction of events with no jet activity in addition to the two $b$-tagged jets above a given \pt\ threshold in a ``veto region'' defined as a rapidity region in the detector. The transverse momentum threshold is defined in two ways, and the gap fraction in two ways accordingly. First, the gap fraction is measured as the fraction of events without any additional jet in that rapidity region above a given \pt\ threshold $Q_0$:
\begin{equation}
\label{eq:fgap}
f_{\mathrm{gap}}(Q_0)=\frac{n(Q_0)}{N_{t\overline t}},
\end{equation}
where $N_{t\overline t}$ is the total number of selected events, $Q_0$ is the \pt\ threshold for any additional jet in the veto region of these events, and $n(Q_0)$ represents the subset of events with no additional jet with \pt$>Q_0$.

The second type of gap fraction is defined as the fraction of events in which the scalar \pt\ sum of all additional jets in the given veto region does not exceed a given threshold $Q_{\mathrm{sum}}$:
\begin{equation} \label{e:fgapsum}
f_{\mathrm{gap}}(Q_{\mathrm{sum}})=\frac{n(Q_{\mathrm{sum}})}{N_{t\overline t}}.
\end{equation}
Here, $n(Q_{\mathrm{sum}})$ represents the subset of events in which the scalar \pt\ sum of all additional jets in the veto region is less than $Q_{\mathrm{sum}}$. The gap fraction defined using $Q_0$ is mainly sensitive to the leading \pt\  emission accompanying the \ttbar\ system, whereas the gap fraction defined using $Q_{\mathrm{sum}}$ is sensitive to all hard emissions
accompanying the \ttbar\ system. 
In the following descriptions of the gap fraction measurement process, the same procedure is followed for $Q_{\mathrm{sum}}$ as for $Q_0$.

Both types of gap fraction are measured in four veto regions: $|y|<0.8$, $0.8<|y|<1.5$, $1.5<|y|<2.1$ and the full central region $|y|<2.1$, where $y$ is calculated as 
\begin{equation}
y=\frac{1}{2}\ln\left(\frac{E+p_z}{E-p_z}\right).
\end{equation}
Furthermore, the gap fraction is measured considering jet activity in the full central region ($|y|<2.1$) for four different subsamples specified by the mass of the $\ensuremath{\emu+2~b\textrm{-tagged jets}}$ system, \memubb. Both the rapidity region and the \memubb\ subsamples are chosen to correspond to those used in earlier publications at lower energies~\cite{TOPQ-2011-21,TOPQ-2015-04}. 

The gap fraction \fqzero\ (and analogously for \fqsum\ in the following) is measured as defined in Equation~(\ref{e:fgapm}) by counting the number of selected data events $N_{\textrm data}$ and the number $n_{\mathrm{data}}(\qzero)$ of those that had no additional jets with $\pt>\qzero$ within the veto region, where the sets of \qzero\ and \qsum\ threshold values correspond approximately to one standard deviation of the jet energy resolution and are the same as in the earlier publications~\cite{TOPQ-2011-21, TOPQ-2015-04}. The number of background events, $N_{\mathrm{bg}}$ and $n_{\mathrm{bg}}(Q_0)$, are then subtracted from these events:
\begin{equation} \label{e:fgapm}
f^{\mathrm{data}}(Q_0)=\frac{n_{\mathrm{data}}(Q_0)- n_{\mathrm{bg}}(Q_0)}{N_{\mathrm{data}} -N_{\mathrm{bg}}}
\end{equation}
and similarly for \fqsum. 
The measured gap fraction \fdata\ is then corrected for detector effects to particle level by multiplying it by a correction factor \corqz\ to obtain \fqzero. The correction factor \corqz\ is determined from the baseline \powpy\ \ttbar\ sample using the simulated gap fraction values at reconstruction level \frecoqz, and at particle level \fpartqz\ : 
\begin{equation}
\corqz = \frac{\fpartqz}{\frecoqz} .
\end{equation}
The values of the correction factors \corqz\ and \corqsm\ deviate by less than 4\% from unity at low \qzero\ and \qsum\ values in the rapidity regions (less than 8\% in the \memubb\ subsamples), and approach unity at higher threshold values. The small corrections reflect the high selection efficiency and high purity of the event samples. At each threshold \qzero, the baseline simulation predicts that around 80\% of the selected reconstructed events that do not have a jet with $\pt>\qzero$ also have no particle-level jet with $\pt>\qzero$. Therefore, a simple bin-by-bin correction method is considered adequate, rather than a full unfolding as used in Section~\ref{sec:jets}.

\begin{figure}[tp]
\centering
\subfloat[][]{\includegraphics[width=0.5\textwidth]{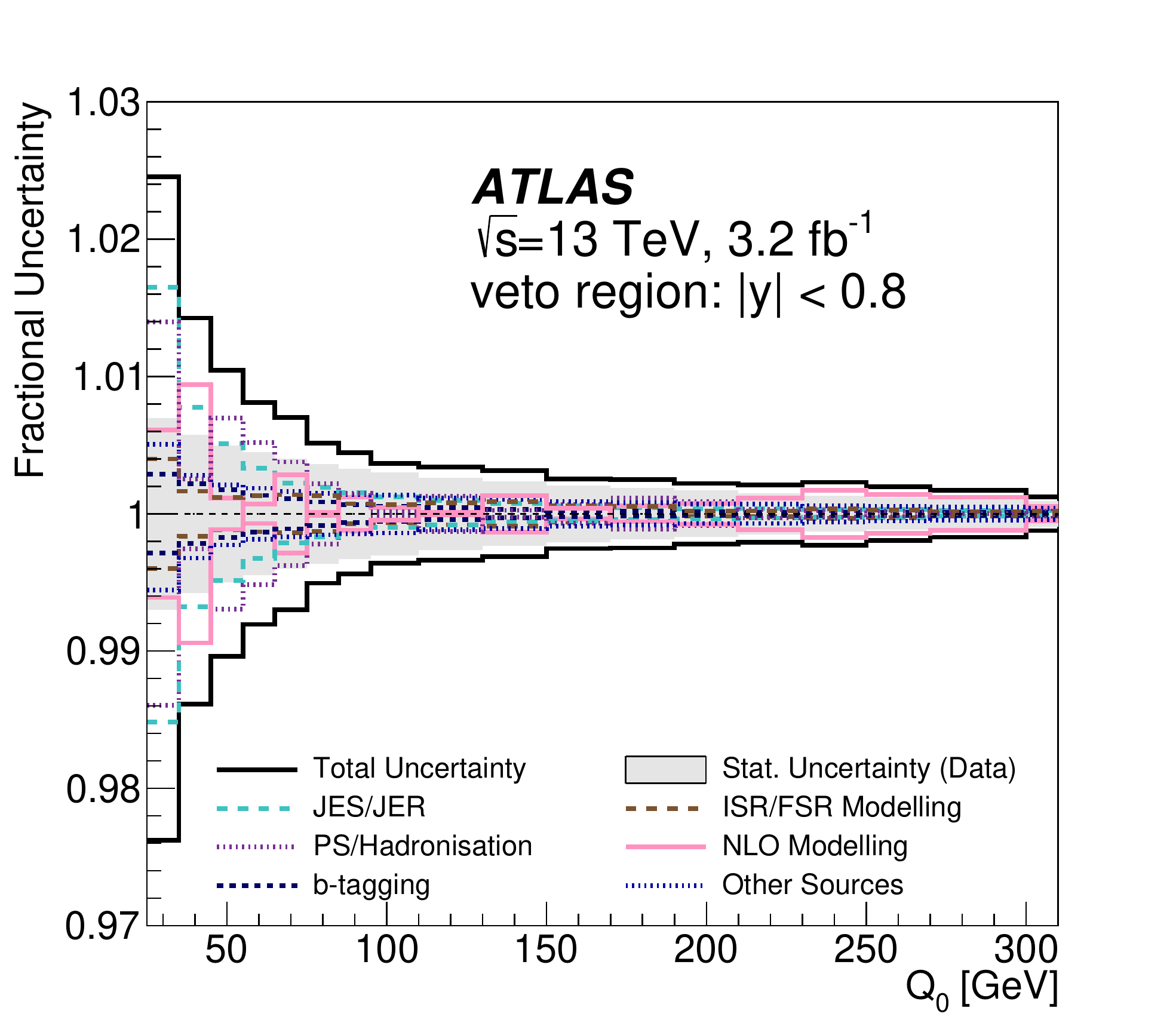}}
\subfloat[][]{\includegraphics[width=0.5\textwidth]{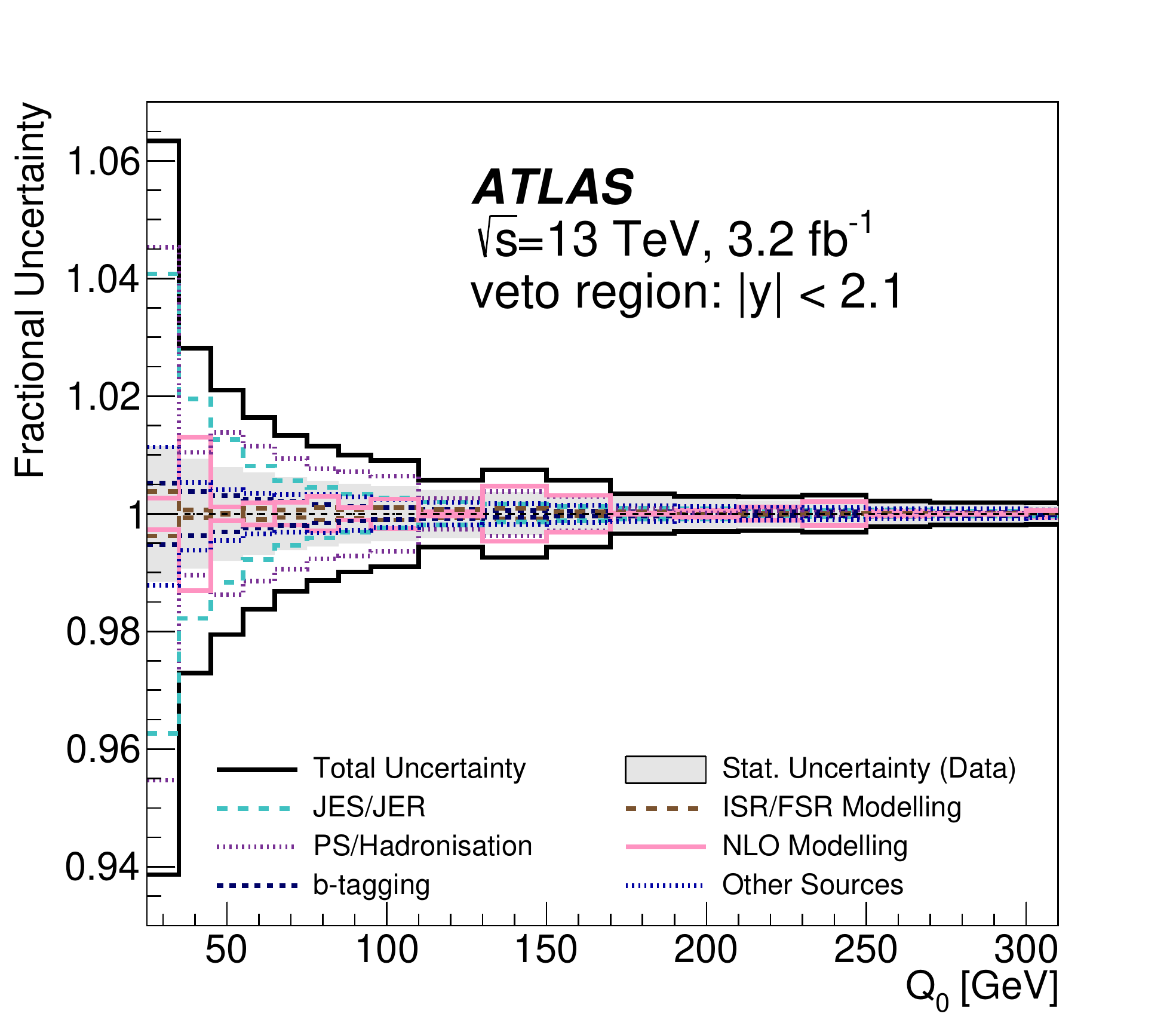}} \\
\caption{Envelope of fractional uncertainties $\Delta f/f$ in the gap fraction \fqzero, centred around unity, for (a) $|y|<0.8$ and (b) $|y|<2.1$. The statistical uncertainty is shown by the shaded area, and the total uncertainty by the solid black line. The systematic uncertainty is shown broken down into several groups, each of which includes various individual components. }
\label{fig:gfsyst}
\end{figure}

Systematic uncertainties arise in this procedure from the uncertainties in \corqz\ and the subtracted backgrounds. The uncertainties, as described in Section~\ref{sec:systematics}, are used to recalculate \fdata\ and \corqz\ to obtain the gap fraction \fqzero. The corresponding quantities for \qsum\ are calculated accordingly. Figure\,\ref{fig:gfsyst} and Table\,\ref{table:systq0} list the resulting relative uncertainty in \fqzero, $\Delta f/f$, for the different sources of uncertainty in the full central rapidity region.

\begin{table}
\centering
\footnotesize
\begin{tabular}{|l|c|c|c|c|c|c|c|}
  \hline
  & \multicolumn{7}{c|}{Uncertainty in  $f_{\textrm gap}(Q_{0})$ in [\%]}\\
  &  \multicolumn{7}{c|}{jet \pt\ threshold} \\
  & 25  & 45  & 65  & 95  & 110 & 150 & 250 \\
  Sources  &  GeV & GeV & GeV& GeV& GeV&GeV&GeV\\  \hline \hline
Data statistics          & 1.2 & 0.8 & 0.6 & 0.5 & 0.4 & 0.3 & 0.2 \\
JES/JER             & 4.1 & 1.3 & 0.6 & 0.3 & 0.2 & 0.1 & 0.1 \\
$b$-tagging           & 0.5 & 0.3 & 0.2 & 0.1 & 0.1 & 0.0 & 0.0 \\
ISR/FSR modelling   & 0.4 & 0.0 & 0.1 & 0.1 & 0.1 & 0.0 & 0.0 \\
Signal Modelling    & 4.5 & 1.4 & 1.0 & 0.7 & 0.3 & 0.4 & 0.1 \\
Other       & 1.1 & 0.4 & 0.3 & 0.2 & 0.2 & 0.1 & 0.1 \\ \hline
{\textbf Total }              & {\textbf 6.3 }& {\textbf 2.1} & {\textbf 1.3} & {\textbf 0.9} & {\textbf 0.6} & {\textbf 0.6} & {\textbf 0.2} \\ \hline
\end{tabular}
\caption{Sources of uncertainty in the gap fraction measurement as a function of $Q_0$ for the full central region $|y|<2.1$, for a selection of $Q_0$ thresholds.  "Signal modelling" sources  of systematic uncertainty includes the hadronisation, parton shower and NLO modelling uncertainties. "Other" sources of systematic uncertainty refer to lepton and jet selection efficiencies, background (including pile-up jets) estimations, and the PDF.}
\label{table:systq0}
\end{table}

\subsection{Gap fraction results in rapidity regions}\label{sec:gfracresrap}

\begin{figure}[tp]
\centering
\subfloat[][]{\includegraphics[width=0.425\linewidth]{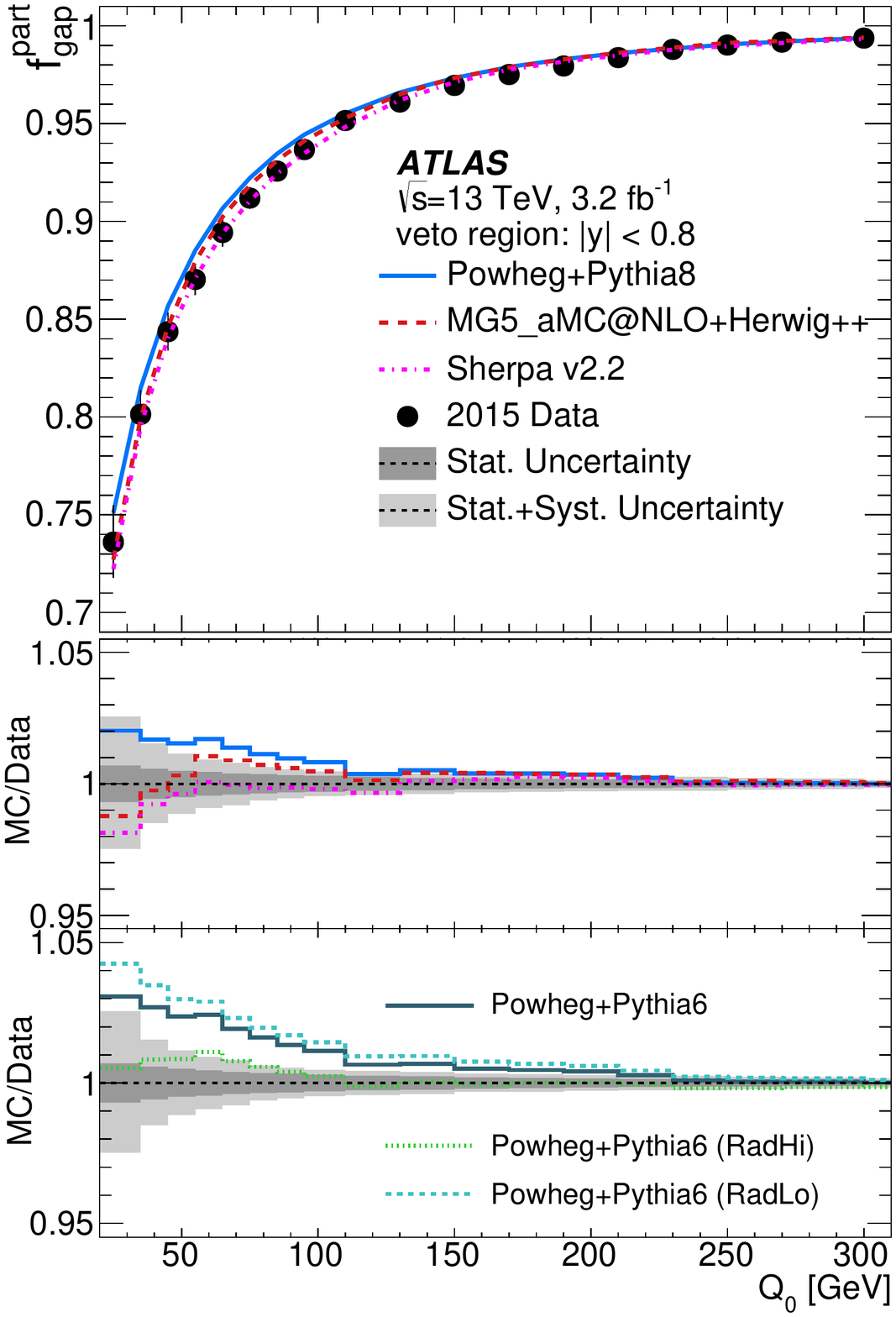}}
\subfloat[][]{\includegraphics[width=0.425\linewidth]{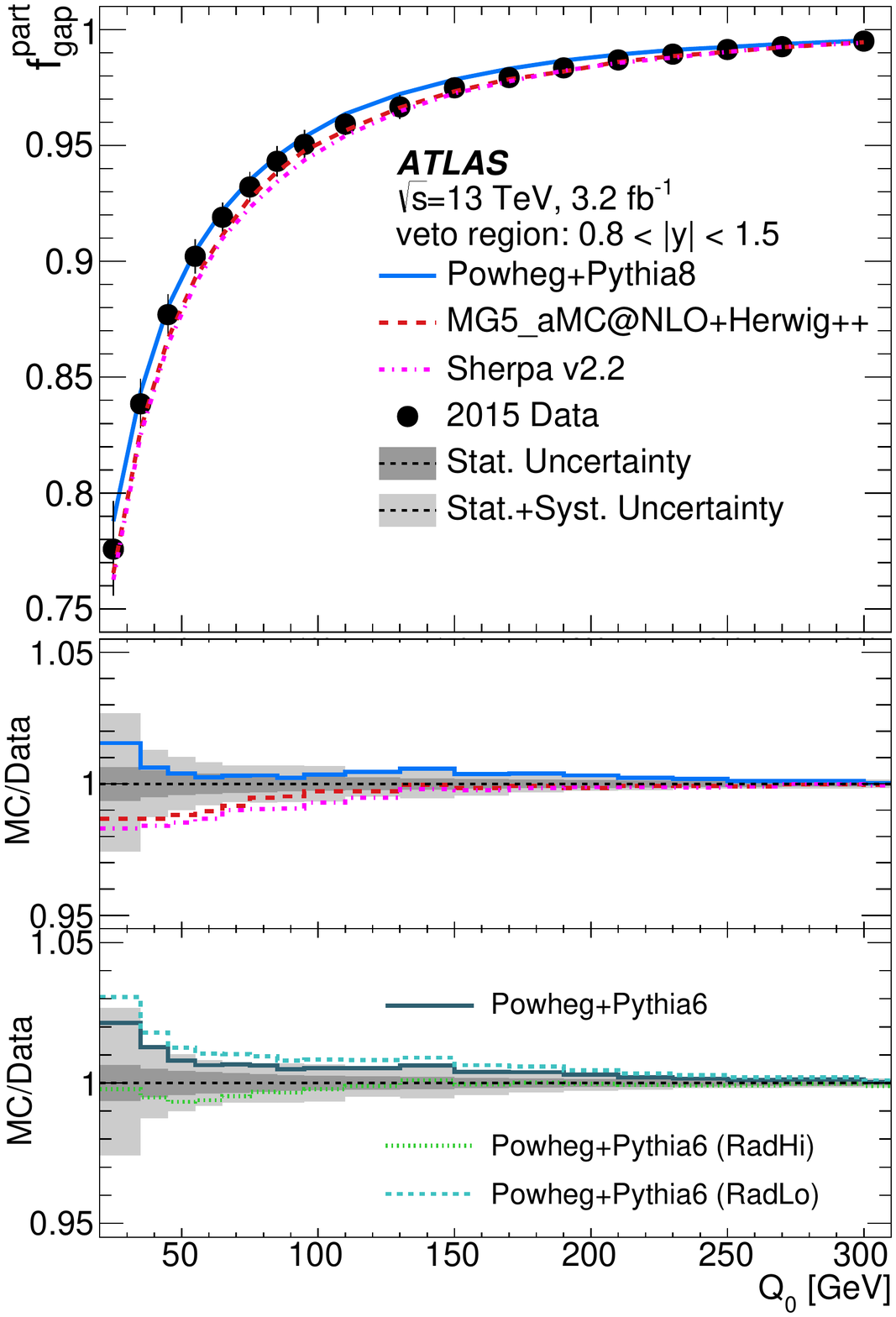}}\\
\subfloat[][]{\includegraphics[width=0.425\linewidth]{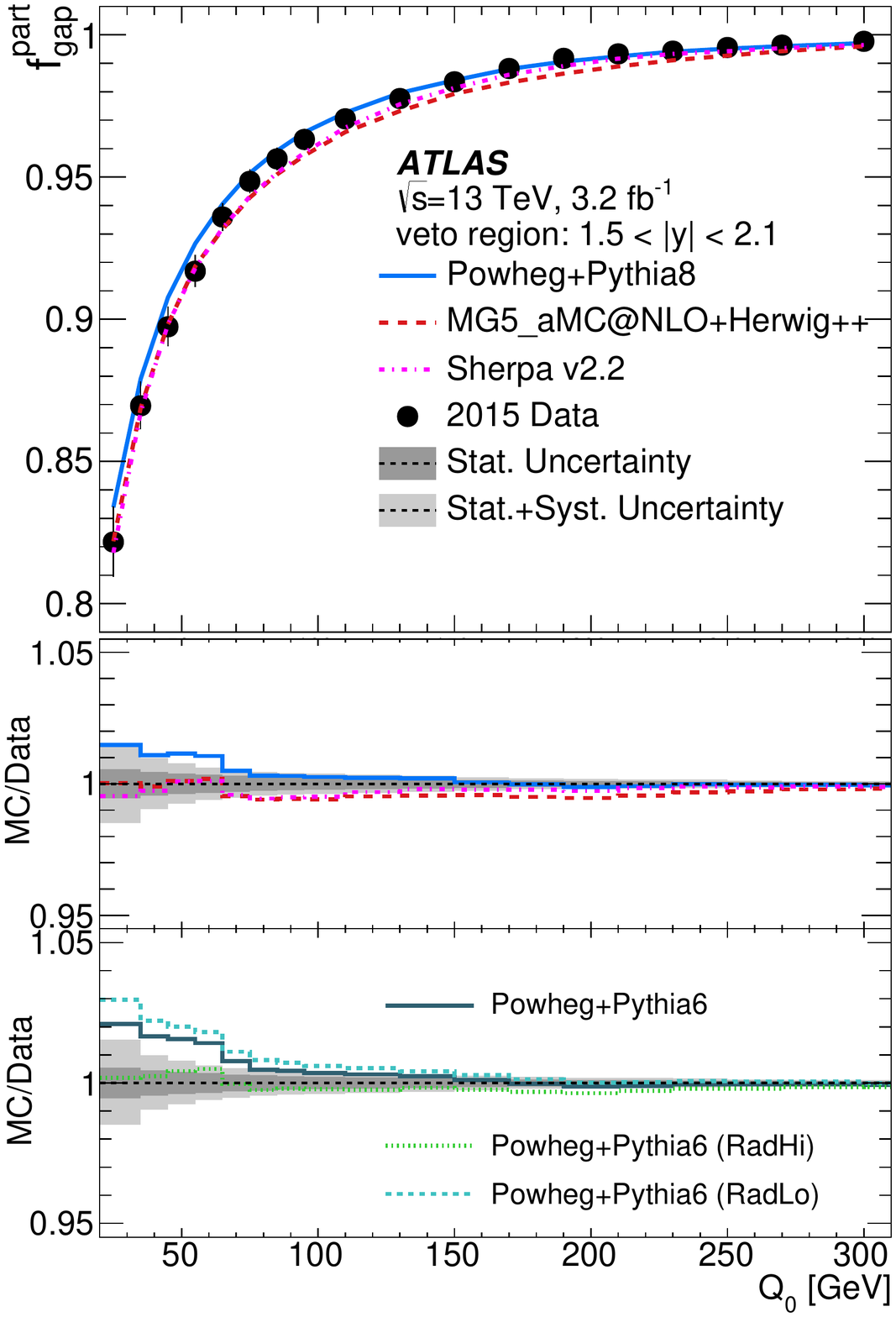}}
\subfloat[][]{\includegraphics[width=0.425\linewidth]{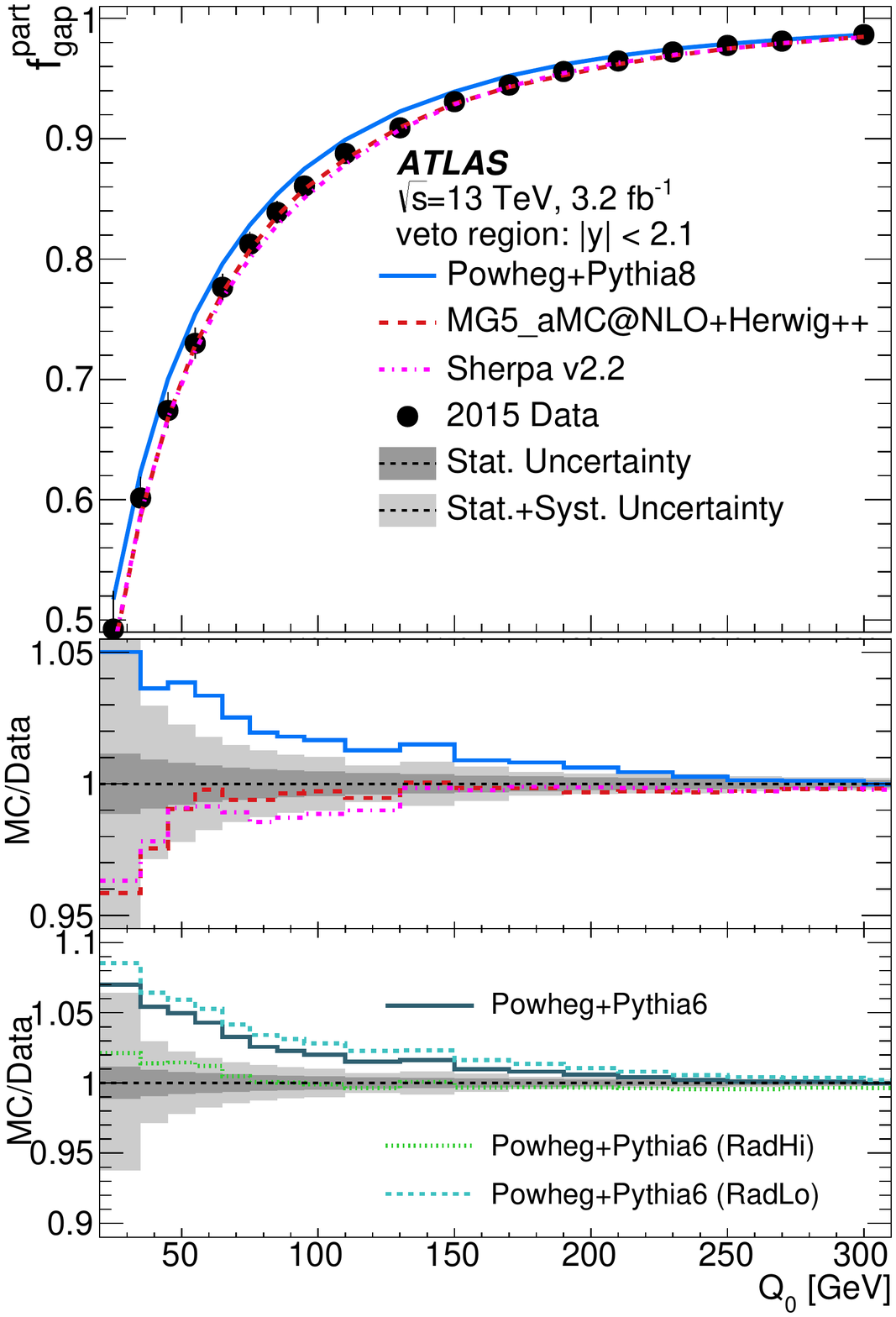}}
\caption{The measured gap fraction \fqzero\ as a function of \qzero\ in different rapidity veto regions, (a) $|y|<0.8$, (b) $0.8<|y|<1.5$, (c) $1.5<|y|<2.1$ and (d) $|y|<2.1$. The data are shown by the points with error bars indicating the total uncertainty, and compared to the predictions from various \ttbar\ simulation samples shown as smooth curves. The lower plots show the ratio of predictions to data, with the data uncertainty indicated by the shaded band, and the \qzero\ thresholds corresponding to the left edges of the histogram bins, except for the first bin.}
\label{fig:gapfracpart}
\end{figure}

\begin{figure}[tp]
\centering
\subfloat[][]{\includegraphics[width=0.425\linewidth]{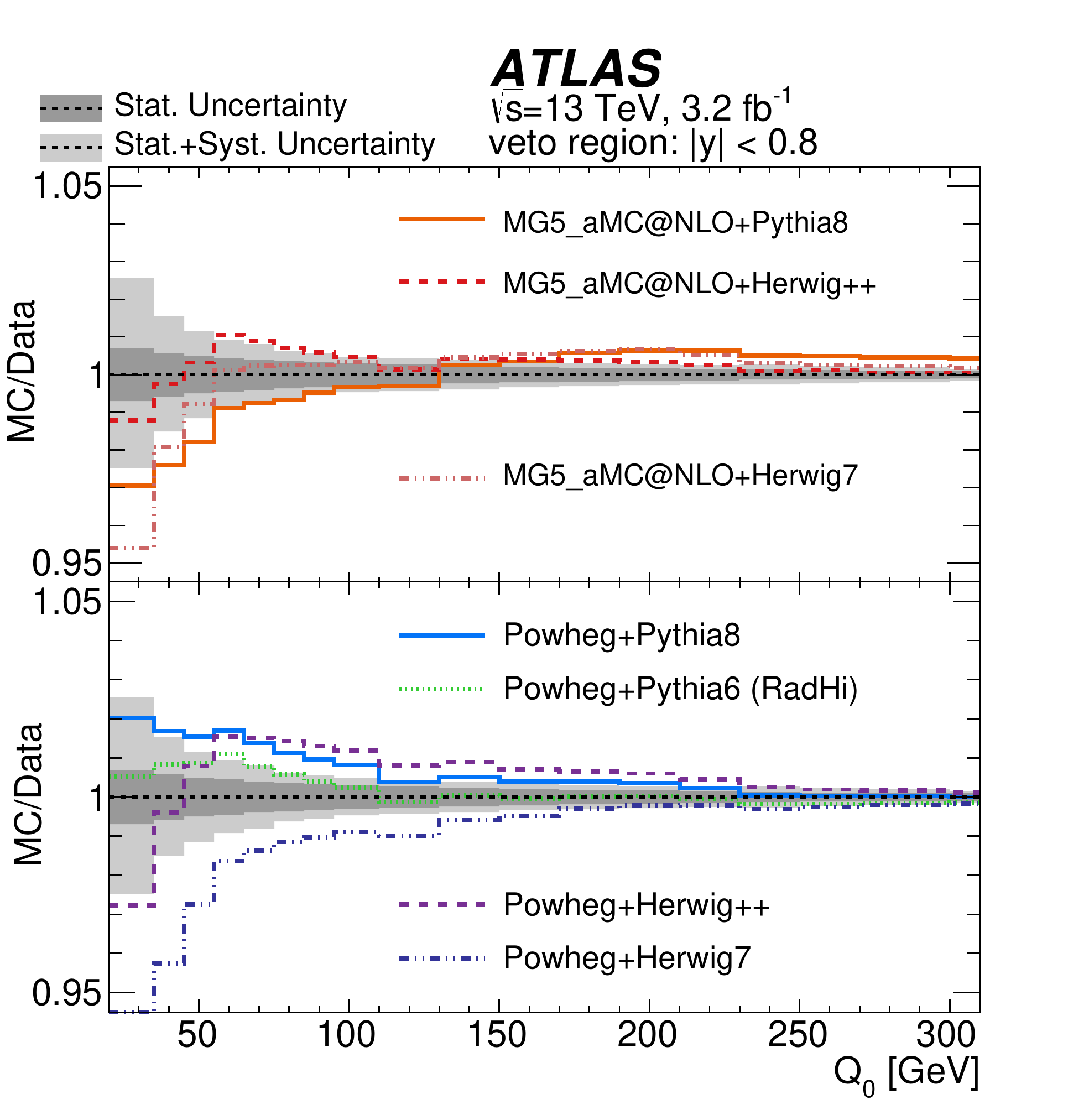}}
\subfloat[][]{\includegraphics[width=0.425\linewidth]{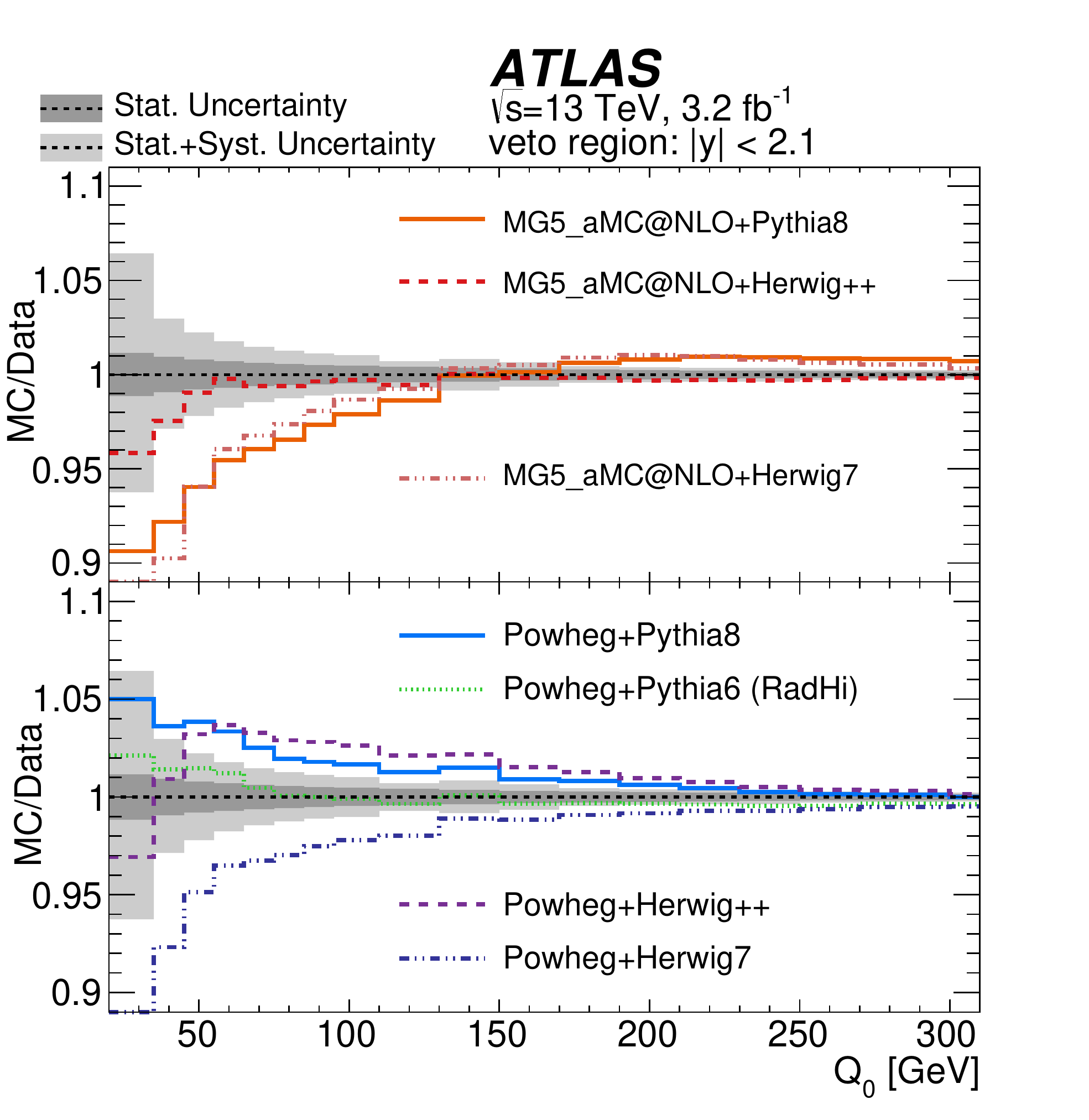}}
\caption{Ratios of prediction to data of the measured gap fraction \fqzero\ as a function of \qzero\ in different rapidity veto regions, (a) $|y|<0.8$ and (b) $|y|<2.1$. The predictions from various \ttbar\ simulation samples are shown as ratios to data, with the data uncertainty indicated by the shaded band, and the \qzero\ thresholds corresponding to the left edges of the histogram bins, except for the first bin.}
\label{fig:gapfracpartrat}
\end{figure}

\begin{figure}[tp]
\centering
\subfloat[][]{\includegraphics[width=0.415\linewidth]{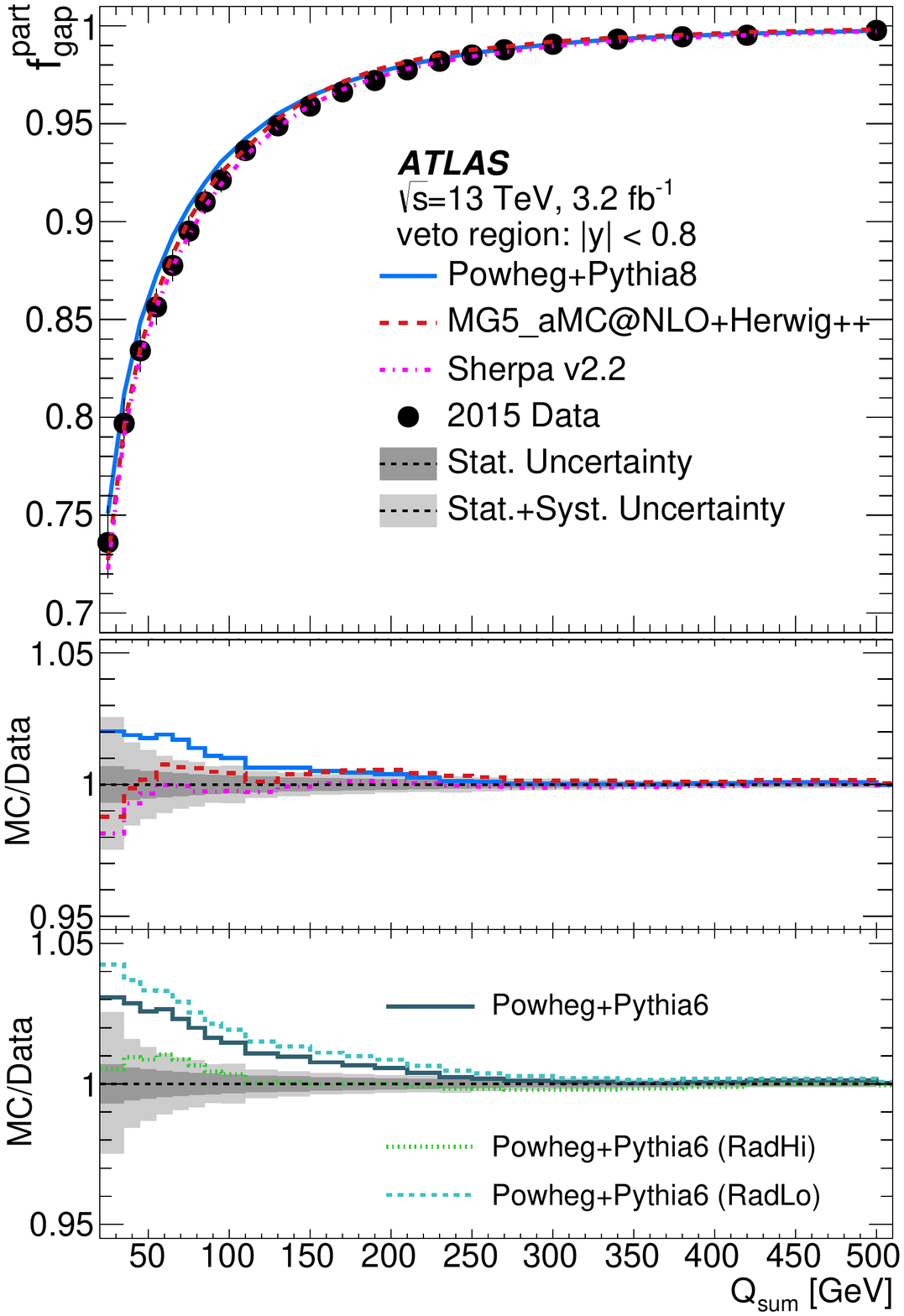}}
\subfloat[][]{\includegraphics[width=0.415\linewidth]{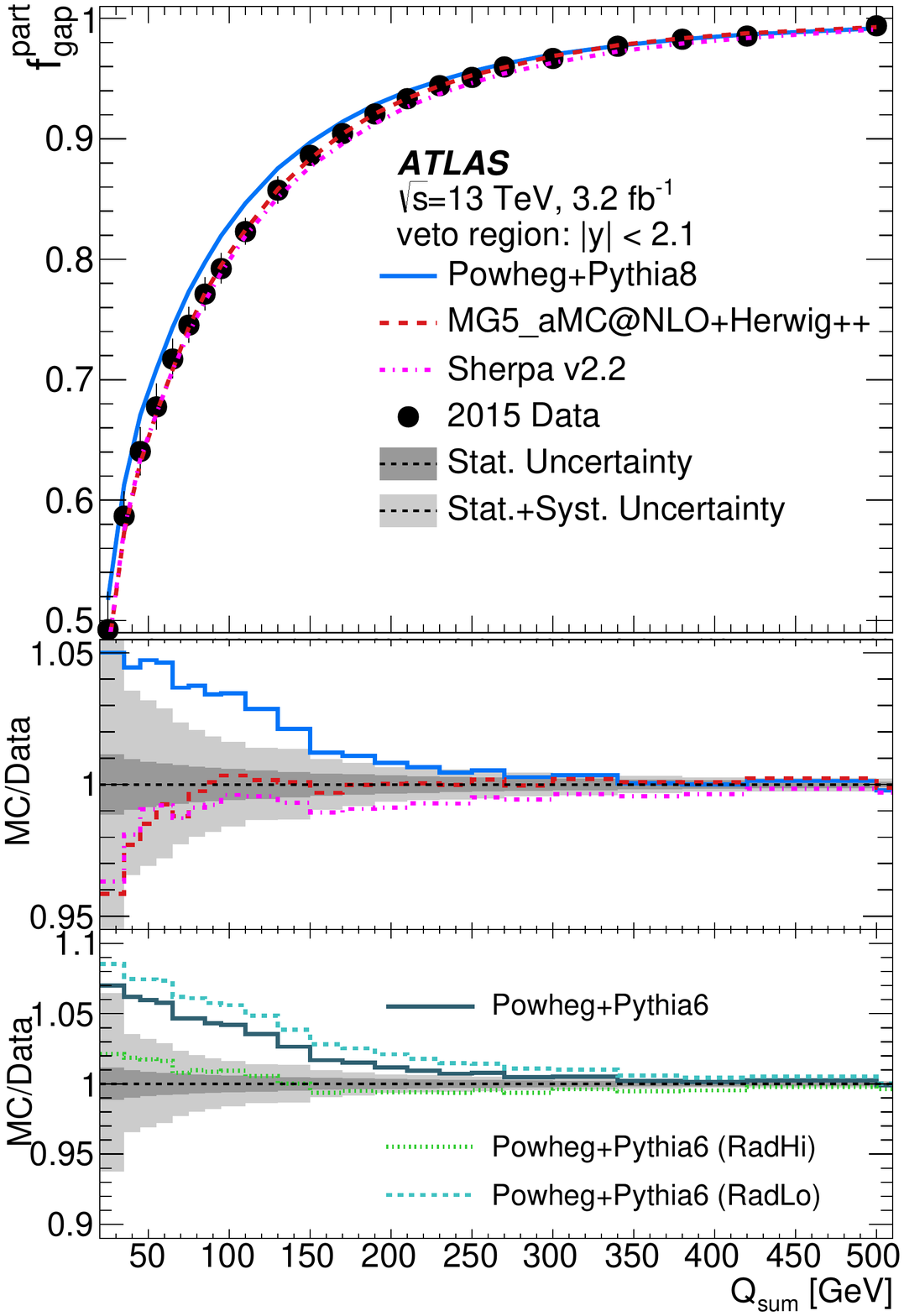}}\\
\subfloat[][]{\includegraphics[width=0.415\linewidth]{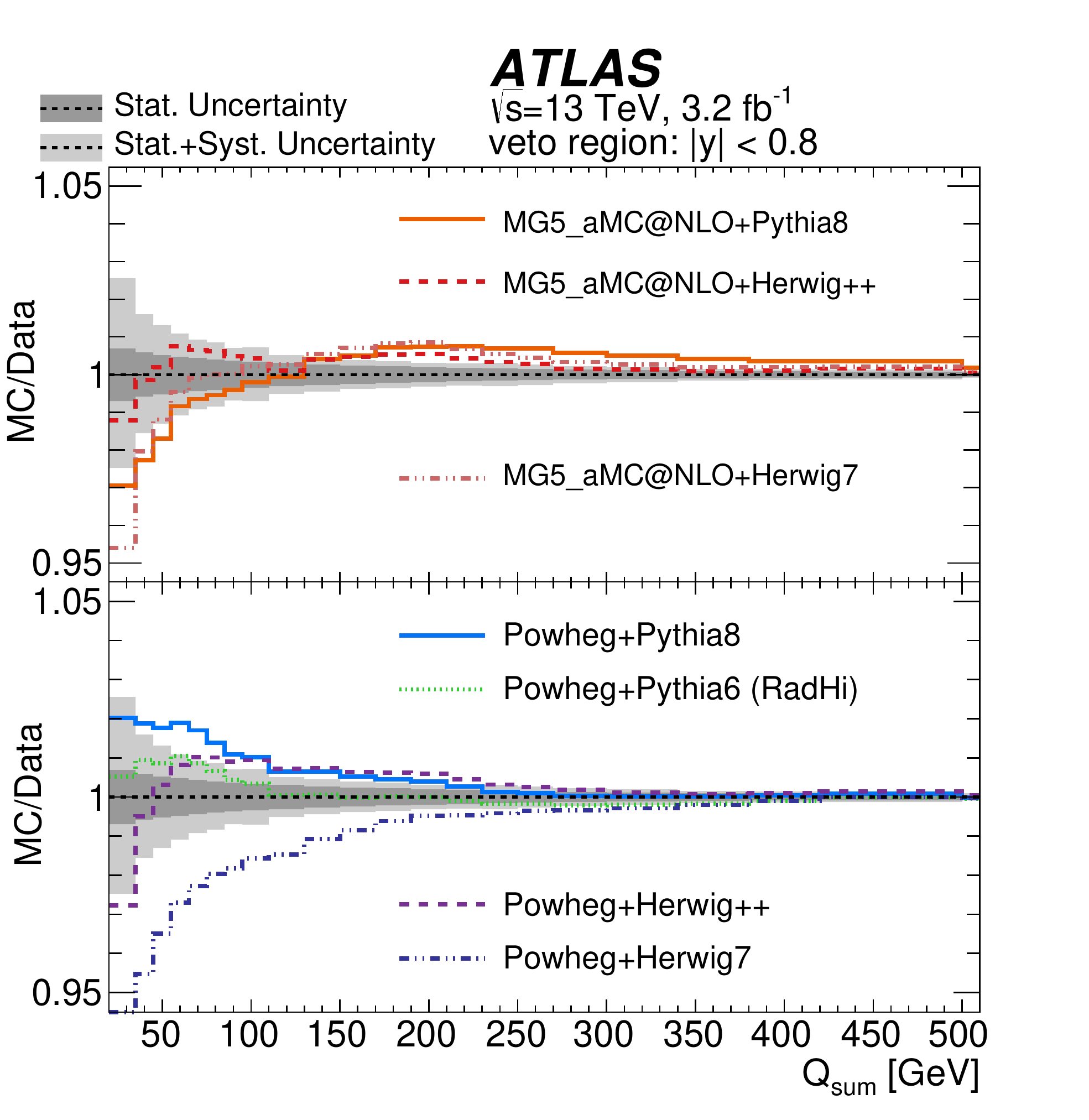}}
\subfloat[][]{\includegraphics[width=0.415\linewidth]{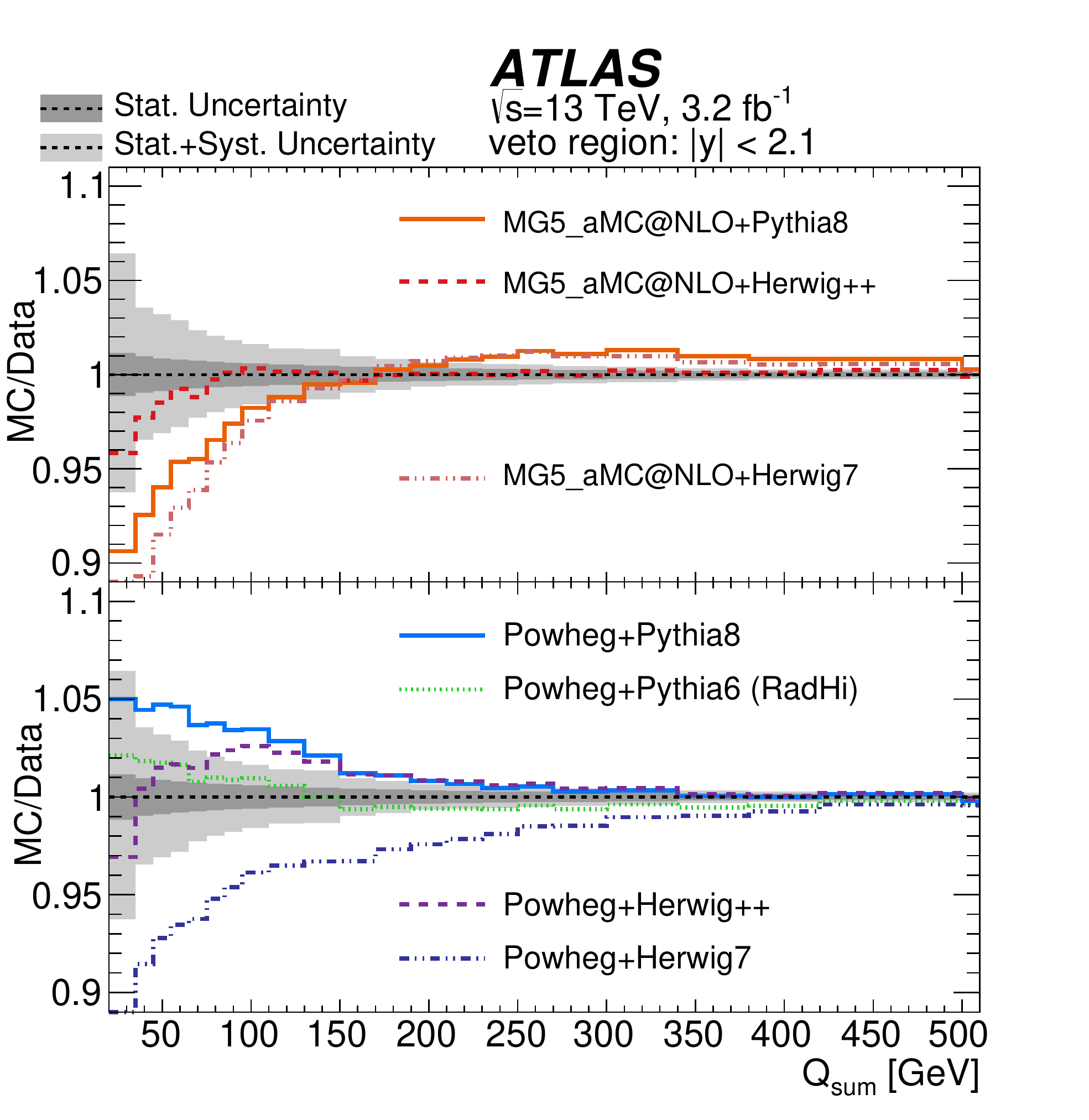}}
\caption{The measured gap fraction \fqsum\ as a function of \qsum\ in different rapidity veto regions, (a) $|y|<0.8$ and (b) $|y|<2.1$, followed by ratios of prediction to data of the measured gap fraction \fqsum\ as a function of \qsum\ in the same two rapidity regions. The data in (a) and (b) are shown by the points with error bars indicating the total uncertainty, and compared to the predictions from various \ttbar\ simulation samples shown as smooth curves. The lower plots in (a) and (b) and the set of ratio plots in (c) and (d) show the ratio of predictions to data, with the data uncertainty indicated by the shaded band, and the \qsum\ thresholds corresponding to the left edges of the histogram bins, except for the first bin.}
\label{fig:gapfracparttot}
\end{figure}

Figure~\ref{fig:gapfracpart} shows the measured gap fractions \fqzero\ in data, corrected to the particle level. The gap fraction \fqzero\ is compared to various MC generator predictions in Figure~\ref{fig:gapfracpartrat}, and Figure~\ref{fig:gapfracparttot} shows the measured gap fractions \fqsum\ compared to various MC generators, corrected to the particle level. The predictions of \sherpa\ and \mcnlo\ +{\textsc Herwig}++ agree well with each other and are within the uncertainties of the data, while \peight\ has slightly higher gap fractions, i.e., predicts too little radiation. Similarly to the jet multiplicity measurements, \powpy\ (RadHi) agrees well with data, while the nominal and the \powpy\ (RadLo) samples give similar but too high predictions compared to data. The results in Fig.~\ref{fig:gapfracpart} (d) can directly be compared with the jet multiplicity results in Fig.~\ref{fig:resultmult}  and ~\ref{fig:multratios} in the one additional jet bin. Here the \peight\ predictions are below 
data for all distributions which proves the consistency of the measurements.  The \pt\ distribution of the first additional jet  shown in Fig.~\ref{fig:resultpt} contains only events with at least one additional jet and   differs in this respect from    the gap fraction distribution which includes events 
with no additional jet.  However, the  results are also consistent as \peight\ predicts a slightly softer \pt\ spectrum for the additional jet which  leads to the observed effect that  less jets above the 25\,GeV threshold are observed.

\begin{figure}[tp]
\centering
\includegraphics[width=0.5\linewidth]{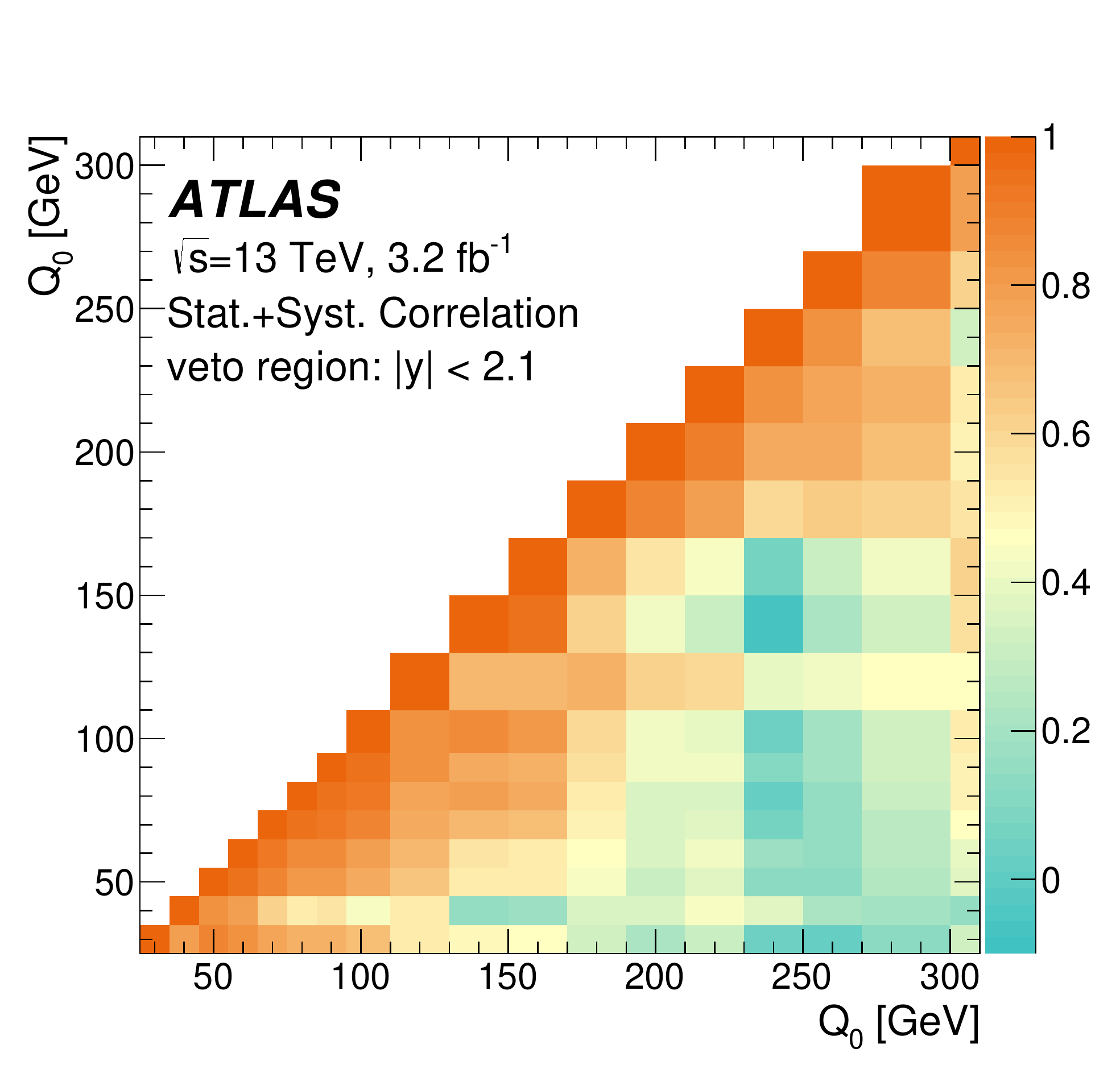}
\caption{The correlation matrix (including statistical and systematic correlations) for the gap fraction measurement at different values of $Q_0$ for the full central rapidity region $|y|<2.1$.}
\label{fig:corrcomb}
\end{figure}

The matrix of statistical and systematic correlations is shown in Figure\,\ref{fig:corrcomb} for the gap fraction measurement at different values of $Q_0$ for the full central $|y|<2.1$ rapidity region. Nearby points in $Q_0$ are highly correlated, while well-separated $Q_0$ points are less correlated. The full covariance matrix, including correlations, is used to calculate a $\chi^2$ value for the compatibility of each of the NLO generator predictions with the data in each veto region. The results are given in Tables~\ref{table:chirapQ0} and~\ref{table:chirapQsum}. 
An analysis of the $p$-values confirms that \powhw, \mcnlohws, \mcnlopye\ and \powpy\ (RadLo) are not consistent with the data. \powpy\ (RadHi) has the best 
$p$-values among the QCD shower variations of \powpy.

\FloatBarrier
\clearpage
\subsection{Gap fraction results in \memubb\ subsamples}\label{sec:gfracresmass}
\begin{figure}[tp]
\centering
\includegraphics[width=0.6\textwidth]{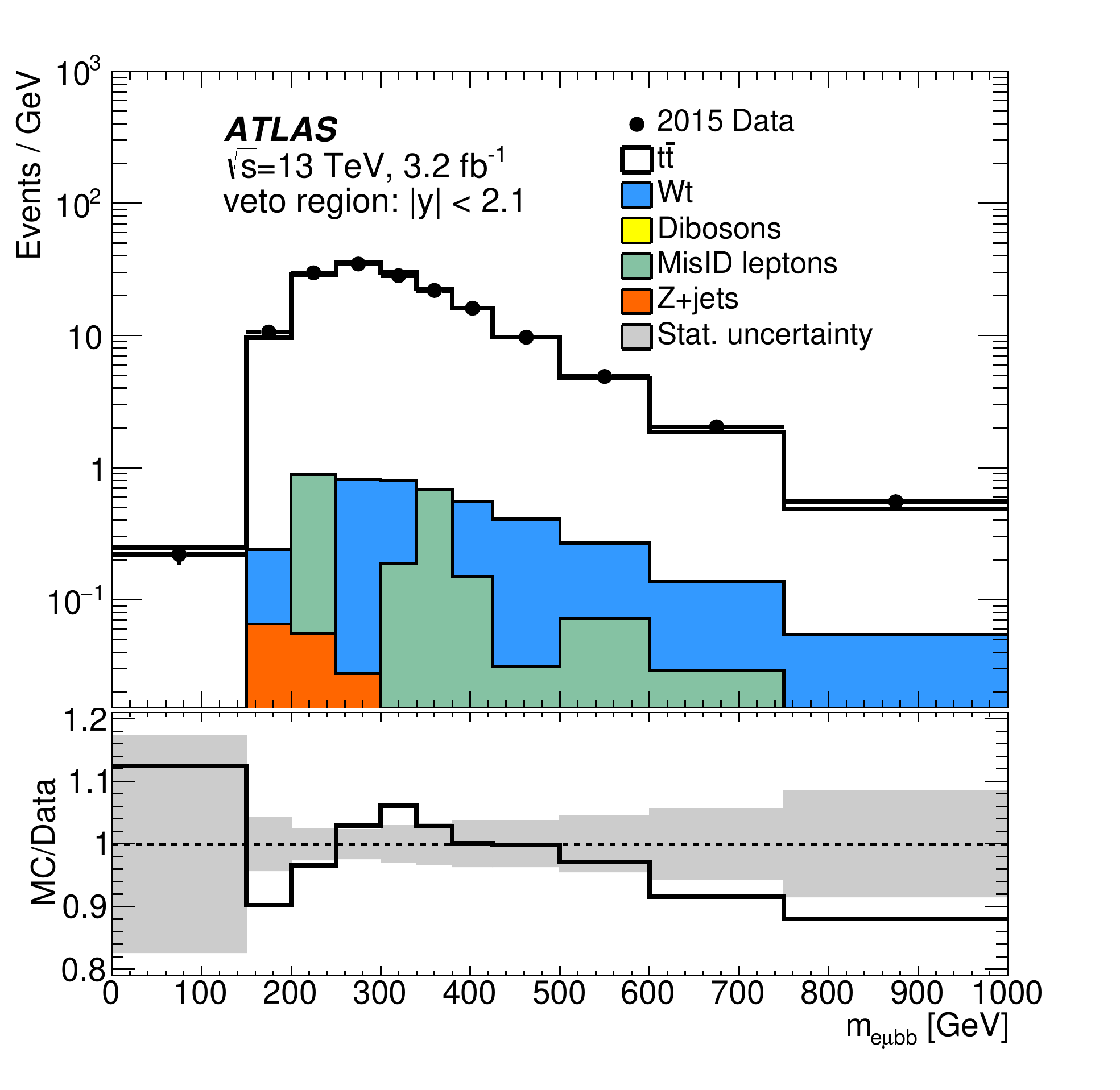} 
\caption{Distribution of the reconstructed invariant mass of the \emubb\ system \memubb\ in data, compared to simulation. The shaded band represents the statistical uncertainty in data. The lower plot shows the ratio of the distribution of invariant mass in simulation compared to data.}
\label{fig:massdist}
\end{figure}

The gap fraction is also measured over the full central veto region $|y|<2.1$ after dividing the data sample into four regions of \memubb. The distribution of reconstructed \memubb\ in the selected $\ensuremath{\emu+2~b\textrm{-tagged jets}}$ events is reasonably well-reproduced by the nominal \ttbar\ simulation sample, as shown in Figure~\ref{fig:massdist}. The distribution is divided into four regions at both reconstruction and particle level: \memubb$<300$~\gev, $300~\gev\ <\memubb<425$~\gev, $425~\gev\ <\memubb<600$~\gev\ and $\memubb>600$~\gev. These boundaries are chosen to minimise migration between the regions. In the baseline simulation, around 85\% of the reconstructed events in each \memubb\ region belong to the corresponding region at particle level. The corresponding correction factors \cormqz\ which translate the measured gap fraction in the reconstruction-level \memubb\ region to the corresponding particle-level gap fractions \fmqzero\, are of similar size to \corqz, with the exception of the highest \memubb\ region, in which they reach about 1.1 at low \qzero.

\begin{table}[]
\centering
\footnotesize
\begin{tabular}{|l|c|c|c|c|c|c|c|c|}
\hline
$Q_0$&\multicolumn{2}{c|}{$|y|<0.8$}&\multicolumn{2}{c|}{$0.8<|y|<1.5$}&\multicolumn{2}{c|}{$1.5<|y|<2.1$}&\multicolumn{2}{c|}{$|y|<2.1$}\\
Generator&$\chi^2$ & $p$-value & $\chi^2$ & $p$-value & $\chi^2$    & $p$-value      & $\chi^2$  & $p$-value    \\ \hline\hline
\pps & 18.5 & 0.42 & 8.0 & 0.98 & 14.8 & 0.67 & 17.4 & 0.50 \\
\peight & 13.3 & 0.77 & 8.7 & 0.97 & 11.8 & 0.86 & 15.0 & 0.66 \\
\powhw & 24.4 & 0.14 & 10.2 & 0.93 & 19.6 & 0.36 & 30.8 & 0.03 \\
\powhws & 18.5 & 0.42 & 14.1 & 0.72 & 14.6 & 0.69 & 18.7 & 0.41 \\
\mcnlo+\pye & 33.7 & 0.01 & 26.2 & 0.09 & 18.0 & 0.45 & 58.6 & $3.4\times 10^{-6}$ \\
\mcnlohw & 14.1 & 0.72 & 8.5 & 0.97 & 18.9 & 0.40 & 8.4 & 0.97 \\
\mcnlohws & 22.2 & 0.22 & 25.7 & 0.11 & 20.3 & 0.32 & 44.6 & $4.7\times 10^{-4}$ \\
\sherpa\ v2.2 & 12.1 & 0.84 & 11.6 & 0.87 & 14.5 & 0.70 & 14.2 & 0.71 \\ \hline
\pps\ RadHi & 10.7 & 0.91 & 6.8 & 0.99 & 13.0 & 0.79 & 11.1 & 0.89 \\
\pps\ RadLo & 23.1 & 0.19 & 12.6 & 0.82 & 17.4 & 0.50 & 24.6 & 0.14 \\
\hline
\end{tabular}
\caption{Values of $\chi^2$ for the comparison of the measured gap fraction distributions with the predictions from various \ttbar\ generator configurations, for the four rapidity regions as a function of \qzero. The $\chi^2$ and $p$-values correspond to 18 degrees of freedom.}
\label{table:chirapQ0}
\end{table}
\begin{table}[]
\centering
\footnotesize
\begin{tabular}{|l|c|c|c|c|c|c|c|c|}
\hline
$Q_{\textrm{sum}}$&\multicolumn{2}{c|}{$|y|<0.8$}&\multicolumn{2}{c|}{$0.8<|y|<1.5$}&\multicolumn{2}{c|}{$1.5<|y|<2.1$}&\multicolumn{2}{c|}{$|y|<2.1$}\\
Generator&$\chi^2$ & $p$-value & $\chi^2$ & $p$-value & $\chi^2$    & $p$-value      & $\chi^2$  & $p$-value    \\ \hline\hline
\pps & 17.5 & 0.74 & 8.6 & 1.00 & 19.0 & 0.64 & 29.0 & 0.15 \\
\peight & 12.4 & 0.95 & 9.7 & 0.99 & 17.7 & 0.72 & 30.8 & 0.10 \\
\powhw & 17.4 & 0.74 & 11.5 & 0.97 & 21.9 & 0.46 & 34.6 & 0.04 \\ 
\powhws & 15.3 & 0.85 & 14.0 & 0.90 & 16.4 & 0.79 & 32.8 & 0.06 \\ 
\mcnlo+\pye & 30.5 & 0.11 & 22.1 & 0.45 & 20.7 & 0.54 & 55.7 & $9.4\times 10^{-5}$ \\
\mcnlohw & 17.8 & 0.72 & 10.4 & 0.98 & 18.4 & 0.68 & 23.6 & 0.37 \\
\mcnlohws & 21.2 & 0.51 & 27.3 & 0.20 & 24.4 & 0.32 & 54.7 & $1.3\times 10^{-4}$ \\
\sherpa\ v2.2 & 6.6 & 1.00 & 9.5 & 0.99 & 14.4 & 0.89 & 19.1 & 0.64 \\ \hline
\pps\ RadHi & 10.3 & 0.98 & 8.8 & 0.99 & 15.4 & 0.85 & 26.5 & 0.23 \\
\pps\ RadLo & 23.0 & 0.40 & 12.6 & 0.94 & 21.6 & 0.49 & 40.7 & $8.9\times 10^{-3}$ \\\hline
\end{tabular}
\caption{Values of $\chi^2$ for the comparison of the measured gap fraction distributions with the predictions from various \ttbar\ generator configurations, for the four rapidity regions as a function of \qsum. The $\chi^2$ and $p$-values correspond to 22 degrees of freedom.}
\label{table:chirapQsum}
\end{table}

\begin{table}[]
\centering
\footnotesize
\begin{tabular}{|l|c|c|c|c|c|c|c|c|}
\hline
$Q_0$&\multicolumn{2}{c|}{$m<300$~\gev}&\multicolumn{2}{c|}{$300<m<425$~\gev}&\multicolumn{2}{c|}{$425<m<600$~\gev}&\multicolumn{2}{c|}{$m>600$~\gev}\\
Generator&$\chi^2$ & $p$-value & $\chi^2$ & $p$-value & $\chi^2$    & $p$-value      & $\chi^2$  & $p$-value    \\ \hline\hline
\pps & 5.1 & 1.00 & 21.1 & 0.28 & 6.7 & 0.99 & 10.4 & 0.92 \\
\peight & 4.4 & 1.00 & 16.7 & 0.55 & 5.9 & 1.00 & 13.6 & 0.76 \\
\powhw & 14.6 & 0.69 & 19.8 & 0.35 & 5.0 & 1.00 & 15.0 & 0.66 \\
\powhws & 9.1 & 0.96 & 16.5 & 0.56 & 8.1 & 0.98 & 13.2 & 0.78 \\
\mcnlo+\pye & 27.5 & 0.07 & 27.6 & 0.07 & 20.4 & 0.31 & 17.8 & 0.47 \\
\mcnlohw & 4.6 & 1.00 & 18.2 & 0.44 & 14.0 & 0.73 & 18.1 & 0.45 \\
\mcnlohws & 20.9 & 0.29 & 23.6 & 0.17 & 10.9 & 0.90 & 15.9 & 0.60 \\
\sherpa\ v2.2 & 7.8 & 0.98 & 11.3 & 0.88 & 5.4 & 1.00 & 13.1 & 0.78 \\ \hline
\pps\ RadHi & 4.1 & 1.00 & 15.5 & 0.63 & 6.2 & 1.00 & 10.3 & 0.92 \\
\pps\ RadLo & 7.4 & 0.99 & 24.9 & 0.13 & 7.9 & 0.98 & 13.0 & 0.79 \\\hline
\end{tabular}
\caption{Measurements of $\chi^2$ comparing the measured gap fraction distributions  with predictions from various \ttbar\ generator configurations, for the four invariant mass \memubb\ regions as a function of \qzero. The $\chi^2$ and $p$-values correspond to 18 degrees of freedom.}
\label{table:chimassQ0}
\end{table}

\begin{table}[]
\centering
\footnotesize
\begin{tabular}{|l|c|c|c|c|c|c|c|c|}
\hline
$Q_{\textrm{sum}}$&\multicolumn{2}{c|}{$m<300$~\gev}&\multicolumn{2}{c|}{$300<m<425$~\gev}&\multicolumn{2}{c|}{$425<m<600$~\gev}&\multicolumn{2}{c|}{$m>600$~\gev}\\
Generator&$\chi^2$ & $p$-value & $\chi^2$ & $p$-value & $\chi^2$    & $p$-value      & $\chi^2$  & $p$-value    \\ \hline\hline
\pps & 18.3 & 0.69 & 27.7 & 0.18 & 19.3 & 0.63 & 11.6 & 0.97 \\
\peight & 18.3 & 0.69 & 28.2 & 0.17 & 17.1 & 0.76 & 12.1 & 0.96 \\
\powhw & 22.9 & 0.41 & 19.7 & 0.60 & 12.5 & 0.95 & 12.8 & 0.94 \\ 
\powhws & 23.7 & 0.36 & 23.2 & 0.39 & 17.5 & 0.73 & 11.1 & 0.97 \\ 
\mcnlo+\pye & 40.8 & 0.01 & 32.2 & 0.07 & 27.6 & 0.19 & 20.2 & 0.57 \\
\mcnlohw & 19.7 & 0.60 & 27.1 & 0.21 & 21.9 & 0.47 & 11.9 & 0.96 \\
\mcnlohws & 39.9 & 0.01 & 37.1 & 0.02 & 20.9 & 0.53 & 16.3 & 0.80 \\
\sherpa\ v2.2 & 16.3 & 0.80 & 18.0 & 0.71 & 14.6 & 0.88 & 9.3 & 0.99 \\ \hline
\pps\ RadHi & 17.4 & 0.74 & 21.4 & 0.50 & 16.0 & 0.82 & 9.8 & 0.99 \\
\pps\ RadLo & 22.2 & 0.45 & 33.4 & 0.06 & 21.3 & 0.05 & 15.9 & 0.82 \\\hline
\end{tabular}
\caption{Measurements of $\chi^2$ comparing the measured gap fraction distributions  with predictions from various \ttbar\ generator configurations, for the four invariant mass \memubb\ regions as a function of \qsum. The $\chi^2$ and $p$-values correspond to 22 degrees of freedom.}
\label{table:chimassQsum}
\end{table}

\begin{figure}[htp]
\centering
\subfloat[][]{\includegraphics[width=0.425\linewidth]{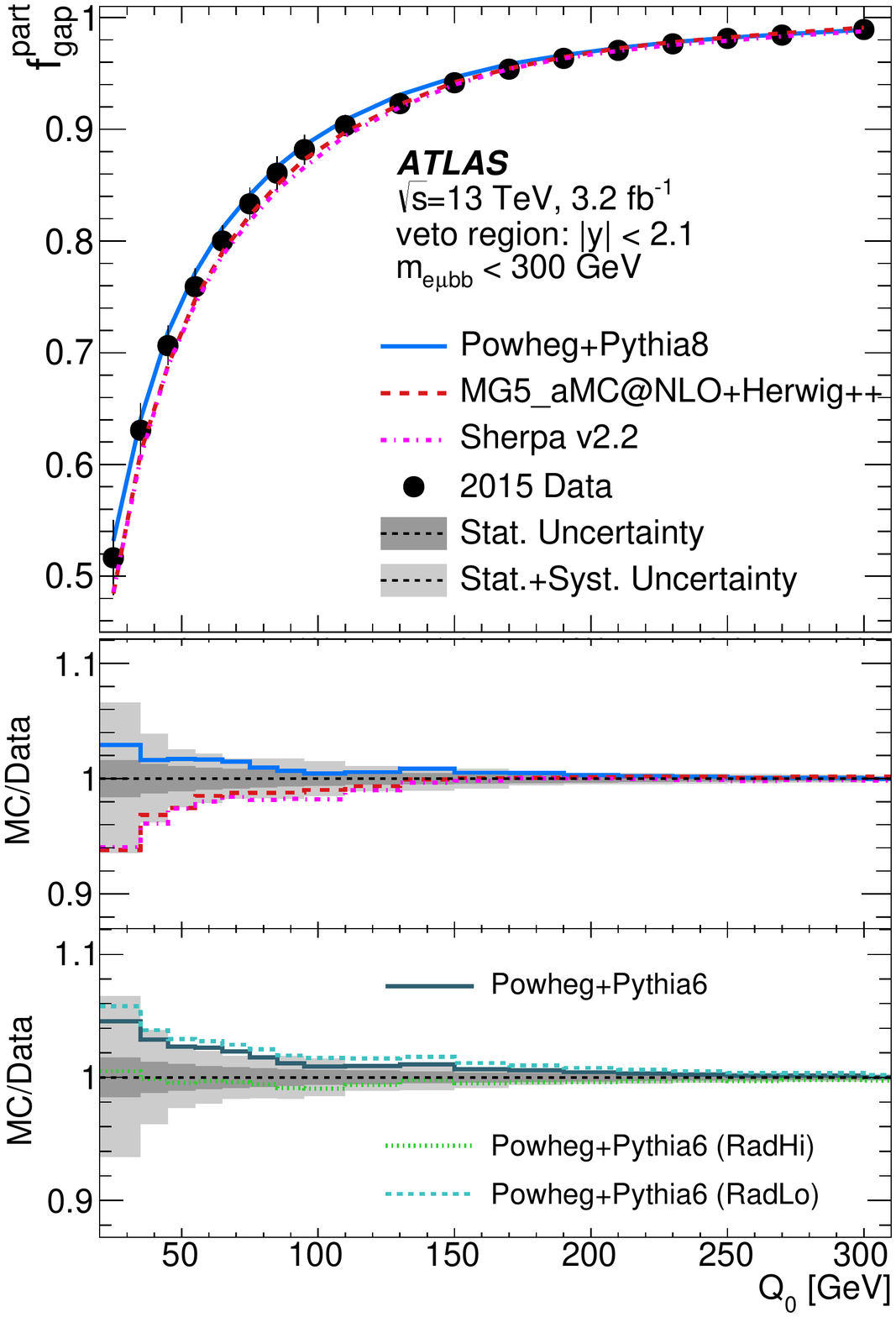}}
\subfloat[][]{\includegraphics[width=0.425\linewidth]{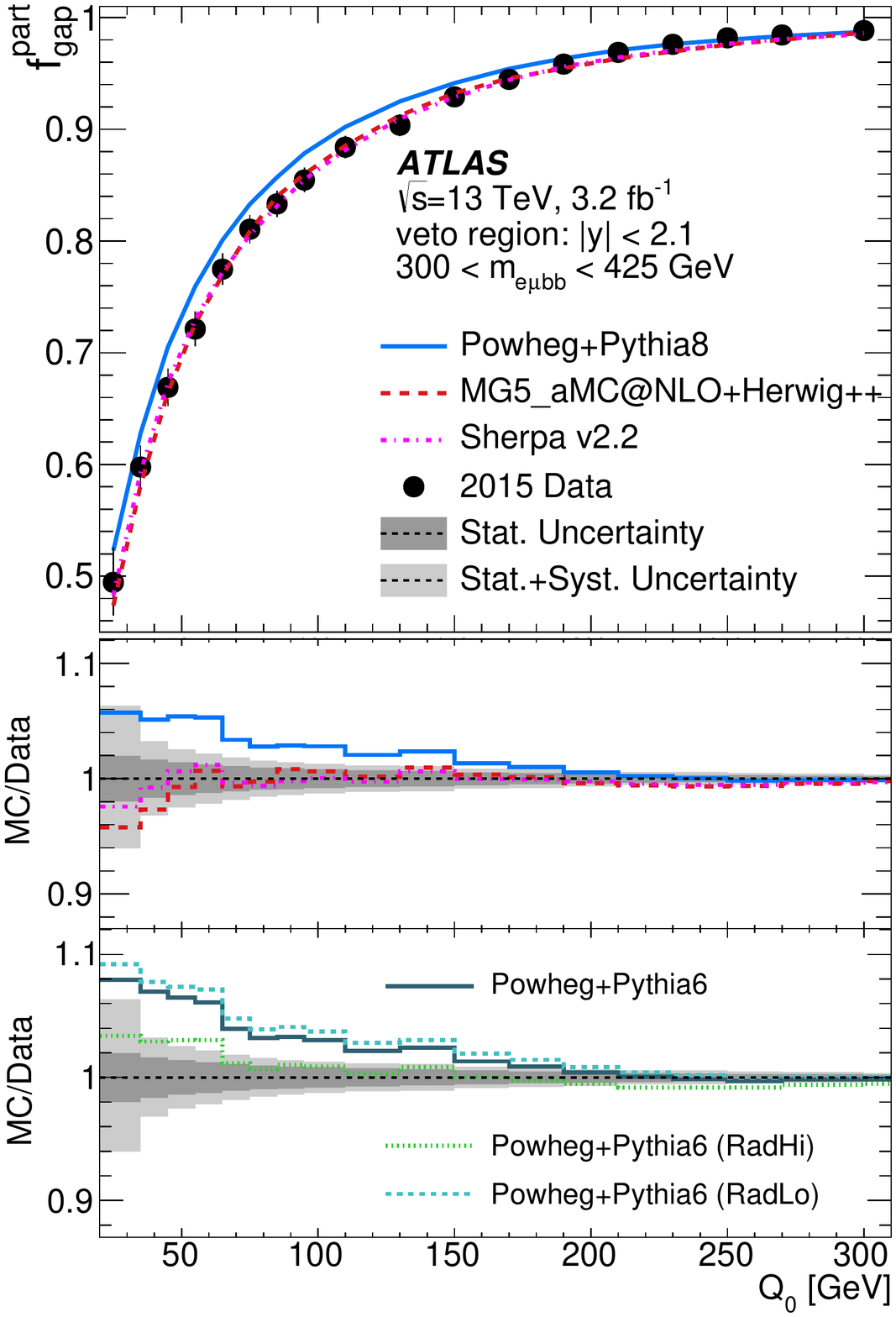}}\\
\subfloat[][]{\includegraphics[width=0.425\linewidth]{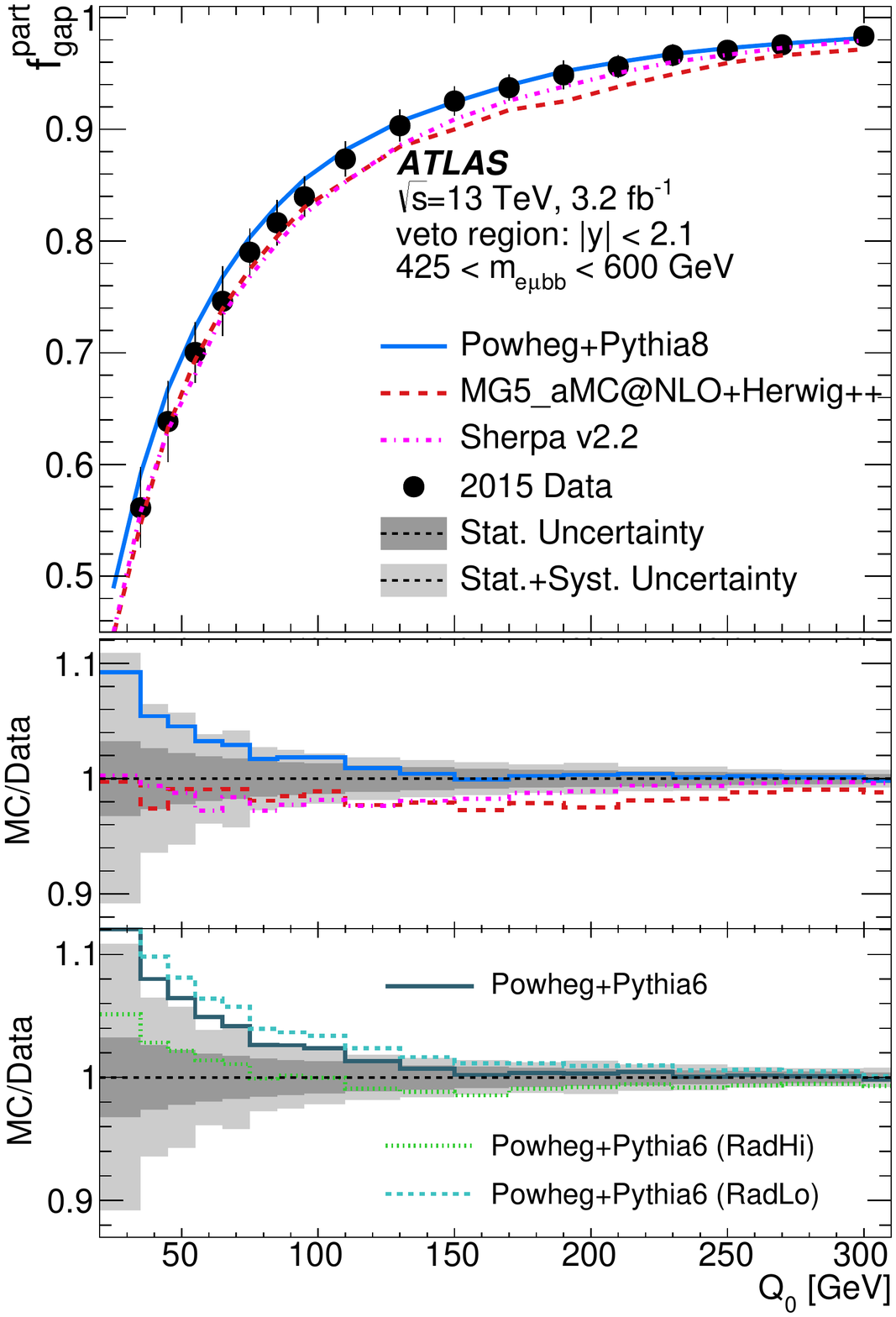}}
\subfloat[][]{\includegraphics[width=0.425\linewidth]{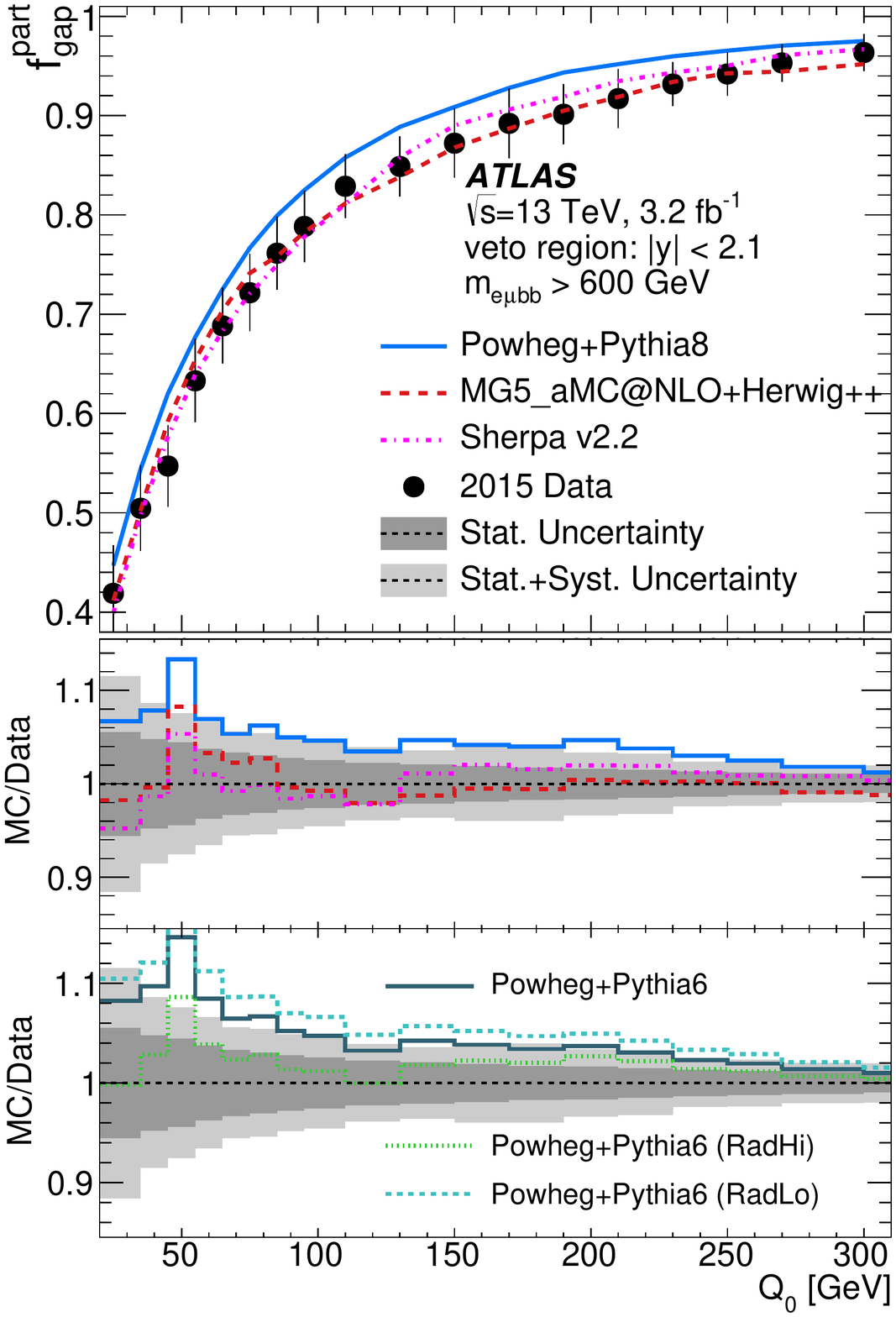}}
\caption{The measured gap fraction \fmqzero\ as a function of \qzero\ in the full central veto region $|y|<2.1$ for the invariant mass regions (a) $\memubb<300$~\gev, (b) $300~\gev\ <\memubb<425$~\gev, (c) $425~\gev\ <\memubb<600$~\gev\ and (d) $\memubb>600$~\gev. The data are shown by the points with error bars indicating the total uncertainty, and compared to the predictions from various \ttbar\ simulation samples shown as smooth curves. The lower plots show the ratio of predictions to data, with the data uncertainty indicated by the shaded band, and the \qzero\ thresholds corresponding to the left edges of the histogram bins, except for the first bin.}
\label{fig:gapfracpartmass}
\end{figure}

\begin{figure}[htp]
\centering
\subfloat[][]{\includegraphics[width=0.425\linewidth]{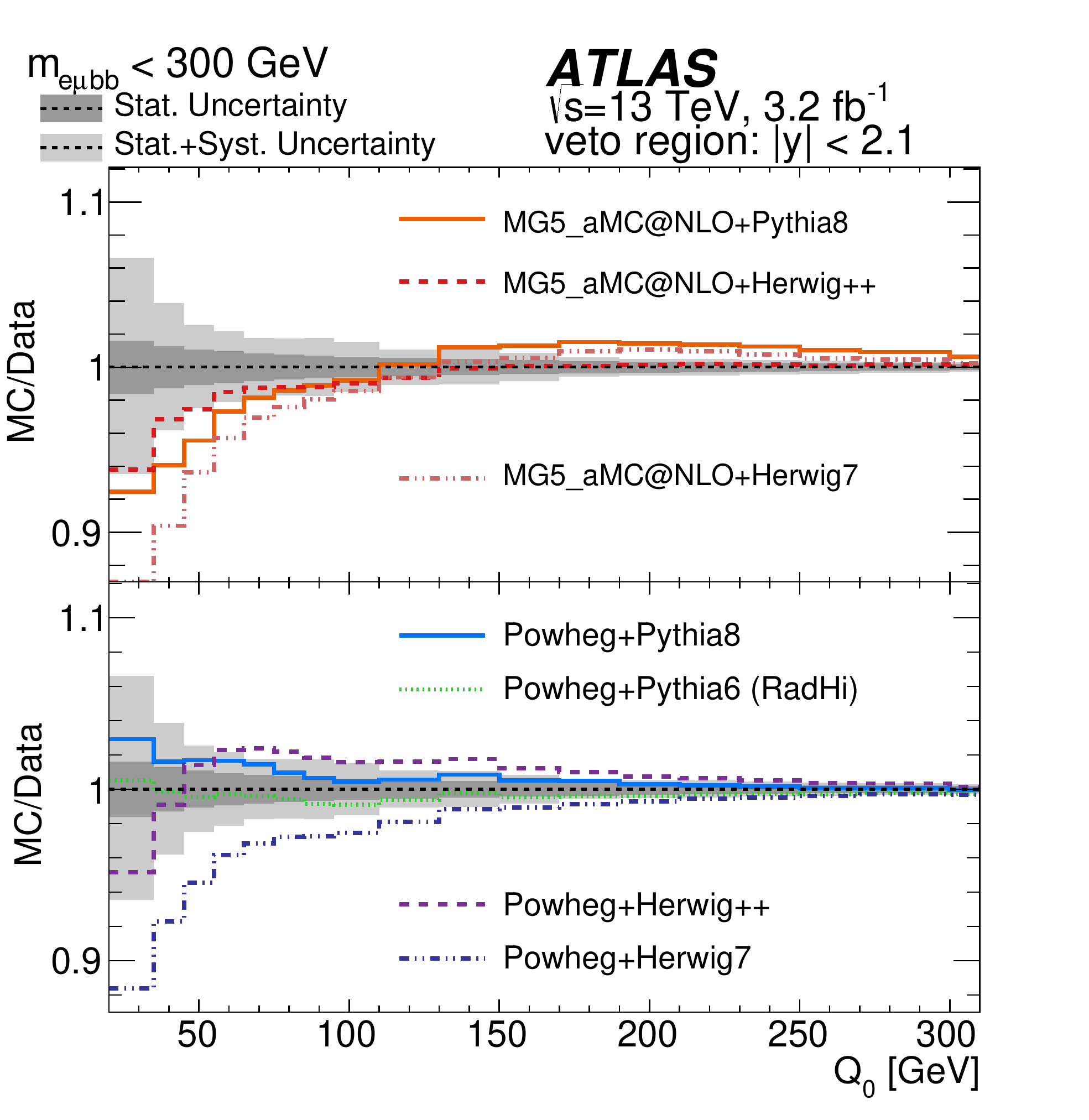}}
\subfloat[][]{\includegraphics[width=0.425\linewidth]{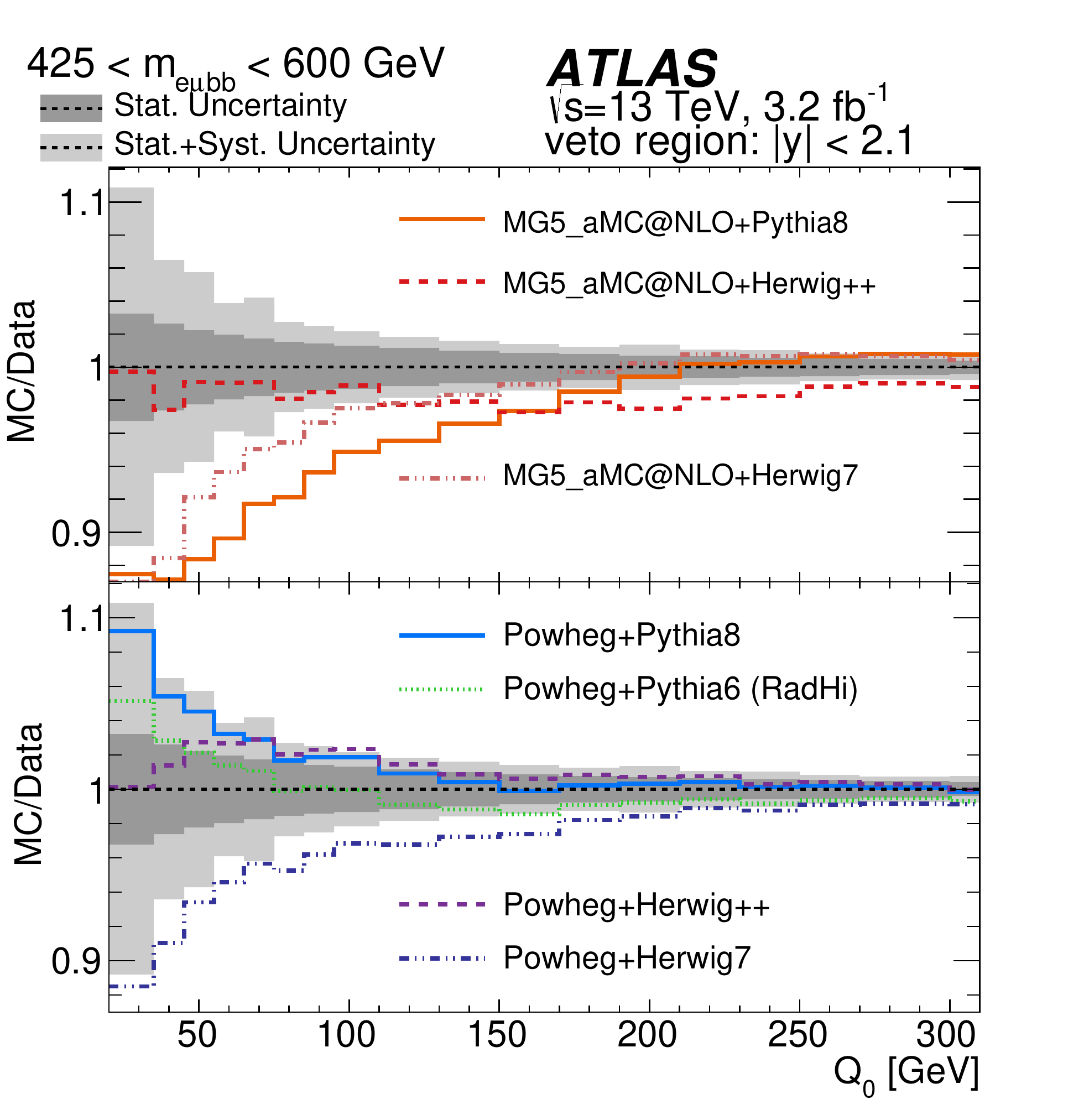}}
\caption{Ratios of prediction to data of the measured gap fraction \fqzero\ as a function of \qzero\ in the full central veto region $|y|<2.1$ for the invariant mass regions (a) $\memubb<300$~\gev\ and (b) $425$~\gev$<\memubb<600$~\gev. The predictions from various \ttbar\ simulation samples are shown as ratios to data, with the data uncertainty indicated by the shaded band, and the \qzero\ thresholds corresponding to the left edges of the histogram bins, except for the first bin.}
\label{fig:gapfracpartratmass}
\end{figure}

\begin{figure}[htp]
\centering
\includegraphics[width=0.6\linewidth]{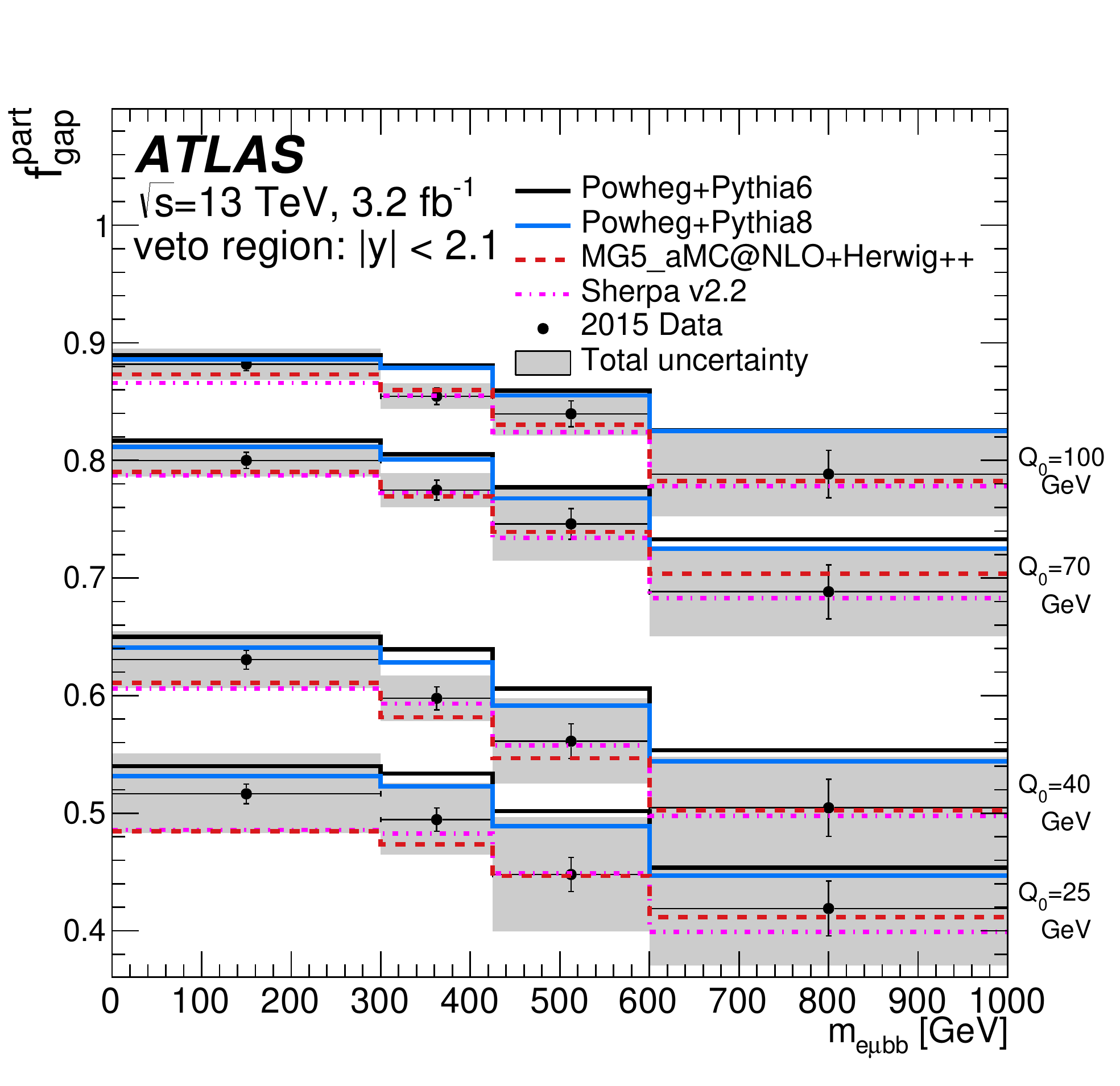}
\caption{The gap fraction measurement \fmqzero\ as a function of the invariant mass \memubb, for several different values of \qzero. The data are shown as points with error bars indicating the statistical uncertainties and shaded boxes the total uncertainties. The data are compared to the predictions from various \ttbar\ simulation samples.}
\label{fig:altmass}
\end{figure}

Figures \ref{fig:gapfracpartmass} and \ref{fig:gapfracpartratmass} show the measured gap fractions as a function of \qzero\ in the four \memubb\ regions in data, compared to the same set of predictions as shown in Figures\,\ref{fig:gapfracpart} and \ref{fig:gapfracpartrat}. Tables~\ref{table:chimassQ0} and~\ref{table:chimassQsum} give the $\chi^2$ and $p$-values taking into account bin-by-bin correlations of the gap fractions compared to the predictions from the different generators. Figure \ref{fig:altmass} gives an alternative presentation of the gap fraction \fmqzero\ as a function of \memubb\ for four different \qzero\ values. The level of agreement between the data and the various predictions is consistent with the results of the gap fraction in rapidity bins. Only in the lowest mass region the \peight\ prediction agrees very well, while \mcnlohw\ and \sherpa\ are at the lower edge of the uncertainties.

\label{sec:gapfraction}
\clearpage

\section{Conclusions}
Studies of additional jet activity, using differential cross-section and gap fraction measurements, are presented for dileptonic $t\bar{t}$ events identified by the presence of an opposite-sign $e\mu$ pair and at least two $b$-tagged jets. These measurements are performed using 3.2~\ifb\ of $\sqrt{s}=13$~\tev\ $pp$ collision data collected by the ATLAS detector in 2015 at the LHC. The measurements are corrected back to the particle level using full unfolding or correction factors, for well-defined fiducial regions and various \pt\ thresholds for the additional jets.

The different measurements are compared to various Monte Carlo predictions and give consistent results. Even though many predictions are within the uncertainty band of the measurements, the proper evaluation of the compatibility of the models, taking into account the bin-by-bin correlations within each measurement, revealed that \powpy\ (RadHi), \mcnlohw\ and \sherpa\ describe the data best for all observables. \powpy\ (RadLo), \mcnlopye\ and all predictions involving \hws\ do not describe the data well.

All studied combinations of the matrix element generators \mcnlo\ and \pow\ with the shower generators  \hw, \py 6 and \py 8 provided 
no systematic trend indicating that one of the matrix element generators describes the data better for all parton shower  generators. We also have no indication that one of the parton shower generators describes  the data systematically better for both matrix element generators.  This observation suggests that the matching between the parton shower and matrix element calculation plays an important role, and motivates further study in this area. The predictions of Sherpa which use NLO matrix elements consistently matched with up to four additional jets at LO  show  similar good agreement with data as the best of the \mcnlo\ and \pow\ predictions.
\label{sec:conclusion}

\section*{Acknowledgements}

We thank CERN for the very successful operation of the LHC, as well as the
support staff from our institutions without whom ATLAS could not be
operated efficiently.

We acknowledge the support of ANPCyT, Argentina; YerPhI, Armenia; ARC, Australia; BMWFW and FWF, Austria; ANAS, Azerbaijan; SSTC, Belarus; CNPq and FAPESP, Brazil; NSERC, NRC and CFI, Canada; CERN; CONICYT, Chile; CAS, MOST and NSFC, China; COLCIENCIAS, Colombia; MSMT CR, MPO CR and VSC CR, Czech Republic; DNRF and DNSRC, Denmark; IN2P3-CNRS, CEA-DSM/IRFU, France; GNSF, Georgia; BMBF, HGF, and MPG, Germany; GSRT, Greece; RGC, Hong Kong SAR, China; ISF, I-CORE and Benoziyo Center, Israel; INFN, Italy; MEXT and JSPS, Japan; CNRST, Morocco; FOM and NWO, Netherlands; RCN, Norway; MNiSW and NCN, Poland; FCT, Portugal; MNE/IFA, Romania; MES of Russia and NRC KI, Russian Federation; JINR; MESTD, Serbia; MSSR, Slovakia; ARRS and MIZ\v{S}, Slovenia; DST/NRF, South Africa; MINECO, Spain; SRC and Wallenberg Foundation, Sweden; SERI, SNSF and Cantons of Bern and Geneva, Switzerland; MOST, Taiwan; TAEK, Turkey; STFC, United Kingdom; DOE and NSF, United States of America. In addition, individual groups and members have received support from BCKDF, the Canada Council, CANARIE, CRC, Compute Canada, FQRNT, and the Ontario Innovation Trust, Canada; EPLANET, ERC, ERDF, FP7, Horizon 2020 and Marie Sk{\l}odowska-Curie Actions, European Union; Investissements d'Avenir Labex and Idex, ANR, R{\'e}gion Auvergne and Fondation Partager le Savoir, France; DFG and AvH Foundation, Germany; Herakleitos, Thales and Aristeia programmes co-financed by EU-ESF and the Greek NSRF; BSF, GIF and Minerva, Israel; BRF, Norway; CERCA Programme Generalitat de Catalunya, Generalitat Valenciana, Spain; the Royal Society and Leverhulme Trust, United Kingdom.

The crucial computing support from all WLCG partners is acknowledged gratefully, in particular from CERN, the ATLAS Tier-1 facilities at TRIUMF (Canada), NDGF (Denmark, Norway, Sweden), CC-IN2P3 (France), KIT/GridKA (Germany), INFN-CNAF (Italy), NL-T1 (Netherlands), PIC (Spain), ASGC (Taiwan), RAL (UK) and BNL (USA), the Tier-2 facilities worldwide and large non-WLCG resource providers. Major contributors of computing resources are listed in Ref.~\cite{ATL-GEN-PUB-2016-002}.

\printbibliography

\clearpage
\FloatBarrier
\newpage
\begin{flushleft}
{\Large The ATLAS Collaboration}

\bigskip

M.~Aaboud$^\textrm{\scriptsize 137d}$,
G.~Aad$^\textrm{\scriptsize 88}$,
B.~Abbott$^\textrm{\scriptsize 115}$,
J.~Abdallah$^\textrm{\scriptsize 8}$,
O.~Abdinov$^\textrm{\scriptsize 12}$,
B.~Abeloos$^\textrm{\scriptsize 119}$,
O.S.~AbouZeid$^\textrm{\scriptsize 139}$,
N.L.~Abraham$^\textrm{\scriptsize 151}$,
H.~Abramowicz$^\textrm{\scriptsize 155}$,
H.~Abreu$^\textrm{\scriptsize 154}$
% ...
\bigskip
\\
% ...
$^{8}$ Department of Physics, The University of Texas at Arlington, Arlington TX, United States of America\\
% ...
$^{12}$ Institute of Physics, Azerbaijan Academy of Sciences, Baku, Azerbaijan\\
% ...
$^{88}$ CPPM, Aix-Marseille Universit{\'e} and CNRS/IN2P3, Marseille, France\\
% ...
$^{115}$ Homer L. Dodge Department of Physics and Astronomy, University of Oklahoma, Norman OK, United States of America\\
% ...
$^{119}$ LAL, Univ. Paris-Sud, CNRS/IN2P3, Universit{\'e} Paris-Saclay, Orsay, France\\
% ...
$^{137}$ $^{(a)}$ Facult{\'e} des Sciences Ain Chock, R{\'e}seau Universitaire de Physique des Hautes Energies - Universit{\'e} Hassan II, Casablanca; $^{(b)}$ Centre National de l'Energie des Sciences Techniques Nucleaires, Rabat; $^{(c)}$ Facult{\'e} des Sciences Semlalia, Universit{\'e} Cadi Ayyad, LPHEA-Marrakech; $^{(d)}$ Facult{\'e} des Sciences, Universit{\'e} Mohamed Premier and LPTPM, Oujda; $^{(e)}$ Facult{\'e} des sciences, Universit{\'e} Mohammed V, Rabat, Morocco\\
% ...
$^{139}$ Santa Cruz Institute for Particle Physics, University of California Santa Cruz, Santa Cruz CA, United States of America\\
% ...
$^{151}$ Department of Physics and Astronomy, University of Sussex, Brighton, United Kingdom\\
% ...
$^{154}$ Department of Physics, Technion: Israel Institute of Technology, Haifa, Israel\\
$^{155}$ Raymond and Beverly Sackler School of Physics and Astronomy, Tel Aviv University, Tel Aviv, Israel\\
% ...
$^{a}$ Also at Department of Physics, King's College London, London, United Kingdom\\
$^{b}$ Also at Institute of Physics, Azerbaijan Academy of Sciences, Baku, Azerbaijan\\
$^{c}$ Also at Novosibirsk State University, Novosibirsk, Russia\\
$^{d}$ Also at TRIUMF, Vancouver BC, Canada\\
$^{e}$ Also at Department of Physics {\&} Astronomy, University of Louisville, Louisville, KY, United States of America\\
% ...
\end{flushleft}


\end{document}